\documentclass[prb,aps,amssymb,twocolumn,nofootinbib,superscriptaddress]{revtex4}
\usepackage{amsmath}
\usepackage{amssymb}
\usepackage{amsthm}
\usepackage{amsfonts}
\usepackage{listings}
\usepackage[hidelinks]{hyperref}
\usepackage{enumerate}
\usepackage{latexsym}

\usepackage{float} 

\usepackage{psfrag}

\usepackage{bm}
\usepackage{graphicx}
\usepackage[caption=false]{subfig}
\usepackage{color}

\newcommand{\beq}{\begin{equation}}
\newcommand{\eneq}{\end{equation}}

\input{epsf}

\newcommand{\bpm}{\begin{pmatrix}}
\newcommand{\epm}{\end{pmatrix}}

\newcommand{\bal}{\begin{align}}

\DeclareMathOperator{\sgn}{sgn}

\begin{document}


\title{Wallpaper Fermions and the Nonsymmorphic Dirac Insulator}

\author{Benjamin J. Wieder$^\dag$}
\thanks{These authors contributed equally to this work.}
\affiliation{Department of Physics,
Princeton University,
Princeton, NJ 08544, USA
            }
\affiliation{Nordita, Center for Quantum Materials, KTH Royal Institute of Technology and Stockholm University,
Roslagstullsbacken 23, SE-106 91 Stockholm, Sweden
}
\affiliation{Department of Physics and Astronomy, University of Pennsylvania, Philadelphia, PA 19104--6323, USA}
            
\author{Barry Bradlyn}
\thanks{These authors contributed equally to this work.}
\affiliation{Princeton Center for Theoretical Science, Princeton University, Princeton, NJ 08544, USA}

\author{Zhijun Wang}
\thanks{These authors contributed equally to this work.}
\affiliation{Department of Physics,
Princeton University,
Princeton, NJ 08544, USA
            }

\author{Jennifer Cano}
\thanks{These authors contributed equally to this work.}
\affiliation{Princeton Center for Theoretical Science, Princeton University, Princeton, NJ 08544, USA}

\author{Youngkuk Kim}
\affiliation{Department of Chemistry, University of Pennsylvania, Philadelphia, Pennsylvania 19104--6323, USA}
\affiliation{Department of Physics, Sungkyunkwan University (SKKU), Suwon 16419,  Korea}

\author{Hyeong-Seok D. Kim}
\affiliation{Department of Physics and Astronomy, University of Pennsylvania, Philadelphia, PA 19104--6323, USA}
\affiliation{Department of Chemistry, University of Pennsylvania, Philadelphia, Pennsylvania 19104--6323, USA}

\author{Andrew M. Rappe}
\affiliation{Department of Chemistry, University of Pennsylvania, Philadelphia, Pennsylvania 19104--6323, USA}

\author{C. L. Kane$^\dag$}
\affiliation{Department of Physics and Astronomy, University of Pennsylvania, Philadelphia, PA 19104--6323, USA}

\author{B. Andrei Bernevig$^\dag$}
\affiliation{Department of Physics,
Princeton University,
Princeton, NJ 08544, USA
	}
\affiliation{Dahlem Center for Complex Quantum Systems and Fachbereich Physik,
Freie Universit{\"a}t Berlin, Arnimallee 14, 14195 Berlin, Germany
	}
\affiliation{Max Planck Institute of Microstructure Physics, 
06120 Halle, Germany
}

\date{\today}
\begin{abstract}
Recent developments in the relationship between bulk topology and surface crystalline symmetries have led to the discovery of materials whose gapless surface states are protected by crystal symmetries, such as mirror topological crystalline insulators and nonsymmorphic hourglass fermion insulators.  In fact, there exists only a very limited set of possible surface crystal symmetries, captured by the 17 ``wallpaper groups.''  Here, we show that a consideration of symmetry-allowed band degeneracies in the wallpaper groups can be used to understand previous topological crystalline insulators, as well as to predict phenomenologically new examples.  In particular, the two wallpaper groups with multiple glide lines, $pgg$ and $p4g$, allow for a new topological insulating phase, whose surface spectrum consists of only a single, \emph{fourfold-degenerate}, true Dirac fermion.  Like the surface state of a conventional topological insulator, the surface Dirac fermion in this ``nonsymmorphic Dirac insulator'' provides a theoretical exception to a fermion doubling theorem. Unlike the surface state of a conventional topological insulator, it can be gapped into topologically distinct surface regions while keeping time-reversal symmetry, giving rise to a network of topological surface quantum spin Hall domain walls with Luttinger liquid character.  We report the theoretical discovery of new topological crystalline phases in the A$_2$B$_3$ family of materials in space group 127 ($P4/mbm$), with Sr$_2$Pb$_3$ hosting the new topological surface Dirac fermion.  Furthermore, $(100)$-strained Au$_2$Y$_3$  and Hg$_2$Sr$_3$ host related topological surface hourglass fermions.  We also report the presence of this new topological hourglass phase in Ba$_5$In$_2$Sb$_6$ in space group 55 ($Pbam$).  For orthorhombic, tetragonal, and cubic crystals with two perpendicular glides and strong spin-orbit coupling, we catalog all possible time-reversal-symmetric bulk topological phases by performing an analysis of the allowed non-abelian Wilson loop connectivities, and provide topological invariants to distinguish them.  Finally, we show how in a particular limit of these systems, the crystalline phases reduce to copies of the Su-Schrieffer-Heeger model.
\end{abstract}

\maketitle

\section{Introduction} 
\label{intro}

Topological phases stabilized by crystal symmetries have already proven to be both a theoretically and an experimentally rich set of systems.  The first class of these proposed materials, rotation or mirror topological crystalline insulators (TCIs), host surface fermions protected by the projection of a bulk mirror plane or rotation axis onto a particular surface\cite{Teo2008,Fu2011,Kin15p086802}.  They have been observed in SnTe~\cite{Hsieh2012,Tanaka2012} and related compounds~\cite{Dziawa2012,Xu2012}.  Recent efforts to expand this analysis to nonsymmorphic systems with surface glide mirrors -- operations composed of a mirror and a half-lattice translation -- have yielded additional exotic free fermion topological phases, which can exhibit the so-called surface gapless ``hourglass fermions,'' and the glide spin Hall effect~\cite{Liu2014,Wang16,Alexandradinata16,Shiozaki16,Chang2016}.  The theoretical proposal of~\onlinecite{Wang16,Alexandradinata16} has recently also seen incipient experimental support~\cite{HourglassObserve}.

In addition, topological insulators (TIs) -- crystalline or otherwise -- provide exceptions to fermion doubling theorems. These theorems impose fundamental bounds on phenomena in condensed matter physics.  For example, in 2D, a single Kramers degeneracy in momentum space must always have another partner Kramers crossing elsewhere in the Brillouin Zone (BZ), otherwise the Berry phase of a loop enclosing the degeneracy suffers from ambiguity~\cite{Haldane2004}.  The discovery of the topological insulator provided the first exception to this theorem: in these systems 2D Kramers pairs are allowed to be isolated on a single 2D surface because they are connected to a 3D topological insulating bulk and have their partners on the opposite surface~\cite{Fu07,Moore07}.

Higher-fold-degenerate bulk fermions, such as Dirac points, which are stabilized by crystal symmetry, may come with their own fermion doubling theorems~\cite{SteveDirac,JuliaDirac,DDP,NewFermions}.  As noted in~\onlinecite{Steve2D}, a single fourfold-degenerate Dirac fermion cannot be stabilized by 2D crystal symmetries as the only nodal feature at a given energy; it must always have at least one partner or accompanying hourglass nodal points.   This is because a single Dirac point in 2D represents the quantum critical point (QCP) separating a trivial insulator (NI) from a topological insulator.  Shown in more detail in Appendix~\ref{sec:Diracdouble}, stabilizing just one of these Dirac points with crystal symmetries would therefore force the broken-symmetry NI and TI phases to be related by just a unitary transformation, violating their $\mathbb{Z}_{2}$ topological inequivalence.  In this manuscript, we report a new class of symmetry-protected topological materials which, like the topological insulator before it, circumvents this restriction by placing a single, stable Dirac point \emph{on the surface of a 3D material}.

To realize this, the crucial requirement is that the surface preserves \emph{multiple} nonsymmorphic crystal symmetries (Appendix~\ref{sec:Diracdouble}).  Until now, most attention has been paid to crystal systems with surfaces that preserve only a single glide mirror.  However, two of the 17 two-dimensional surface symmetry groups, called wallpaper groups, host two intersecting glide lines~\cite{ConwayWallpaper}.   As we show in Appendix~\ref{sec:wallsyms}, the algebra of the two glides requires that bands appear with fourfold degeneracy at a single time-reversal-invariant momentum (TRIM) at the edge of the BZ.

In this work, we study the non-interacting topological phases allowed in bulk crystals with surfaces invariant under the symmetries of these two wallpaper groups, $pgg$ and $p4g$.  We show that, in addition to generalizations of the hourglass fermions introduced in~\onlinecite{Wang16}, they host a novel topological phase characterized by a single, symmetry-enforced \emph{fourfold} Dirac surface fermion, i.e., \emph{twice} the degeneracy of a traditional topological insulator surface state.  This Dirac fermion is nonsymmorphic symmetry-pinned to the QCP between a TI and an NI, allowing for controllable topological phase transitions of the 2D surface under spin-independent glide-breaking strain. 

We classify the allowed topological phases for orthorhombic, tetragonal, and cubic crystals with two perpendicular glides that are preserved on a single surface by considering the possible connectivities of the non-abelian Wilson loop eigenvalues~\cite{Fu06,Ryu10,Soluyanov11,Yu11,Taherinejad14,Alexandradinata14,Alexandradinata16}.  We demonstrate that these systems allow for three classes of topological phases: an hourglass phase with broken $C_{4z}$ symmetry, a previously uncharacterized ``double-glide spin Hall'' phase, and the novel topological nonsymmorphic Dirac insulating phase mentioned above.  We present topological invariants to distinguish these phases and use these invariants to predict material realizations of our new phases.  Using density functional theory (DFT) to calculate bulk Wilson loop eigenvalues and surface Green's functions, we report the existence of the nonsymmorphic Dirac insulating phase in Sr$_2$Pb$_3$ in space group (SG) 127.  We also report the discovery of a related topological hourglass fermion phase in the narrow-gap material Ba$_5$In$_2$Sb$_6$ in SG 55.  In Appendix~\ref{sec:YKmatdeets}, we show that two additional members of the SG 127 A$_2$B$_3$ family of materials, Au$_2$Y$_3$ and Hg$_2$Sr$_3$, can also realize this topological hourglass phase under $(100)$-strain, giving unique promise for strain-engineering topological phase transitions in these materials.  Finally, we show in Appendix~\ref{sec:myTB} how these crystalline phases reduce in a particular limit to copies of the Su-Schrieffer-Heeger (SSH) model~\cite{SSH}.
 
\section{Wallpaper Groups $pgg$ and $p4g$}

\begin{figure}[t]
\includegraphics[width=\linewidth]{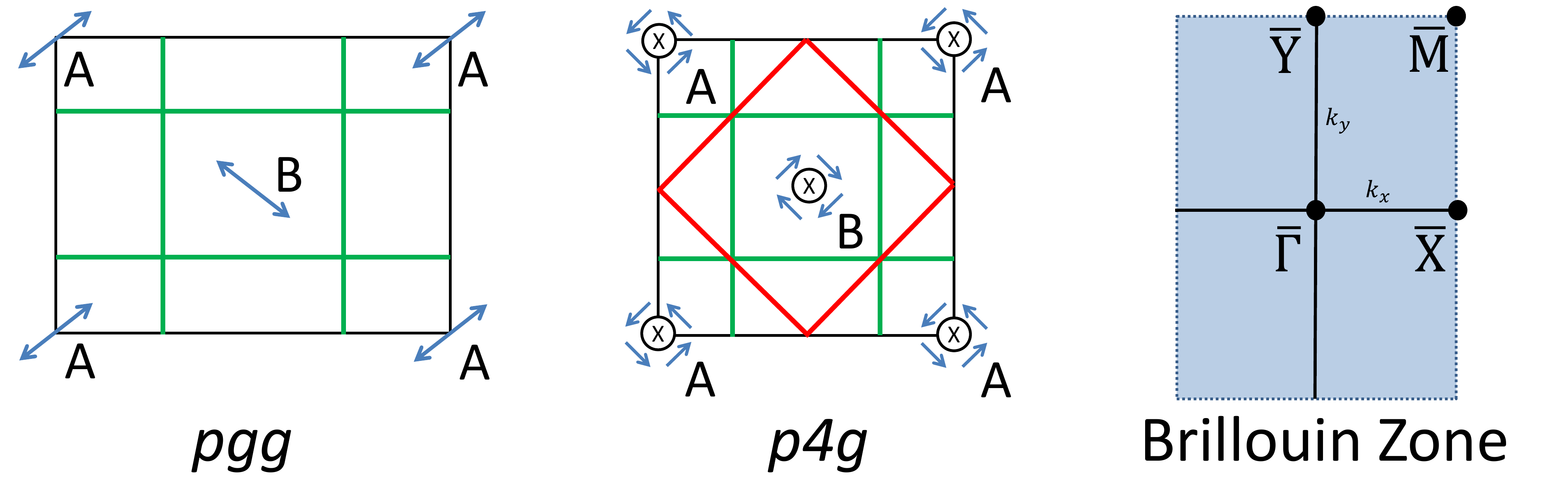}
\caption{Unit cells and Brillouin Zone (BZ) for two-site realizations of wallpaper groups $pgg$ and $p4g$, the only two wallpaper groups with multiple nonsymmorphic symmetries.  The A and B sites are characterized by $\mathcal{T}$-symmetric internal degrees of freedom (blue arrows) that are transformed under crystal symmetry operations. These correspond physically to nonmagnetic properties which transform as vectors, such as atom displacements or local electric dipole moments.  Glide lines (green) exchange the sublattices through fractional lattice translations.  In $p4g$, there is an extra $C_{4}$ symmetry ($\otimes$) about the surface normal with axes located on the sites.  The combination of this $C_{4}$ and the glides produces additional diagonal symmorphic mirror lines (red) in $p4g$.  In the BZ of $p4g$, $C_{4}$ relates $\bar{Y}$ to $\bar{X}$.}  
\label{fig:combinedWall}
\end{figure}

The surface of a nonmagnetic crystal is itself a lower-dimensional crystal, which preserves a subset of the bulk crystal symmetries, and all 2D nonmagnetic surfaces are geometrically constrained to be invariant under the action of the 17 wallpaper groups.  The set of spatial wallpaper group symmetries is restricted to those 3D space group symmetries which preserve the surface normal vector: rotations about that vector and in-plane lattice translations, mirror reflections, and glide reflections.  Surface band features will therefore be constrained by the irreducible corepresentations of these symmetries and their momentum-space compatibility relations~\cite{PSUTCI,QuantumChemistry} (Appendix~\ref{sec:wallsyms}).  Focusing on the 17 wallpaper groups which describe the surfaces of 3D crystals with strong spin-orbit coupling (SOC) and time-reversal symmetry, we designate all topological 2D surface nodes formed from symmetry-enforced band degeneracies ``wallpaper fermions.''  In the language of group theory, wallpaper fermions are therefore fully captured by the irreducible corepresentations of the wallpaper groups, such that, for example, the fourfold-rotation-protected quadratic topological node introduced in~\onlinecite{Fu2011} is a ``spinless wallpaper fermion,'' whereas the magnetic twofold surface degeneracy in~\onlinecite{FangFuNSandC2}, for which both bands have the \emph{same} irreducible representations and are instead prevented from gapping by an additional rotation- and time-reversal-enforced ($C_{2}\times\mathcal{T}$) 1D loop invariant, is not a wallpaper fermion. 

Though other works have focused on the mathematical classification of topological phases protected by wallpaper group symmetries~\cite{ShiozakiWallpaper,LiangPointGroup}, we here, in addition to providing a further topological classification, search for topological insulating phases with \emph{phenomenologically distinct} surface states.  For the symmorphic wallpaper groups, topological rearrangements of the minimal surface band connectivities, recently enumerated for 3D bulk systems in~\onlinecite{QuantumChemistry}, allow for only quantum spin Hall (QSH) phases and rotational variants of the mirror TCI phases, which exhibit twofold-degenerate free fermions along high-symmetry lines in the surface BZ~\cite{Hsieh2012,Tanaka2012,Dziawa2012,Xu2012}.

For the four wallpaper groups with nonsymmorphic glide lines ($pg$, $pmg$, $pgg$, and $p4g$), this picture is enriched.  Even in 2D, glide symmetries require that groups of four or more bands be connected, an effect which frequently manifests itself in hourglass-like band structures~\cite{Steve2D,WPVZ,WiederLayers}.  For the wallpaper groups with only a single glide line, $pg$ and $pmg$, surface bands can be connected in topologically-nontrivial patterns of interlocking hourglasses~\cite{Wang16,Alexandradinata16}, or  in ``M\"{o}bius strip'' connectivities if time-reversal-symmetry is relaxed~\cite{SSGMobius,FangFuNSandC2}, both of which exhibit twofold-degenerate surface fermions along some of the momentum-space glide lines.  In the remaining two wallpaper groups with multiple glide symmetries, $pgg$ and $p4g$, higher-degenerate wallpaper fermions are uniquely allowed.

We consider a $z$-normal surface with glides $g_{x,y} \equiv \{m_{x,y}|\frac{1}{2}\frac{1}{2}0\}$, i.e., a mirror reflection through the $x,y$-axis followed by a translation of half a lattice vector in the $\hat{x}$ and $\hat{y}$ directions (Fig.~\ref{fig:combinedWall}).
When spin-orbit coupling is present, $g_{x,y}^{2}=-e^{ik_{y,x}}$.  At the corner point, $\bar{M}$ of the surface BZ ($k_{x}=k_{y}=\pi$),  $g_{x}^{2}=g_{y}^2=+1$, and $\{g_{x},g_{y}\}=0$.  When combined with time-reversal, $\mathcal{T}^2=-1$, this symmetry algebra requires that all states at $\bar{M}$ are fourfold-degenerate.  Furthermore, wallpaper groups with two glides are the only nonmagnetic surface groups that admit this algebra, and therefore the only surface groups that can host protected fourfold degeneracies on strong-SOC crystals~\cite{WiederLayers}.  The examination of symmetry-allowed terms reveals that fourfold points in these wallpaper groups will be linearly-dispersing (Appendix~\ref{sec:wallsyms}), rendering them true surface Dirac fermions, more closely related by symmetry algebra and quantum criticality to the bulk nodes in nonsymmorphic 3D Dirac semimetals~\cite{Fu07,SteveDirac,JuliaDirac} than to the surface states of a conventional TI.  In Appendix~\ref{sec:wallsyms}, we provide proofs relating this algebra to Dirac degeneracy and dispersion.

\begin{figure}[t]
\includegraphics[width=\linewidth]{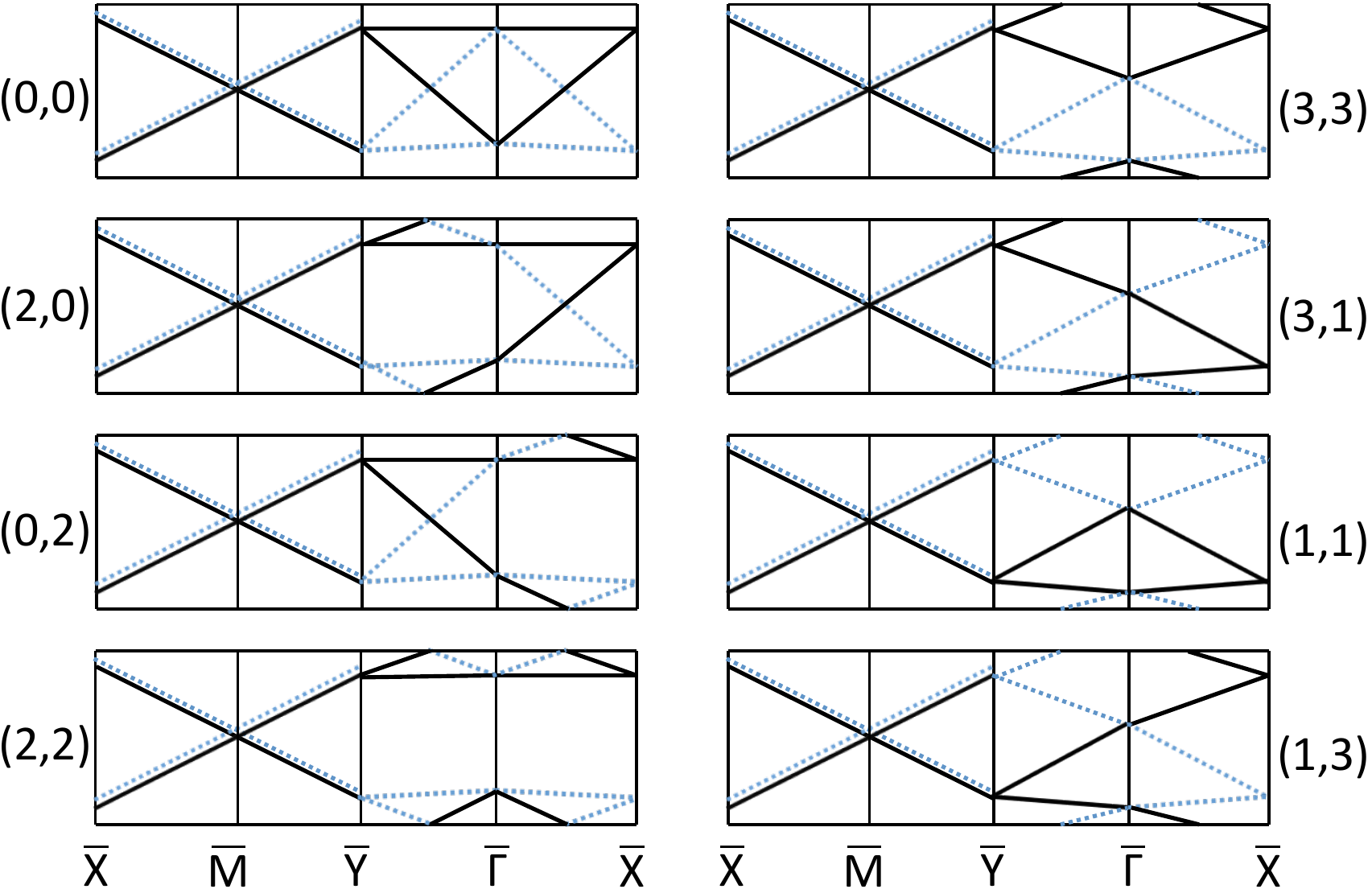}
\caption{The eight topologically distinct Wilson band connectivities for bulk insulators with crystal surfaces which preserve 2D glide reflection on the projections of two orthogonal bulk glide planes. Each band structure is labeled by its two $\mathbb{Z}_4$ indices, $(\chi_x,\chi_y)$, subject to the constraint that $\chi_x+\chi_y=0\mod 2$.  Under the imposition of $C_{4z}$ symmetry in wallpaper group $p4g$, connectivities are excluded for which $\chi_{x}\neq\chi_{y}$.  Solid black (dashed blue) lines in the regions $\bar{X}\bar{M}$ and $\bar{\Gamma}\bar{Y}$ indicate bands with $g_x$ eigenvalue $\pm ie^{ik_y/2}$, while the solid (dashed) lines in the regions $\bar{Y}\bar{M}$ and $\bar{\Gamma}\bar{X}$ indicate bands with $g_y$ eigenvalue $\pm ie^{ik_x/2}$. When bulk inversion symmetry is present, the spectra will be particle-hole symmetric.  Bands along $\bar{X}\bar{M}\bar{Y}$ are doubly-degenerate and meet at $\bar{M}$ in a fourfold-degenerate point, and bands along $\bar{Y}\bar{\Gamma}\bar{X}$ display either the hourglass (left column) or ``double-glide spin Hall'' (right column) flows. The (2,0) and (0,2) phases are relatives of the hourglass topologies proposed in~\onlinecite{Wang16,Alexandradinata16}.  The novel (2,2) nonsymmorphic Dirac insulating phase can host a surface state consisting of a single, \emph{fourfold-degenerate} Dirac fermion.}
\label{fig:SurfaceBands}
\end{figure}

For bulk insulators, the glide-preserving bulk topological phase and, consequently, $z$-normal surface states, can be determined without imposing a surface by classifying the allowed connectivities of the $z$-projection Wilson loop holonomy matrix~\cite{Fidkowski2011,Alexandradinata14,Alexandradinata16}, a bulk quantity defined by:
\begin{align}
 \left[ \mathcal{W}_{(k_x,k_y,k_{z0})}\right]_{ij}  & \equiv  \langle u^i(k_x,k_y,k_{z0}+2\pi) |  \nonumber \\
& \hat{\Pi}(k_x,k_y,k_{z0})  |u^j(k_x,k_y,k_{z0} )\rangle,  
\label{eq:wilsoncont}
\end{align}
where we have defined the product of projectors,
\begin{align}
\hat{\Pi}(k_x,k_y,k_{z}) \equiv\ & \hat{\mathcal{P}}(k_x,k_y, k_{z} + 2\pi ) \nonumber \\
 \hat{\mathcal{P}}(k_x,k_y, k_{z}+&\frac{2\pi (N-1)}{N})\cdots \hat{\mathcal{P}}(k_x,k_y,k_z+\frac{2\pi }{N}) 
\end{align}
and $\hat{\mathcal{P}}(\mathbf{k})$ is the projector onto the occupied bands at $\mathbf{k}$.  The rows and columns of $\mathcal{W}$ correspond to filled bands, where $|u^j(\mathbf{k})\rangle$ is the cell-periodic part of the Bloch wavefunction at momentum $k$ with band index $j$. The eigenvalues of $\mathcal{W}$ are gauge invariant and of the form $e^{i\theta(k_x,k_y)}$.  As detailed in Appendices~\ref{sec:TBnot} and~\ref{sec:Wilson}, the Wilson bands inherit the symmetries of the $z$-normal wallpaper group, and therefore must also exhibit the required degeneracy multiplets of wallpaper groups $pgg$ and $p4g$.  In particular, both surface and Wilson bands are twofold-degenerate along $\bar{X}\bar{M}$ ($\bar{Y}\bar{M}$) by the combination of time-reversal and $g_{y}$ ($g_{x}$) and meet linearly in fourfold degeneracies at $\bar{M}$.

\begin{figure}[t]
\includegraphics[width=\linewidth]{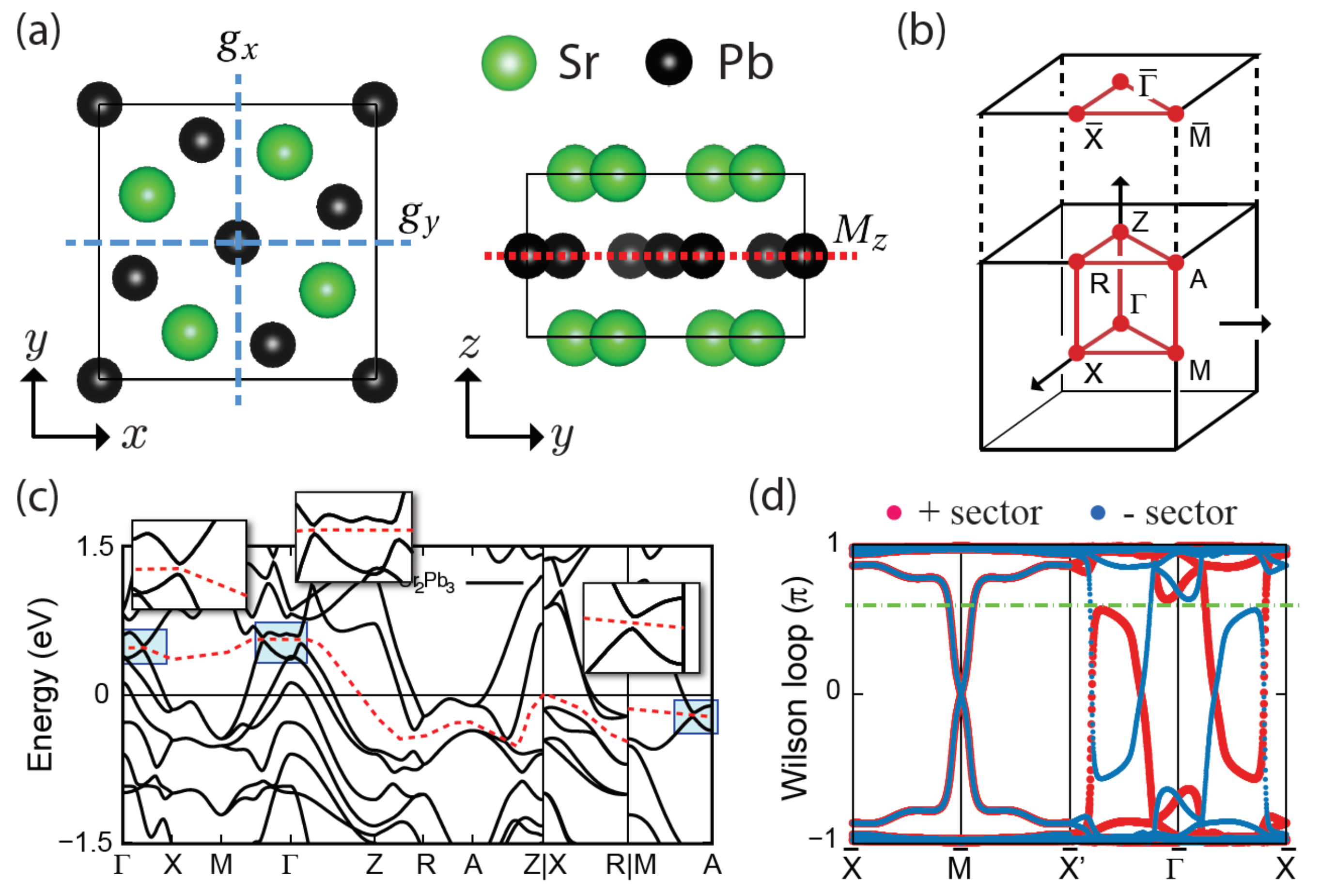}
\caption{The crystal and electronic structures of Sr$_2$Pb$_3$ in SG 127 ($P4/mbm$). (a) The unit cell of Sr$_2$Pb$_3$. (b) The bulk Brillouin zone (BZ) and the $(001)$-surface BZ (wallpaper group $p4g$). (c) The electronic bands obtained using DFT; the Fermi level is set to 0 eV.  At each point in the BZ, there is a band gap near the Fermi energy, indicated by the dashed red line; insets show magnified images of the boxed regions. (d) The $(001)$-directed Wilson bands; red (blue) points indicate Wilson bands with positive (negative) surface glide eigenvalues $\lambda_{x,y}^{+(-)}$.  By counting the Wilson bands within each glide sector that cross the dashed green line, we find that Sr$_2$Pb$_3$ has the bulk topology of a $(2,2)$ nonsymmorphic Dirac insulator (Appendix~\ref{sec:YKmatdeets}).}
\label{Fig_SRBand}
\end{figure}

By generalizing the $\mathbb{Z}_{4}$ invariant defined in~\onlinecite{Shiozaki16} for the single-glide wallpaper groups~\cite{AlexandradinataConvo,AlexandradinataPub}, we define topological invariants for double-glide systems using the (001)-directed Wilson loop eigenvalues.  For $g_{y}$ in a surface BZ in wallpaper group $pgg$, the quantized invariant $\chi_{y}$ is defined in~\onlinecite{Shiozaki16} by integrating the Wilson phases, $\theta_{j}(k_{x},k_{y})$, along the path $\bar{M}\bar{Y}\bar{\Gamma}\bar{X}$:
\begin{eqnarray}
\chi_y &\equiv& \frac{1}{\pi}\sum_{j=1}^{n_{\rm occ}/2}\bigg[\theta^{+}_{j}(\bar{M}) - \theta^{+}_{j}(\bar{X}) + \int_{\bar{M}\bar{Y}}d\theta_{j}^{+} + \int_{\bar{\Gamma}\bar{X}}d\theta_{j}^{+}\bigg] \nonumber \\
&+& \sum_{j=1}^{n_{\rm occ}}\frac{1}{2\pi}\int_{\bar{Y}\bar{\Gamma}}d\theta_{j} \mod 4, 
\label{eq:chi}
\end{eqnarray}
where $n_{\rm occ}$ is the number of occupied bands, the superscript $\pm$ indicates the glide sector, and the absence of a superscript indicates the line where $g_{y}$ is not a symmetry and the sum is over all bands.  In the presence of an additional glide, $g_x$, one can obtain $\chi_{x}$ by the transformation $x\leftrightarrow y, \bar{X} \leftrightarrow \bar{Y}$ in Eq.~(\ref{eq:chi}).  Though Eq.~(\ref{eq:chi}) appears complicated, $(\chi_{x},\chi_{y})$ can be easily evaluated by considering the bands within each glide sector which cross an arbitrary horizontal line in the Wilson spectrum (Appendix~\ref{sec:Z4}).  Wallpaper group $p4g$ also has $C_{4z}$ symmetry, which requires $\chi_{x}=\chi_{y}$ and implies the existence of the symmorphic mirrors, $\{m_{110}|\frac{1}{2}\frac{1}{2}0\}$ and $\{m_{1\bar{1}0}|\frac{1}{2}\bar{\frac{1}{2}}0\}$.  These mirrors yield $\mathbb{Z}$ mirror Chern numbers, $n_{110}, n_{1\bar{1}0}$, respectively, with values constrained by the glide invariant $\chi_{x}$, $(-1)^{n_{110}}= (-1)^{n_{1\bar{1}0}} = (-1)^{\chi_x}$ (Appendix~\ref{sec:mirrorcherncalc}). 

To enumerate the allowed topological phases shown in Fig.~\ref{fig:SurfaceBands}, we consider possible restrictions on $(\chi_{x},\chi_{y})$.  Though $\chi_{x,y}$ can individually take on values $0,1,2,3$; only pairs that satisfy $\chi_{x}+\chi_{y} \mod 2 = 0$ are permitted in bulk insulators; this can be understood as follows: if $\chi_{x}+\chi_{y}$ is odd, the 2D surface consisting of the four partial planes $(0\leq k_x\leq \pi,0(\pi),k_z)$ and $(0(\pi),0\leq k_y\leq \pi,k_z)$ possesses an overall Chern number, which implies the existence of a gapless point~\cite{Fang15,Alexandradinata16b}, contradicting our original assumption that the system is insulating.  We present a rigorous proof in Appendix~\ref{sec:doubleZ4}, and show that the remaining collection of eight insulating phases is indexed by the group $\mathbb{Z}_{4}\times\mathbb{Z}_{2}$. 

For $\chi_{x,y} = 1,3$, the system is a strong topological insulator (STI).  These four ``double-glide spin Hall'' phases possess the usual twofold-degenerate Kramers pairs at $\bar{\Gamma},\bar{X},$ and $\bar{Y}$ as well as a fourfold-degenerate Dirac point at $\bar{M}$.  The four STI phases are topologically distinct, but will appear indistinguishable in glide-unpolarized ARPES experiments.  However, if two double-glide spin Hall systems with differing $\chi_{x,y}$ are coupled together, the resulting surface modes will distinguish between $\chi_{x,y}=1,3$~\cite{Shiozaki16} (Appendix~\ref{sec:doubleZ4}). 

When $\chi_{x,y}=0,2$, the system is in a topological crystalline phase.  For $(\chi_{x},\chi_{y})=(0,2)$ or $(2,0)$, which is only permitted in a $C_{4z}$-broken surface $pgg$, a variant of the hourglass insulating phase~\cite{Wang16} is present on the surface.  For example, when $(\chi_{x},\chi_{y})=(0,2)$, either time-reversed partners of twofold-degenerate free fermions live along $\bar{\Gamma}\bar{X}$ or both twofold-degenerate fermions live along $\bar{\Gamma}\bar{Y}$ and a fourfold-degenerate Dirac fermion exists at $\bar{M}$.  

Finally, but most interestingly, for $\chi_{x}=\chi_{y}=2$, we find that the system exists in a previously uncharacterized ``nonsymmorphic Dirac insulating'' phase, capable of hosting just a single fourfold-degenerate Dirac surface fermion at $\bar{M}$.  

\begin{figure}[t]
\includegraphics[width=\linewidth]{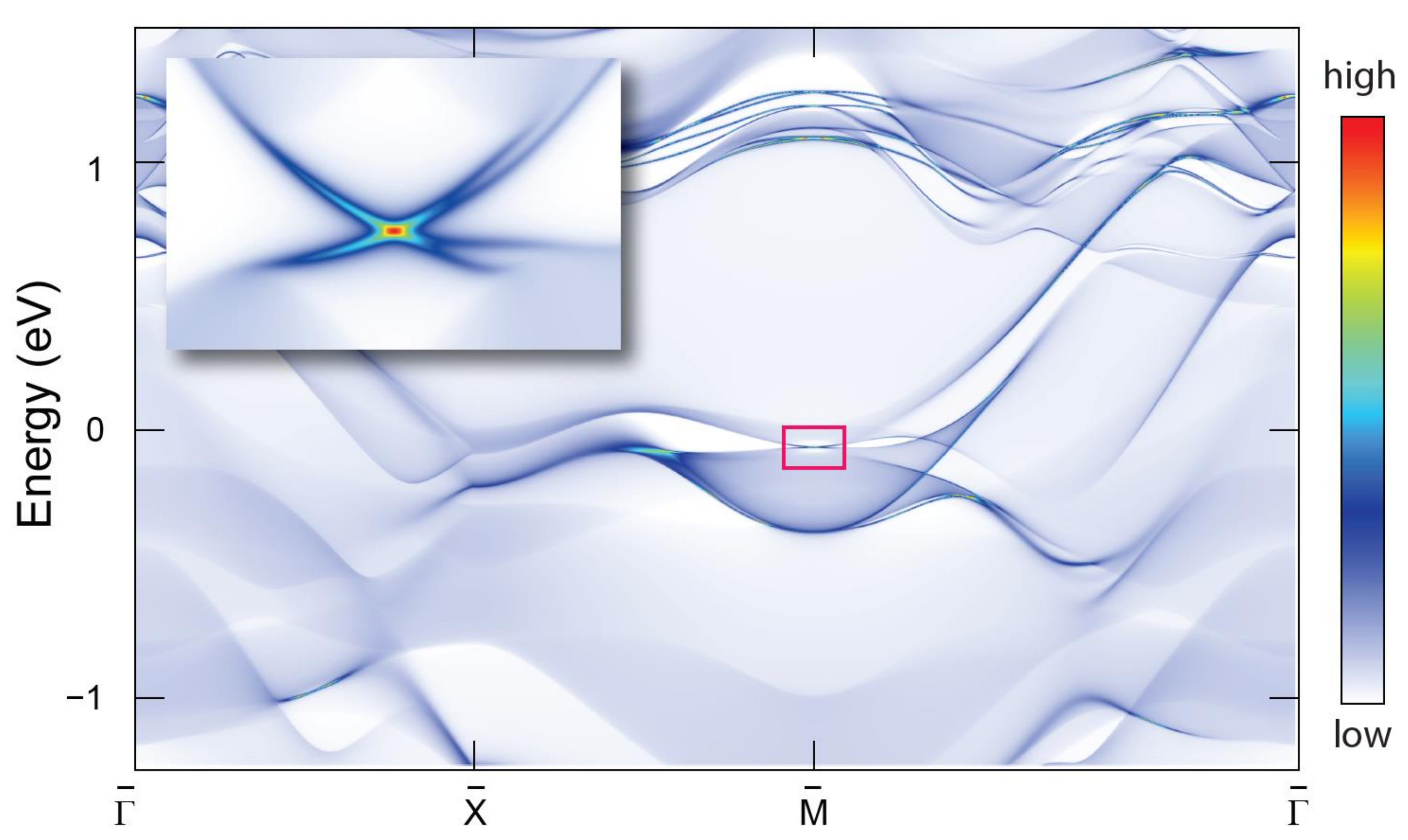}
\caption{The $(001)$-surface band structure of Sr$_2$Pb$_3$ (wallpaper group $p4g$). The Fermi level is set to zero. The fourfold surface Dirac fermion appears at $\bar{M}$ in the region indicated by the red rectangle, and is shown in more detail in the inset rectangle.  Unlike in graphene, the cones of this Dirac point are nondegenerate, except along $\bar{X}\bar{M}$.  The four dark blue surface bands dispersing from $\bar{M}$ confirm that the red surface-localized point is fourfold-degenerate.}
\label{Fig_SR_surface}
\end{figure}

\section{Materials Realizations}

\begin{figure}[t]
\includegraphics[width=\linewidth]{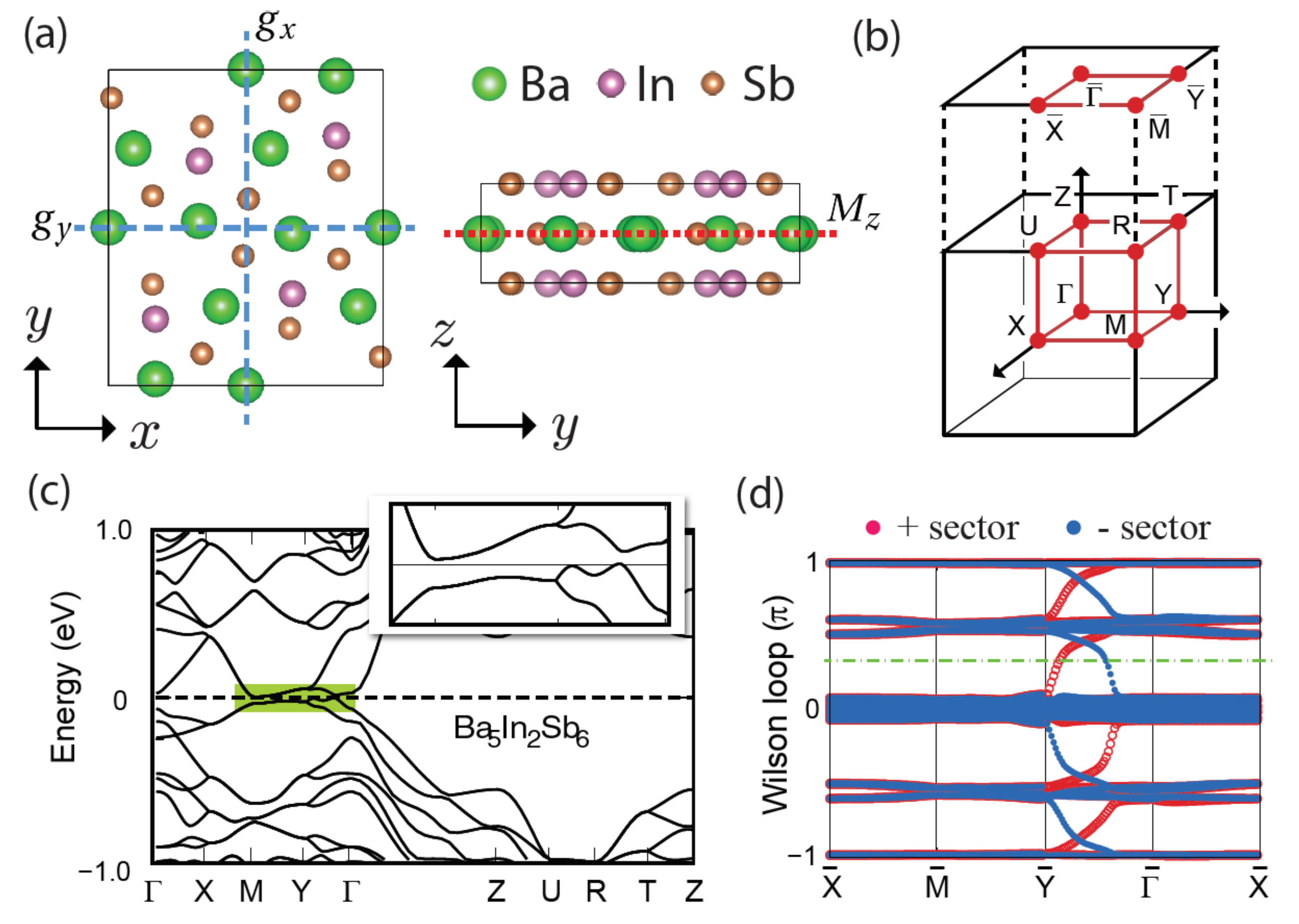}
\caption{The crystal and electronic structures of Ba$_5$In$_2$Sb$_6$ in SG 55 ($Pbam$). (a) The unit cell of Ba$_5$In$_2$Sb$_6$. (b) The bulk Brillouin zone (BZ) and the $(001)$-surface BZ (wallpaper group $pgg$). (c) The electronic bands obtained using DFT; the Fermi level is set to 0 eV.  There is an insulating gap at the Fermi energy, indicated by the dashed black line. (d) The $(001)$-surface Wilson bands; red (blue) points indicate Wilson bands with positive (negative) surface glide eigenvalues $\lambda_{x,y}^{+(-)}$.  Counting the Wilson bands crossing the dashed green line, we find that the bulk displays a $(2,0)$ topological hourglass connectivity (Appendix~\ref{sec:ZJmatdeets}).}
\label{Fig_BABand}
\end{figure}

We apply DFT including the effects of SOC to predict topological phases stabilized by wallpaper groups $pgg$ and $p4g$ in previously synthesized, stable materials.  The details of these large-scale calculations are provided in Appendix~\ref{sec:DFTstuff}.  We find double-glide topological phases on the $(001)$-surface (wallpaper group $p4g$) of three members of the SG 127 ($P4/mbm$) A$_2$B$_3$ family of materials: Sr$_2$Pb$_3$~\cite{Merlo84p78, Bruzzone81p155}, Au$_2$Y$_3$~\cite{Chai11pi53}, and Hg$_2$Sr$_3$~\cite{Druska96p401, Gumiski05p81}.  Shown in Fig.~\ref{Fig_SRBand}, we find that Sr$_2$Pb$_3$ has at each crystal momentum a gap at the Fermi energy, in spite of the presence of electron and hole pockets.  A Wilson loop calculation of the bands up to this gap (Fig.~\ref{Fig_SRBand}(d)) indicates that this material possesses the bulk topology of a $(2,2)$ nonsymmorphic Dirac insulator (Appendix~\ref{sec:YKmatdeets}).  Calculating the surface spectrum through surface Green's functions (Fig.~\ref{Fig_SR_surface}), we find that the $(001)$-surface of Sr$_2$Pb$_3$, while displaying an overall metallic character, develops a gap of 45 meV at the Fermi energy at $\bar{M}$.  Inside this gap, we observe a single, well-isolated, fourfold-degenerate surface Dirac fermion.  

Unlike Sr$_2$Pb$_3$,  Au$_2$Y$_3$ and Hg$_2$Sr$_3$ in SG 127 ($P4/mbm$) are gapless, with bulk $C_{4z}$-protected Dirac nodes~\cite{Wang13} present near the Fermi energy.  In Appendix~\ref{sec:YKmatdeets}, we show that under weak $(100)$-strain, these Dirac nodes can be gapped to induce the $(0,2)$ topological hourglass phase in these two materials.  

\begin{figure}[t]
\includegraphics[width=\linewidth]{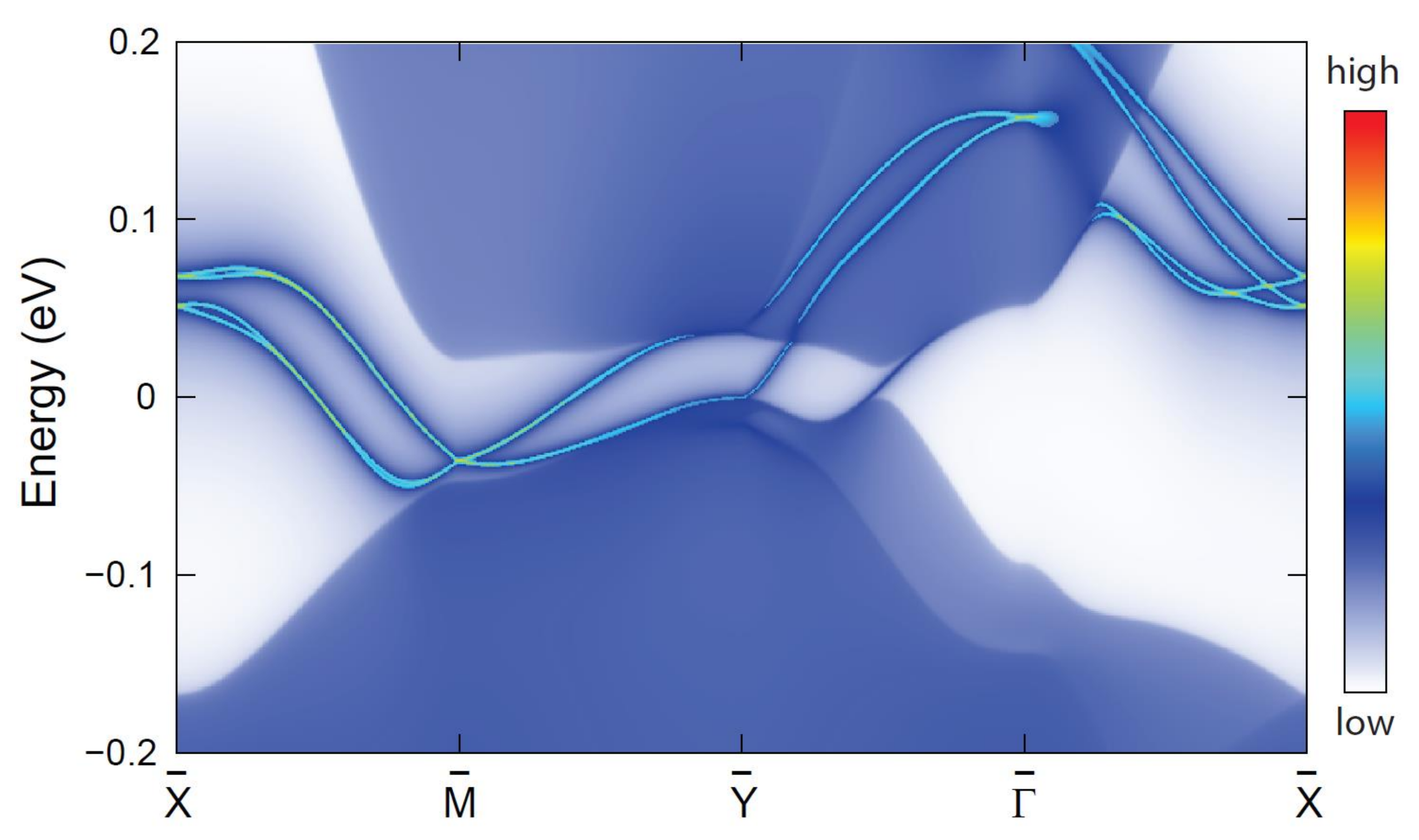}
\caption{Spectral function of the $(001)$-directed surface of Ba$_5$In$_2$Sb$_6$ (wallpaper group $pgg$). The Fermi level is set to zero.  Surface bands from the top of one hourglass fermion and the bottom of another connect the valence and conduction manifolds along $\bar{Y}\bar{\Gamma}$;  the hourglass fermions themselves are buried in the bulk manifolds along this line.  The surface bands also display other signatures of the $(2,0)$ topological hourglass connectivity, including a fourfold Dirac fermion at $\bar{M}$ connected to a clearly distinguishable hourglass fermion along $\bar{\Gamma}\bar{X}$.  The maximally localized Wannier functions obtained numerically from ab initio calculations (Appendix~\ref{sec:ZJmatdeets}) are only approximately symmetric under the surface wallpaper group, and therefore bands along $\bar{X}\bar{M}$ here appear weakly split.}
\label{Fig_BA_surface}
\end{figure}

We additionally find that the $(001)$-surface (wallpaper group $pgg$) of the narrow-gap insulator Ba$_5$In$_2$Sb$_6$ in SG 55 ($Pbam$)~\cite{ZJGermanBA} hosts a double-glide topological hourglass fermion.  Shown in Fig.~\ref{Fig_BABand}, we find that Ba$_5$In$_2$Sb$_6$ develops an indirect band gap of 5 meV (direct band gap: 17 meV).  The Wilson loop spectrum obtained from the occupied bands, shown in (Fig.~\ref{Fig_BABand}(d)), demonstrates that this material is a $(2,0)$ double-glide topological hourglass insulator (Appendix~\ref{sec:ZJmatdeets}).  We find that the $(001)$-surface of Ba$_5$In$_2$Sb$_6$ has a projected insulating bulk gap which is spanned along $\bar{Y}\bar{\Gamma}$ by the top and bottom bands of two different topological hourglass fermions, which themselves are degenerate with states in the bulk spectrum (Fig.~\ref{Fig_BA_surface}).  However, these fermions are topologically connected to a clearly distinguishable hourglass fermion along $\bar{\Gamma}\bar{X}$ and a fourfold-degenerate surface Dirac fermion at $\bar{M}$, both of which could in-principle be observed through ARPES.  

\section{Discussion}

We have demonstrated the existence of a nonsymmorphic Dirac insulator -- a topological crystalline material with a single fourfold-degenerate surface Dirac point stabilized by two perpendicular glides.  After an exhaustive study of the 17 time-reversal-symmetric, strong-SOC wallpaper groups, only $pgg$ and $p4g$ are revealed to be capable of supporting this fourfold fermion.  This phase is one of eight topologically distinct phases that can exist in insulating orthorhombic crystals with surfaces that preserve two perpendicular glides; we have classified all eight phases by topological indices $(\chi_{x},\chi_{y})$ that characterize the connectivity of the $z$-projection Wilson loop spectrum.  We report the discovery of the nonsymmorphic Dirac insulating phase in Sr$_2$Pb$_3$ and of related double-glide topological hourglass phases in  Ba$_5$In$_2$Sb$_6$, as well as in $(100)$-strained Au$_2$Y$_3$ and Hg$_2$Sr$_3$.  We also report the theoretical prediction of a set of novel double-glide spin Hall phases.  Though their surface Kramers pairs and fourfold Dirac fermions should be distinctive in ARPES experiments, a characterization of transport in the double-glide spin Hall phases remains an open question.

We also find that there exists a simple intuition for the topological crystalline phases $\chi_{x,y}=0,2$.  In Appendix~\ref{sec:TBwC4}, we present an eight-band tight-binding model which, when half-filled, can be tuned to realize all $\mathbb{Z}_{4}\times\mathbb{Z}_{2}$ double-glide insulating phases.  
In a particular regime of parameter space, in which SOC is absent at the $\bar{X}$ and $\bar{Y}$ points and bulk inversion symmetry is imposed, the Wilson loop eigenvalues at the edge TRIMs are pinned to $\pm 1$ ($\theta(\bar{M}/\bar{X}/\bar{Y})=0,\pi$)  and each TRIM represents the end of a doubly-degenerate SSH model~\cite{SSH}.  In this limit, when the product of parity eigenvalues at $\bar{\Gamma}$ satisfies $\xi(\bar{\Gamma})=+1$, the bulk topology is fully characterized by the relative SSH polarizations, $\chi_{x,y} = 2\{\frac{1}{\pi}[\theta(\bar{M})-\theta(\bar{Y},\bar{X})]\mod 2\}$.

Finally, as the $(2,2)$ topological surface Dirac point is symmetry-pinned to the QCP between a 2D TI and an NI, we examine its potential for hosting strain-engineered topological physics.  Consider the two-site surface unit cell in wallpaper group $pgg$ from Fig.~\ref{fig:combinedWall}.  In the $(2,2)$ nonsymmorphic Dirac insulating phase, the surface Dirac fermion can be captured by the $k\cdot p$ Hamiltonian near $\bar{M}$:
\begin{equation}
\mathcal{H}_{\bar{M}} = \tau^{x}\left(v_{x}\sigma^{x}k_{x} + v_{y}\sigma^{y}k_{y}\right),
\label{eq:surfaceKP}
\end{equation}
where $\tau$ is a sublattice degree of freedom, $\sigma$ is a $\mathcal{T}$-odd orbital degree of freedom, and $g_{x/y} = \tau^{y}\sigma^{x/y}$ ($g_{x/y}^{2}=+1$).  There exists a single, $\mathcal{T}$-even mass term, $V_{m}=m\tau^{z}$, which satisfies $\{\mathcal{H}_{\bar{M}},V_{m}\}=0$ and is therefore guaranteed to fully gap $\mathcal{H}_{\bar{M}}$.  Therefore, surface regions with differing signs of $m$ will be in topologically distinct gapped phases and must be separated by 1D topological QSH surface domain walls, protected only by time-reversal symmetry~\cite{JackiwRebbi}.  As $pgg$ has point group $2mm$, and $\{V_{m},g_{x,y}\}=0$, $V_{m}$ can be considered an $xy$ $A_{2}$ distortion~\cite{TinkhamBook}, which could be achieved by strain in the $x+y$ direction and compression in the $x-y$ direction.  These domain walls would appear qualitatively similar to those proposed in bilayer graphene~\cite{BLG3,BLG4,BLG5}.  However, whereas those domain walls are protected by sublattice symmetry and are therefore quite sensitive to disorder, domain walls originating from nonsymmorphic Dirac insulators are protected by only time-reversal symmetry, and therefore should be robust against surface disorder.  Under the right interacting conditions or chemical modifications, a nonsymmorphic Dirac insulator surface may also reconstruct and self-induce regions of randomly distributed $\pm m$, separated by a network of 1D QSH domain walls.  These domain walls are closely related to the 1D helical hinge modes of ``second-order'' topological insulators~\cite{HigherOrderTIBernevig,HigherOrderTIChen,HOTIBismuth}.  Furthermore, in the presence of electron-electron interactions these domain walls will form helical Luttinger liquids~\cite{BernevigLuttinger}.  Though Sr$_2$Pb$_3$ is insufficiently insulating to experimentally isolate these effects, future, bulk-insulating, nonsymmorphic Dirac insulators could provide an experimental platform for probing and manipulating the physics of time-reversal-invariant topological Luttinger liquids.

\begin{acknowledgements}
$^\dag$Corresponding author. Email: \url{bwieder@princeton.edu} (BJW), \url{kane@physics.upenn.edu} (CLK), \url{bernevig@princeton.edu} (BAB).  We thank Aris Alexandradinata for a discussion about the invariant in Eq.~(\ref{eq:chi}), and Eugene Mele for fruitful discussions.  BJW and CLK acknowledge support through a Simons Investigator grant from the Simons Foundation to Charles L. Kane and through Nordita under ERC DM 321031.  ZW and BAB acknowledge support from the Department of Energy Grant No. DE-SC0016239, the National Science Foundation EAGER Grant No. DMR-1643312, Simons Investigator Grants No. ONR-N00014-14-1-0330, No. ARO MURI W911NF-12-1-0461, and No. NSF-MRSEC DMR- 1420541, the Packard Foundation, the Schmidt Fund for Innovative Research, and the National Natural Science Foundation of China (No. 11504117).  YK and AMR thank the National Science Foundation MRSEC Program for support under DMR-1120901 and acknowledge the HPCMO of the U.S. DOD and the NERSC of the U.S. DOE for computational support.  BAB wishes to thank \'{E}cole Normale Sup\'{e}rieure, UPMC Paris, and the Donostia International Physics Center for their generous sabbatical hosting during some of the stages of this work.
\end{acknowledgements} 

\clearpage
\onecolumngrid
\begin{appendix}

\section{Fermion Doubling in Quasi-2D Crystals}
In this section, we discuss how fermion doubling theorems in 2D, which apply to both truly 2D wallpaper-group and quasi-2D layer-group systems, constrain band structures, and how apparent exceptions to them manifest on the surfaces of bulk 3D topological phases.  Specifically, we develop fermion doubling constraints for the larger set of 2D Hamiltonians invariant under layer groups, which consequently also apply to the subset of those Hamiltonians also invariant under wallpaper groups.  In this work, we use the phrase ``Dirac fermion'' only to refer to fermions with point-like fourfold degeneracies and linear dispersion in the two in-plane reciprocal lattice directions of a two-dimensional surface or a quasi-two-dimensional material, or in all three independent directions if it is in a three-dimensional bulk.  In this nomenclature, the surface states of a topological insulator are therefore not ``Dirac points,'' but are twofold-degenerate linearly dispersing fermions by another name (in a topological insulator they are linearly dispersing Kramers pairs).  

This is an unfortunate consequence of the competing and contradicting contexts in which the name ``Dirac point'' has been previously applied.  In earlier works, it has been used to describe the twofold degeneracies in 2D in spinless graphene~\cite{Semenoff,MeleDirac}, the spinful twofold degeneracies on the 2D surface of a topological insulator~\cite{Hsieh2008}, and the spinful fourfold degeneracies in the 3D bulks of Na$_3$Bi~\cite{Dirac2} and Cd$_3$As$_2$~\cite{Dirac1}.  Although it differs slightly from the usage in high-energy physics for two dimensions~\cite{Weinberg}, we choose to designate fourfold linear degeneracies in two and three dimensions as condensed matter realizations of ``Dirac fermions.''  In so doing, we keep the classification of the bulk degeneracies in graphene as Dirac points, but are forced to expand the $2\times2$ matrices from~\onlinecite{Semenoff,MeleDirac} to explicitly include an additional spin degree of freedom.  As we show in~\ref{sec:Diracdouble}, this definition has precise symmetry and topology underpinnings; it enforces the relationship of these fourfold degeneracies with crystalline-symmetry-enhanced fermion doubling, and it preserves their role as the symmetry-pinned quantum critical points between trivial and topological insulating phases.

\subsection{2D Fermion Doubling for Twofold-Degenerate Linear Fermions}

In 2D, a system may host linearly-dispersing twofold-degenerate fermions.  Such gapless fermions may exist as fine-tuned points or, if additional symmetries are present to protect them, may exist in pairs in a stable phase (i.~e.~ spinless graphene).  Here, we are interested in stable phases.  Thus, we restrict ourselves to only discussing systems for which all possible symmetry-allowed hopping terms have been included.  For example, the critical point separating two topologically distinct insulating phases in a two-band model features a single twofold-degenerate gapless fermion. However, without imposing additional symmetries, this fermion can be gapped, and, generically, the system will be insulating at half-filling.

The simplest example of a symmetry protecting a twofold linear fermion occurs in a crystal with time-reversal symmetry, $\mathcal{T}$, satisfying $\mathcal{T}^2=-1$. This requires states to be twofold-degenerate at the Time-Reversal-Invariant-Momenta (TRIMs) by Kramers' theorem.  Twofold-degenerate fermions can also appear pairwise along high-symmetry lines and planes, with symmetry stabilization coming from a combination of crystalline and time-reversal symmetries.

In 2D systems, these symmetry-protected gapless points come in pairs, a consequence of the parity anomaly~\cite{AG1984,Redlich1984,Jackiw1984}.  We illustrate this result in the case of a single twofold-degenerate linear fermion at a TRIM protected by time-reversal symmetry, $\mathcal{T}=i\sigma^{y}K$, described by the following 2D $k\cdot p$ theory:
\begin{equation}
\mathcal{H} = v_{x}k_{x}\sigma^{x} + v_{y}k_{y}\sigma^{y}.
\label{TRIM2}
\end{equation}
There is a single remaining Pauli matrix, $\sigma^{z}$, which anticommutes with all of the terms in Eq.~(\ref{TRIM2}) and opens a gap.  The mass term  $V_{m}=m\sigma^{z}$, which could originate from an external magnetic field or mean-field magnetic order, breaks $\mathcal{T}$ symmetry and gaps locally to a $k\cdot p$ theory of a Chern insulator, with two bands of winding number $C$ and $C+1$, respectively, for some $C\in\mathbb{Z}$.  If Eq.~(\ref{TRIM2}) is a complete description of the low-energy physics, a contradiction arises: because $\{V_{m},\mathcal{T}\}=0$, the two gapped phases that result from choosing opposite signs of $m$ are related to each other by transformation under $\mathcal{T}$. In particular, the band with Chern number $C$ when $m>0$ is related to the band with Chern number $C+1$ when $m<0$. Since $C$ is odd under time-reversal, $-C=C+1$. This condition cannot be satisfied by $C\in \mathbb{Z}$. 

Thus, the hypothesis that the stable gapless fermion (Eq.~(\ref{TRIM2})) is a complete description of the low-energy physics cannot be true.  For a system with a single twofold fermion, even at $V_m=0$, time-reversal symmetry must be anomalously broken by terms beyond the $\mathbf{k}\cdot\mathbf{p}$ level, and the system forms an anomalous Hall state with $|C|=1/2$ (here $C$ need not be integer since the system is gapless). In order for time-reversal to remain unbroken, there must be a compensating second degeneracy point somewhere else in the Brillouin zone, also with $|C|=1/2$.  The only details of time-reversal symmetry that entered into the preceding argument are that $\mathcal{T}V_m\mathcal{T}^{-1}=-V_m$, and so the general result remains true for any twofold symmetry that protects a gapless fermion in two dimensions\footnote{This is due to the fact that a Pauli-Villars regulator is a function of $V_m$, and so this regularization \emph{must} break the symmetry, and so we derive the same anomaly-generating functional as in the literature.}.  As the generic Hamiltonian of a twofold-degenerate linear fermion in 2D only contains two linear terms which exhaust two of the Pauli matrices, the statement that the fermion is symmetry-protected also implies that a mass term proportional to the remaining Pauli matrix will be odd under the symmetry it breaks, and that it will anticommute with the Hamiltonian and necessarily open a gap.

We can gain some further intuition about the parity anomaly by noting that the anomalous Hall conductance $C$ is related to the Berry phase at the Fermi surface by~\cite{Haldane2004}:
\begin{equation}
C=\frac{1}{2\pi i}\log \left(P\exp{i\oint_{FS}\mathcal{A}\cdot d\mathbf{k}}\right),
\end{equation}
where $P$ is the path-ordering operator. Let us take a compact Brillouin zone with $N$ gapless fermions, and let the Fermi level be above all bands. By evaluating the Berry phase we find that $2\pi i C=\log((-1)^{N})$. However, with time-reversal symmetry, we also have that $C$ is the total Chern number of all bands, and hence $C=0$. Thus we conclude $N\in 2\mathbb{Z}$. In the context of three-dimensional topological insulators, we note that each surface taken in isolation is a 2D system with a single gapless fermion. From the above discussion we thus recover the well-known result that each surface of a topological insulator has a half-quantized anomalous Hall conductivity, and that only when both surfaces are connected by a bulk can the system be described in an anomaly-free way~\cite{Fu07,Qi2008,Mulligan2012}.  This is the sense in which a topological insulator is sometimes said to ``cheat'' the doubling of twofold-degenerate linear fermions.  Of particular note is that the strongly spin-orbit coupled topological crystalline insulating phases discovered to date, the mirror TCI in SnTe~\cite{Hsieh2012,Tanaka2012,Dziawa2012,Xu2012} and the hourglass insulator in KHgSb~\cite{Wang16,Alexandradinata16,HourglassObserve}, exhibit time-reversed pairs of twofold-degenerate surface fermions, and thus \emph{do not} carry the same relationship with fermion doubling as does a QSH insulator.  Rather their \emph{individual} surfaces exhibit integer-quantized Hall conductivities and conversely do not anomalously violate parity.

The same logic can be applied to a 3D material by replacing the mass term $m\sigma^z$ by $v_{z}k_z\sigma^z$; in this case the gapless point would be a Weyl point of Chern number $+1$ and the two gapped Hamiltonians of opposite-signed mass lie in the planes above and below it. This expresses the so-called ``descent relation'' between the parity anomaly in two dimensions and the \emph{chiral anomaly} in three dimensions~\cite{AG1985}.  If $\mathbf{k}_z$ is periodic, this would imply that, absent another Weyl point, two systems with different Chern number (above and below the Weyl point) could be adiabatically connected through the BZ boundary, which is impossible. To avoid this contradiction, the doubling theorem then requires that the low-energy physics cannot be described by only a single Weyl point: there must be another Weyl point or other band crossing at the Fermi level. In 3D, this is the celebrated Nielsen-Ninomiya theorem~\cite{NielsenNinomiya1}. 

\subsection{Quasi-2D Fermion Doubling for Dirac Fermions}
\label{sec:Diracdouble}

We now extend these arguments to show why fourfold-degenerate quasi-2D Dirac fermions cannot be stabilized as the only nodal features in a metallic phase with time-reversal symmetry at a given energy.  As in the twofold-degenerate case, while many models might display Dirac fermions upon fine-tuning, here we are interested in robust Dirac fermion phases. Thus, we only consider systems that display Dirac fermions when all symmetry-allowed hopping terms are present.  The crux of the arguments in this section was originally highlighted in~\onlinecite{Steve2D,Kane3DTI}.  

When Dirac points occur off of the TRIMs, which can only occur in quasi-2D systems invariant under nonsymmorphic layer groups~\cite{WiederLayers}, time-reversal requires that they come in pairs~\cite{Yang2014}.  The remaining Dirac points which occur in 2D or quasi-2D systems are filling-enforced and pinned to the TRIMs by crystalline symmetries and time-reversal.  As shown in~\onlinecite{WiederLayers}, these filling-enforced, high-symmetry Dirac points can only occur in layer or wallpaper groups where either inversion anticommutes with a twofold nonsymmorphic symmetry or where two perpendicular nonsymmorphic symmetries anticommute, such as the two glides in $pgg$.  

Consider first a quasi-2D spinful system with inversion symmetry $\mathcal{I}$ and a screw rotation, $s_{2y}=t_{y/2}C_{2y}$.  At $k_{y}=\pi$, $\mathcal{I}^2=s_{2y}^{2}=+1$ and $\{\mathcal{I},s_{2y}\}=0$.  In a time-reversal-invariant system, these symmetries require a four-dimensional corepresentation: more generally, for any Hamiltonian invariant under two symmetries, $A$ and $B$, in addition to $\mathcal{T}$, such that $\{A,B\}=[ A,\mathcal{T} ]=[B,\mathcal{T} ]=0$ and $A^2=1$, any eigenstate of the Hamiltonian, $\psi$, which is also an eigenstate of $A$, is part of a fourfold-degenerate quartet of orthogonal states, $\psi, \mathcal{T}\psi, B\psi, \mathcal{T}B\psi$.  In our example, there is a fourfold degeneracy at each TRIM with $k_y=\pi$.  In general, the combination of $\mathcal{I}$ and a twofold nonsymmorphic symmetry will always mandate that there are no fewer than two Dirac points for a given filling: since $\mathcal{I}$ anticommutes with the nonsymmorphic symmetry at two of the TRIMs in 2D and $\mathcal{I}^2=+1$, there are always at least two TRIM points with fourfold degeneracies.

In the case where there are two perpendicular nonsymmorphic symmetries, but no inversion symmetry, the obstruction to forming an isolated, stable Dirac fermion takes a slightly different form.  Consider the trivial phase of wallpaper group $pgg$, for example, which is characterized by $\mathcal{T}$ and two glides, $g_{x,y}=\{m_{x,y}|\frac{1}{2}\frac{1}{2}\}$.  At $k_{x}=k_{y}=\pi$, $\{g_{x},g_{y}\}=0$ and $g_{x}^{2}=g_{y}^{2}=+1$, and therefore a Dirac point exists.  However, this condition is only met at this corner TRIM, and no other fourfold degeneracies are allowed elsewhere in this system.  In this case, the Dirac point is obstructed from being alone by the filling-enforced \emph{hourglass} structures also required to exist by the presence of singly degenerate eigenstates of $g_{x,y}$.  In these systems, the Dirac point occurs at the same filling as four 2D twofold-degenerate linear fermions, and is also prevented from being alone and stable at any filling.  

We pause to briefly note here that the expression ``nonsymmorphic symmetry,'' is a slight abuse of terminology, though one rampant in this field.  For precision, and for consistency with the terminology employed in~\onlinecite{Wang16,Alexandradinata16}, we define for this work a nonsymmorphic symmetry to be a generating element $g$ of the maximal fixed-point-free subgroup(s), or Bieberbach subgroup(s)~\cite{WPVZ,WiederLayers}, of a nonsymmorphic wallpaper or space group $\mathcal{G}$, modulo full lattice translations $T$.  Consider, for example, wallpaper group $pgg$, generated by $g_{x}=t_{y/2}M_{x}$ and $g_{y}=t_{x/2}M_{y}$.  There are two maximal fixed-point-free subgroups of $pgg$: one generated by $g_{x}$ and $t_{x}$ and one generated by $g_{y}$ and $t_{y}$, where $t_{x,y}$ is a full lattice translation.  Both of these groups are isomorphic to the Bieberbach wallpaper group~\cite{WiederLayers} $pg$, and when the generating elements are taken modulo the group of full lattice translations respectively give $g_{x}$ and $g_{y}$.  For a more complicated example, consider space group 14 $P2_{1}/b$, generated by $s_{2x}=t_{x/2}t_{z/2}C_{2x}$, $t_{y}$, $t_{z}$, and spatial inversion $\mathcal{I}$ about the origin~\cite{BigBook}.  This group also has two maximal fixed-point-free subgroups, which when taken modulo the group of full lattice translations respectively give $s_{2x}$ (isomorphic to Bieberbach space group 4 $P2_{1}$ modulo $T$) and $g_{x}=\mathcal{I}\times s_{2x}$ (isomorphic to Bieberbach space group 7 $Pb$ modulo $T$).

We pose an explanation for this obstruction, which is similar to the resolution of the parity anomaly in the previous section.  Suppose a combination of symmetries were to allow a single stable Dirac point.  The $k\cdot p$ model Hamiltonian at a corner TRIM in our previous system with $\mathcal{I}$ and $s_{2y}$ is described by a linear $k\cdot p$ model:
\begin{equation}
\mathcal{H}= v_{x}k_{x}\tau^{x}\sigma^{y} + k_{y}[v_{y1}\tau^{y}+v_{y2}\tau^{x}\sigma^{x} +v_{y3}\tau^{x}\sigma^{z}], 
\end{equation}
where $\mathcal{I}=\tau^{z}$, $s_{2y}=\tau^{y}\sigma^{y}$, and $\mathcal{T}=i\sigma^{y}K$.  The four Dirac matrices present in $\mathcal{H}$ span the space of symmetry-allowed matrices.  Thus, the system supports a robust fourfold-degenerate gapless fermion stabilized by crystal-symmetries.

As before, we can examine the consequences of locally breaking one of the symmetries.  To guarantee that the $k\cdot p$ Hamiltonian is gapped everywhere, we seek a mass term to anticommute with all of the terms in $\mathcal{H}$.  Generically, the Clifford algebra of Dirac matrices is spanned by four $\mathcal{T}$-odd matrices which couple to crystal momenta, and one $\mathcal{T}$-even matrix, here $V_{m}=m\tau^{z}$. For either sign of $m$, the resulting phase is 2D, gapped, and $\mathcal{T}$-symmetric with $\mathcal{T}^{2}=-1$; a quantum spin Hall (QSH) index can thus be defined.  Noting that at $\mathbf{k}=0$, $\mathcal{H}=V_{m}=m\mathcal{I}$, a generic feature of Hamiltonians restricted to TRIM points, c.~f.~\onlinecite{bernevigbook}, we see by the Fu-Kane formula~\cite{Fu07} for the QSH index that occupied bands for $m>0$ and $m<0$ have opposite parity $\mathbb{Z}_2$ indices.  This is expected, as it is known that such a Dirac point is the boundary between a trivial and a topological insulator~\cite{bernevigbook}.

To show the need for fermion doubling, we expand to the full BZ of a hypothetical system where this fermion is the only feature at the Fermi energy and show that there is a contradiction. We label the Bloch wavefunctions of the phase when $m>0$ by $|u(k)\rangle$ and those of the phase with $m<0$ by $|\tilde{u}(\tilde{k})\rangle=s_{2y}|u(k)\rangle$, where $\tilde{k}=s_{2y}k$.  The integral of the pfaffian of the matrix $w_{ij}=\langle u(-k)_{i}|\mathcal{T}|u(k)_{j}\rangle$ gives the QSH $\mathbb{Z}_{2}$ topological invariant~\cite{Kane05}.  Consider relating this matrix for one gapped phase to the other by the operation of the broken symmetry, $s_{2y}$: $\tilde{w}_{ij}=\langle u(-k)_{i}|s_{2y}^{\dagger}\mathcal{T}s_{2y}|u(k)_{j}\rangle=\langle u(-k)_{i}|\mathcal{T}s_{2y}^{\dagger}s_{2y}|u(k)_{j}\rangle = w_{ij}$ because $[\mathcal{T},s_{2y}]=0$.  Therefore, having the same $w$ matrix, the two phases have the \emph{same} QSH invariant, contradicting the earlier Fu-Kane requirement that the two insulating phases are topologically distinct.

The resolution of this is a fermion doubling requirement for 2D fourfold-degenerate Dirac points.  Specifically, \emph{a closed 2D crystal, invariant under either just a wallpaper group or both a wallpaper group and its layer supergroup, cannot host a single symmetry-stabilized Dirac point at the Fermi level}.  In the cases where the Fermi surface is gapped except for exactly two Dirac points, each Dirac point can have a single $\mathcal{T}$-symmetric mass term $m_{1,2}$ such that the overall $\mathbb{Z}_{2}$ QSH invariant, $n$, satisfies $n=\frac{1}{2}[\sgn(m_{1})-\sgn(m_{2})]\mod 2$.  Under the action of the broken symmetry operation, the signs of both $m_{1,2}$ are flipped and $n$ is preserved~\cite{Steve2D}. 

Like the topological insulator before it, the nonsymmorphic Dirac insulator that we present in this manuscript ``cheats'' this fermion doubling by placing each of its two Dirac points on opposite surfaces of a 3D bulk.  While they each live alone on a surface, and therefore pin that surface to a 2D QSH transition, the Dirac points (or a Dirac point and four twofold degenerate points) in the \emph{combined} system of two opposing surfaces respect the fermion doubling requirement.

\section{Symmetries and Degeneracies of Wallpaper Groups $pgg$ and $p4g$}
\label{sec:wallsyms}

It has been extensively demonstrated that the bulk electronic degeneracies of a 3D crystal are constrained by the allowed irreducible (co)representations of the space group of that crystal, up to topological features, such as Weyl points~\cite{SteveDirac,JuliaDirac,DDP,NewFermions,QuantumChemistry,BigBook,manes,RhSi}.  Additional degeneracies may be realized in toy models, but those fermions are unlikely to manifest in real materials, as they require symmetry-allowed terms to be manifestly zero.  For example, the threefold symmorphic fermions predicted in the toy model in~\onlinecite{AdyTriple} appear to be realizable using only wallpaper group symmetries, but in fact they additionally require an unphysical anticommuting mirror to artificially constrain their $k\cdot p$ Hamiltonians.  In fact, similar symmorphic realizations of Spin-1 Weyl fermions~\cite{NewFermions} \emph{can} be realized in cubic crystals with point group $23$ ($T$) and weak SOC, such as, for example at the $\Gamma$ point in SG 195~\cite{manes,RhSi}.  But these fermions are inherently three-dimensional, and are completely characterized by the single-valued corepresentations of these cubic space groups.  Therefore, in this work, to find novel surface degeneracies, we explore the allowed band degeneracies of the time-reversal-symmetric wallpaper groups as captured by their double-valued corepresentations.

The wallpaper, or plane, groups describe the 17 possible configurations of symmetries on the two-dimensional surface of a three-dimensional, time-reversal-symmetric crystal~\cite{ConwayWallpaper}. To be mathematically precise, in this work, our use of the term ``wallpaper groups'' is shorthand for the Shubnikov~\cite{BigBook} wallpaper groups $W_{II}$ formed from index-2 supergroups of their type-I counterparts $G_{I}$ by the addition of time-reversal-symmetry $W_{II}=G_{I} + \mathcal{T}G_{I}$.  The irreducible corepresentations of $W_{II}$ are formed by considering the action of $\mathcal{T}$ on the irreducible representations~\cite{QuantumChemistry,BigBook} of $G_{I}$.  Of these groups, only two have multiple nonsymmorphic symmetries~\cite{BigBook} (as defined previously in~\ref{sec:Diracdouble}).  Specifically, wallpaper groups $pgg$ and $p4g$ contain glide lines in the $x$ and $y$ (surface in-plane) directions; $p4g$ has an additional $C_{4z}$ symmetry that is not present in $pgg$.  The group $pgg$ is generated entirely by the two glides:
\begin{equation}
g_{x,y} = \{m_{x,y}|\frac{1}{2}\frac{1}{2}\},
\end{equation} 
while wallpaper group $p4g$ is generated by 
\begin{equation}
g_{x}= \{m_{x}|\frac{1}{2}\frac{1}{2}\},\ C_{4z}=\{C_{4z}|00\}.
\end{equation}
The product of the two generators for $p4g$ yields two additional symmorphic mirror symmetries, $\{ m_{x+y}|\frac{1}{2}\frac{1}{2} \}$ and $\{m_{x-y} | \frac{1}{2}\bar{\frac{1}{2}} \}$.  Though containing translations, these symmetries are not glides, as glides are, by definition, free of fixed points in position space, and $\{ m_{x\pm y}|\frac{1}{2} ,\pm \frac{1}{2}\}$ leaves the line $y=\pm (-x+\frac{1}{2})$ invariant.  In  Fig.~\ref{fig:combinedWall}, we show the locations of the glides, mirrors, and $C_{4z}$ rotation centers for both wallpaper groups realized with a two-site unit cell.
\begin{figure}[t]
\centering
\includegraphics[width=0.52\textwidth]{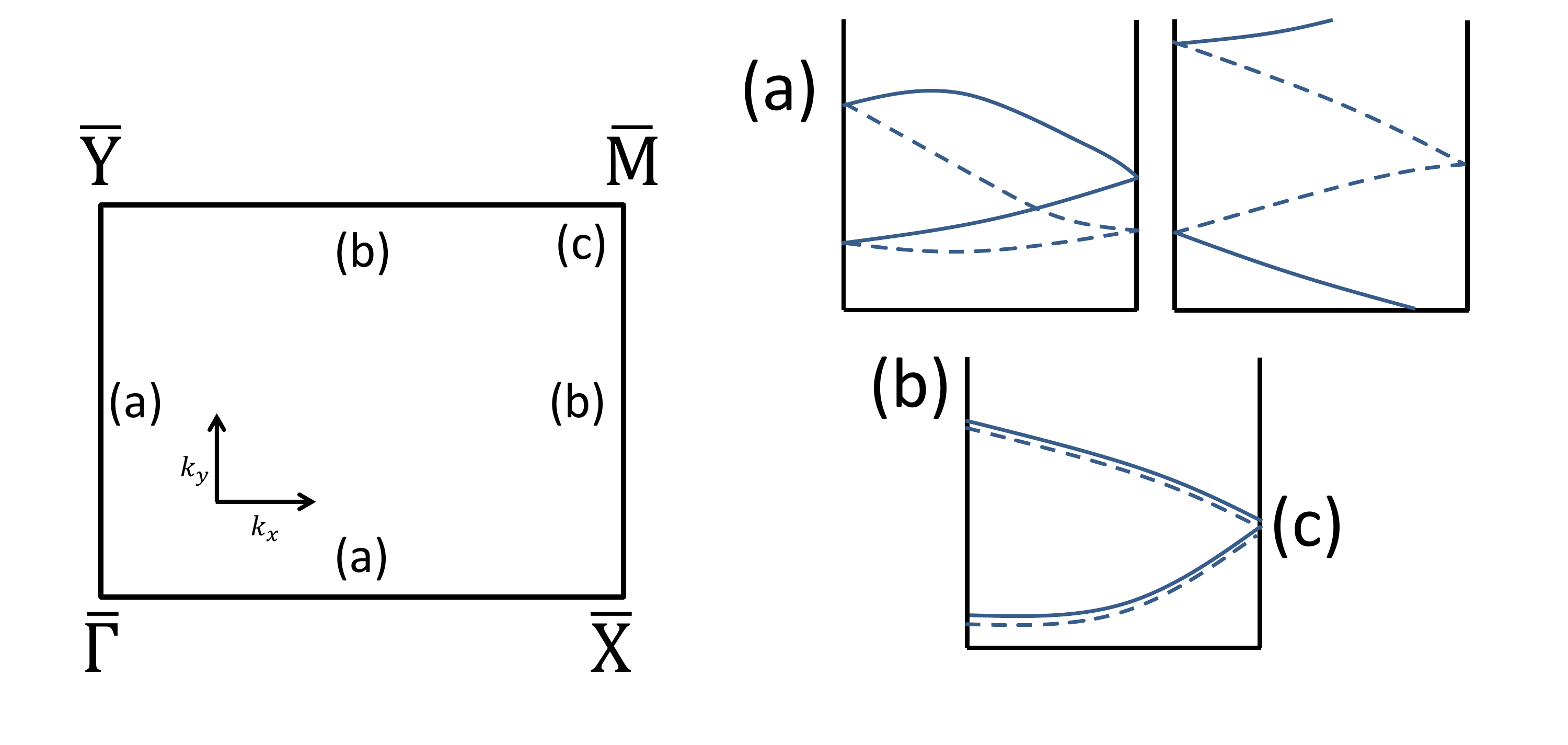}
\caption{The 2D surface Brillouin zone for wallpaper group $pgg$.
Bands with glide eigenvalue $\lambda^{+}$ ($\lambda^{-}$) are drawn as solid (dashed) lines.  Bands along lines of type (a) are singly degenerate eigenstates of $g_{y,x}$ and therefore are restricted to either form hourglass or glide spin Hall connectivities along $\bar{\Gamma}\bar{X}$ and $\bar{\Gamma}\bar{Y}$.  Along lines of type (b), bands are twofold-degenerate because they are invariant under the combined operation $(g_{y,x}\mathcal{T})^2=-1$; since such pairs have opposite $g_{x,y}$ eigenvalues, generically bands cannot cross to form fourfold degeneracies at low-symmetry points along this line.  However, at (c), the $\bar{M}$ point, bands meet and form a \emph{fourfold-degenerate} 2D Dirac point.  Bands in wallpaper group $p4g$ behave the same way, with the additional restriction from $C_{4z}$ symmetry that bands along $\bar{\Gamma}\bar{X}$ and $\bar{\Gamma}\bar{Y}$ form the same connectivities.}  
\label{fig:wallpaperBZ}
\end{figure}

In crystal momentum space, the symmetry generators enforce required band groupings along lines and at points.  In Figure~\ref{fig:wallpaperBZ}, we show one quarter of the surface BZ and identify relevant lines and points with the letters (a), (b), and (c).  Lines sharing the same letter obey the same symmetry restrictions, though for wallpaper group $pgg$, which lacks a $C_{4z}$ symmetry, they may individually display different features.  

The lines designated (a) in Fig.~\ref{fig:wallpaperBZ} are $\bar{\Gamma}\bar{X}$ and $\bar{\Gamma}\bar{Y}$ and host singly degenerate bands which are eigenstates of $g_y$ and $g_x$, respectively.  As detailed extensively in~\onlinecite{Steve2D,WiederLayers,Wang16,Alexandradinata16}, singly degenerate bands along lines invariant under a glide form hourglass or quantum spin Hall flows.

Bands along lines designated (b), $\bar{X}\bar{M}$ and $\bar{Y}\bar{M}$, are also eigenstates of $g_{x}$ and $g_y$, respectively.  However, unlike the bands along (a), they are doubly degenerate, because the symmetry operation $\mathcal{T}g_{y}$ ($\mathcal{T}g_{x}$) leaves the line $\bar{X}\bar{M}\ (\bar{Y}\bar{M})$ invariant and squares to $-1$. 

We now show that all potential band crossings along the line $\bar{X}\bar{M}$ are avoided by utilizing the fact that bands along this line are also eigenstates of $g_x$.  If $g_x|(k_{x}=0,\pi;k_{y}),+ \rangle =g_{x}|+\rangle= ie^{ik_y/2}|+\rangle$, then, using the commutation relation $g_xg_y=-g_yg_xt_xt_{-y}$, and by evaluating the resulting expression along $\bar{X}\bar{M}$ at $k_{x}=\pi$,
\begin{equation} 
g_x(\mathcal{T}g_y|+\rangle) = -\mathcal{T}g_yg_xt_xt_{-y}|+\rangle =  -\mathcal{T}g_yie^{-ik_y/2}e^{ik_{x}}|+\rangle = -ie^{ik_y/2}\left( \mathcal{T}g_y|+\rangle\right).
\label{someDiracAlgebra}
\end{equation}
Thus, the Kramers partners $|+\rangle$ and $\mathcal{T}g_{y}|+\rangle$ have opposite $g_x$ eigenvalues.  We note that eigenstates of $g_{y}$ along $\bar{Y}\bar{M}$ must behave in the same manner by $x\leftrightarrow y$ exchange symmetry.  Since all sets of Kramers pairs along line (b) have the same pair of eigenvalues, they belong to the same corepresentation, and so crossings between these bands will generically be avoided.

Finally, at the point $\bar{M}$, labeled (c), bands are fourfold-degenerate, a unique property of wallpaper groups $pgg$ and $p4g$.  This can be easily seen because the $\bar{M}$ point is invariant under $g_x, g_y$, and $\mathcal{T}$. Thus, if $\psi$ is an eigenstate of $g_x$, $\psi, \mathcal{T}\psi, g_y\psi$, and $\mathcal{T}g_y\psi$ form a degenerate quartet of states; clearly the Kramers pairs are orthogonal and since the first two states have the opposite $g_x$ eigenvalue as the last two states, they are also orthogonal.

This algebra can \emph{only} be satisfied in the strong spin-orbit coupled wallpaper groups by two perpendicular glides.  Furthermore, an examination of two-dimensional filling constraints in~\onlinecite{WiederLayers} confirms that no other algebra in 2D can enforce fourfold point-like band degeneracies.  Therefore a fourfold point degeneracy can \emph{only} be hosted on the surfaces of time-reversal-symmetric, strong spin-orbit, three-dimensional crystals invariant under the symmetries of $pgg$ or $p4g$.  

We now determine the low-energy band dispersion near a fourfold crossing at (c).  Returning to the two-site unit cell depicted in Fig.~\ref{fig:combinedWall}, we construct a $k\cdot p$ theory around $\bar{M}$.  Letting $\tau$ represent the sublattice degree of freedom and $\sigma$ the local spin degree of freedom, we choose the representation $g_{x/y}=\tau^{y}\sigma^{x/y}$ and $\mathcal{T}=i\sigma^{y}\mathcal{K}$.  Then, to linear order in $k_{x,y}$, the most general allowed Hamiltonian is given by,
\begin{equation}
\mathcal{H}_{\bar{M}} = \tau^{x}(v_{x}\sigma^{x}k_{x} + v_{y}\sigma^{y}k_{y}).
\end{equation}
This is the equation of a linearly-dispersing fourfold-degenerate 2D fermion. Therefore bands at $\bar{M}$ generically touch in Dirac points.  It is worth noting that unlike the 3D Dirac fermions characterized in~\onlinecite{Wang13,Wang12}, which are optional degeneracies created by band inversion, \emph{all} band multiplets at $\bar{M}$ in $pgg$ and $p4g$ must be at least fourfold-degenerate, and therefore the Dirac fermions in these wallpaper groups are instead more closely related to the filling-enforced 2D Dirac points proposed in~\onlinecite{Steve2D,WiederLayers,SteveMagnet}.  Like the 2D inversion-broken magnetic Dirac points in~\onlinecite{SteveMagnet}, these surface Dirac points also have generically non-degenerate cones;  the cones here are only degenerate along the glide lines $\bar{X}\bar{M}$ and $\bar{Y}\bar{M}$.  In our $k\cdot p$ model, this behavior emerges upon the introduction of the quadratic term $v_{xy}k_{x}k_{y}\tau^{z}$.

Inducing a $\mathcal{T}$-symmetric mass term $V_{m}=m\tau^{z}$ breaks each surface into a 2D trivial or topological insulator.  On a single surface, as discussed in more detail in~\ref{sec:Diracdouble}, domain walls between regions with different signs of $m$ will host 1D QSH edge modes, which are topologically protected by time-reversal symmetry alone.  These domain walls have been shown to host Luttinger liquid physics~\cite{BernevigLuttinger}, and intersections between them can form effective quantum point contacts with switching behavior characterized by universal scaling functions and critical exponents~\cite{TeoQPC}.  Additionally, these surface domain walls would provide an improvement over recent efforts to test this physics in qualitatively similar domain walls in gapped bilayer graphene~\cite{BLG1,BLG2,BLG3,BLG4,BLG5}, which have also been proposed as Luttinger liquids~\cite{GrapheneLuttinger,BLGWieder}.  Notably, as the domain walls in bilayer graphene are protected by valley index, they are quite sensitive to disorder, and thus far Luttinger liquid physics in bilayer graphene has not been observed~\cite{BLGNature,BLGNano}.  Domain wall modes on gapped nonsymmorphic Dirac insulator surfaces, conversely, should remain robust against nonmagnetic disorder.


\section{Tight-Binding Notation}
\label{sec:TBnot}

Here we provide the tight-binding notation that will be used in subsequent sections, following~\onlinecite{Alexandradinata16}.  In the unit cell labeled by the Bravais lattice vector, $\mathbf{R}$, the wavefunction corresponding to an orbital labeled by $\alpha$ at position $\mathbf{R}+\mathbf{r}_\alpha$ is denoted by $|\phi_{\mathbf{R},\alpha}\rangle$.  The Fourier transformed operators are given by,
\begin{equation} 
|\phi_{\mathbf{k},\alpha}\rangle = \frac{1}{\sqrt{N}}\sum_\mathbf{R} e^{i\mathbf{k}\cdot(\mathbf{R}+\mathbf{r}_\alpha)}|\phi_{\mathbf{R},\alpha}\rangle.
\label{eq:fourier}
\end{equation}
The single-particle Hamiltonian, $\hat{H}$, defines a tight-binding Hamiltonian,
\begin{equation} 
H(\mathbf{k})_{\alpha,\beta} = \langle \phi_{\mathbf{k},\alpha}|\hat{H}|\phi_{\mathbf{k},\beta} \rangle,
\end{equation}
whose eigenstates are denoted $|u^n(\mathbf{k})\rangle$.

The Fourier transform in Eq.~(\ref{eq:fourier}) shows that the Hamiltonian is not necessarily invariant under a shift of a reciprocal lattice vector, $\mathbf{G}$.  Instead,
\begin{equation}
\hat{H}(\mathbf{k}+\mathbf{G})=\hat{V}(\mathbf{G})^{-1}\hat{H}(\mathbf{k})\hat{V}(\mathbf{G}),
\label{eq:defV}
\end{equation}
where $\hat{V}(\mathbf{G})_{\alpha\beta} = \delta_{\alpha\beta}e^{i\mathbf{G}\cdot\mathbf{r}_\alpha}$.  Thus, we can choose the basis of eigenstates so that, 
\begin{equation}|u^n(\mathbf{k}+\mathbf{G})\rangle = \hat{V}(\mathbf{G})^{-1}|u^n(\mathbf{k})\rangle.
\label{eq:vecaddBZ}
\end{equation}

\subsection{Symmetries}

If the lattice is invariant under a spatial symmetry, $\hat{g}$, that acts in real space by taking $\mathbf{r} \rightarrow D_g\mathbf{r} + \mathbf{\delta}$, where $D_g$ is the matrix that enforces a point group operation and $\mathbf{\delta}$ is a (perhaps fractional) lattice translation, then $\hat{g}$ acts on states by: 
\begin{equation} 
\hat{g} |\phi_{\mathbf{R},\alpha} \rangle = |\phi_{\mathbf{R}',\beta} \rangle \left[U_g\right]_{\alpha\beta},
\end{equation}
where $\mathbf{R}'=D_g(\mathbf{R}+\mathbf{r}_\alpha)+\delta-\mathbf{r}_\beta$ is a Bravais lattice vector and $U_g$ is a unitary matrix. 
It is convenient to define the Fourier transformed operator, $\hat{g}_\mathbf{k}$, which can have explicit $\mathbf{k}$ dependence when it acts on Bloch states, by:
\begin{align}
\hat{g}_\mathbf{k}|\phi_{\mathbf{k},\alpha} \rangle &\equiv \frac{1}{\sqrt{N}}\sum_{\mathbf{R}} e^{i\mathbf{k}\cdot (\mathbf{R}+\mathbf{r}_\alpha)} \hat{g}|\phi_{\mathbf{R},\alpha} \rangle \nonumber\\
&=  \frac{1}{\sqrt{N}}\sum_{\mathbf{R}} e^{i\mathbf{k}\cdot (\mathbf{R}+\mathbf{r}_\alpha)} |\phi_{D_g(\mathbf{R}+\mathbf{r}_\alpha)+\delta-\mathbf{r}_\beta,\beta} \rangle \left[U_g\right]_{\alpha\beta} \nonumber\\
&= \frac{1}{\sqrt{N}}e^{-i(D_g\mathbf{k})\cdot \delta}   \sum_{\mathbf{R}} e^{i(D_g\mathbf{k})\cdot (D_g(\mathbf{R}+\mathbf{r}_\alpha)+\delta)} |\phi_{D_g(\mathbf{R}+\mathbf{r}_\alpha)+\delta-\mathbf{r}_\beta,\beta} \rangle \left[U_g\right]_{\beta\alpha} \nonumber\\
&= e^{-i(D_g\mathbf{k})\cdot \delta}|\phi_{D_g\mathbf{k},\beta} \rangle \left[ U_g\right]_{\alpha\beta}.
\label{eq:Fourierg}
\end{align}
Thus, $\hat{g}_\mathbf{k}$ separates into a product of a $\mathbf{k}$-dependent phase and a matrix, $U_g$, that rotates the orbitals:
\begin{equation}
\hat{g}_\mathbf{k} = e^{-i(D_g\mathbf{k})\cdot\delta}U_g.
\end{equation}
If $\hat{H}$ is invariant under $\hat{g}$, then Eq.~(\ref{eq:Fourierg}) shows that:
\begin{equation} 
H(\mathbf{k}) = g_\mathbf{k}^\dagger H(D_g\mathbf{k}) g_\mathbf{k}.
\label{eq:defsymmetry}
\end{equation}

We can follow the same procedure for an antiunitary operator, $\mathcal{T}_g \equiv \mathcal{T}\hat{g}$, to find:
\begin{equation} 
\mathcal{T}_{g,\mathbf{k}} \equiv e^{i(D_g\mathbf{k})\cdot\delta}U_{\mathcal{T}g} \mathcal{K}, 
\label{eq:FourierT}
\end{equation}
where $\mathcal{K}$ is the complex conjugation operator. Similarly to Eq.~(\ref{eq:defsymmetry}),
\begin{equation} 
H(\mathbf{k}) = \mathcal{T}_{g,\mathbf{k}}^{-1} H(-D_g\mathbf{k})\mathcal{T}_{g,\mathbf{k}}.
\label{eq:defantisymmetry}
\end{equation}

\subsection{Projector onto Occupied States}

We define the projector onto the $n_{\rm occ}$ occupied states:
\begin{equation}
\hat{\mathcal{P}}(\mathbf{k}) = \sum_{n=1}^{n_{\rm occ}}|u^n(\mathbf{k})\rangle \langle u^n(\mathbf{k})|,
\label{eq:defproj}
\end{equation}
which satisfies, using Eq.~(\ref{eq:vecaddBZ}),
\begin{equation}
\hat{\mathcal{P}}(\mathbf{k}) = 
\hat{V}(\mathbf{G})\hat{\mathcal{P}}(\mathbf{k}+\mathbf{G})\hat{V}(\mathbf{G})^\dagger.
\label{eq:projG}
\end{equation}
Given a spatial symmetry, $\hat{g}$, Eq.~(\ref{eq:defsymmetry}) shows that $\hat{g}_\mathbf{k}|u^n(\mathbf{k})\rangle$ has the same energy as $|u^n(\mathbf{k})\rangle$ and is a state at momentum $D_g\mathbf{k}$. Hence, the projector onto the occupied states at momentum $D_g \mathbf{k}$ is given by:
\begin{equation}
\hat{\mathcal{P}}(D_g\mathbf{k}) =  \sum_{n=1}^{n_{\rm occ}} \hat{g}_\mathbf{k}| u^n(\mathbf{k})\rangle\langle  u^n(\mathbf{k})|\hat{g}_\mathbf{k}^\dagger = \hat{g}_\mathbf{k} \hat{\mathcal{P}}(\mathbf{k})\hat{g}_\mathbf{k}^\dagger.
\label{eq:projsymm}
\end{equation}

In subsequent sections, we will want to know how $\hat{g}_{\mathbf{k}+\mathbf{G}}$ is related to $\hat{g}_k$. Plugging Eq.~(\ref{eq:defV}) into Eq.~(\ref{eq:defsymmetry}):
\begin{equation}
\hat{g}_\mathbf{k}\hat{V}(\mathbf{G})\hat{H}(\mathbf{k+G})\hat{V}(\mathbf{G})^\dagger \hat{g}_\mathbf{k}^{-1} = \hat{V}(D_g\mathbf{G})\hat{H}(D_g(\mathbf{k+G}))\hat{V}(D_g\mathbf{G})^\dagger.
\end{equation}
Thus,
\begin{equation} 
\hat{g}_{\mathbf{k+G}}=\hat{V}(D_g\mathbf{G})^\dagger \hat{g}_\mathbf{k}\hat{V}(\mathbf{G}).
\label{eq:Vg}
\end{equation}

For an antiunitary symmetry, $\mathcal{T}_g \equiv \mathcal{T} \hat{g}$, the analogous equations are:
\begin{equation}
\hat{\mathcal{P}}(-D_g\mathbf{k}) = \mathcal{T}_{g,\mathbf{k}} \sum_{n=1}^{n_{\rm occ}} |u^n(\mathbf{k})\rangle\langle    u^n(\mathbf{k})| \mathcal{T}_{g,\mathbf{k}}^\dagger
\end{equation}
and
\begin{equation} 
\mathcal{T}_{g,\mathbf{k+G}}=\hat{V}(-D_g\mathbf{G})^\dagger \mathcal{T}_{g,\mathbf{k}}\hat{V}(\mathbf{G}).
\end{equation}


\section{Wilson Loops}
\label{sec:Wilson}

A precise way to distinguish the distinct surface connectivities is obtained through the eigenvalues of Wilson loops, which, as we will elaborate, also give information about the edge state spectrum~\cite{Fu06,Ryu10,Soluyanov11,Yu11,Taherinejad14,Alexandradinata14,Fidkowski2011,Alexandradinata16}.  Here, we are interested in the Wilson loop matrix:
\begin{equation}
\left[ \mathcal{W}_{(k_\perp,k_{z0})}  \right]_{nm}  \equiv \left[ P e^{i \int_{k_{z0}}^{k_{z0}+2\pi} dk_z A_z(k_\perp,k_{z0})}\right]_{nm} ,
\label{eq:SupWilsoncont}
\end{equation} 
where $P$ indicates that the integral is path-ordered, and $A_z(\mathbf{k})_{ij} \equiv  i\langle u^i(\mathbf{k})|\partial_{k_z} u^j(\mathbf{k})\rangle$ is a matrix whose rows and columns correspond to filled bands. The eigenvalues of $\mathcal{W}$ are gauge invariant and of the form $e^{i\theta(k_x,k_y)}$, i.e., they are independent of the `base point,' $k_{z0}$.

In this appendix, we show that a space group with two glides, $g_{x,y}\equiv \{m_{x,y}| \frac{1}{2}\frac{1}{2}0\}$, as well as time-reversal symmetry, $\mathcal{T}$, has the following constraints on its $z$-directed Wilson loop eigenvalues:
\begin{itemize}
\item Along the lines $(k_x,\pi )$ and $(\pi, k_y)$, the Wilson loop eigenvalues are doubly degenerate, due to the antiunitary symmetries $g_x\mathcal{T}$ and $g_y\mathcal{T}$, respectively (see~\ref{sec:WilsonTR}).
\item At the $(\pi,\pi)$ point, the doubly degenerate bands meet at a fourfold band crossing (see~\ref{sec:WilsonUnitary}).
\item Along the $\bar{X}\bar{M}\ (\bar{Y}\bar{M})$ and $\bar{\Gamma}\bar{Y}\ (\bar{\Gamma}\bar{X})$ lines, bands can be labeled by the eigenvalues of $g_x\ (g_y)$ (see~\ref{sec:WilsonUnitary}).
\end{itemize} 
The gauge invariance of the loops allows us to write down a topological invariant that distinguishes the states, as derived in~\ref{sec:Z4}.

\subsection{Discretized Wilson Loop}

For completeness, following~\onlinecite{Alexandradinata16}, we derive a discretized version of the Wilson loop (Eq.~(\ref{eq:SupWilsoncont})), which is useful for clarity when deriving symmetry constraints.  Using the projector onto occupied states, $\hat{\mathcal{P}}$, defined in Eq.~(\ref{eq:defproj}), we derive the discretized Wilson loop matrix, 
\begin{align}
\left[ \mathcal{W}_{(k_\perp,k_{z0})}  \right]_{nm} & \equiv \left[ P e^{i \int_{k_{z0}}^{k_{z0}+2\pi} dk_z A_z(k_\perp,k_{z0})}\right]_{nm}  \nonumber\\
&\approx  \left[ P e^{i \frac{2\pi}{N}\sum_{j=1}^N A_z(k_\perp,k_{z0}+ \frac{2\pi j}{N}) }\right]_{nm} \nonumber\\
&\approx \langle u^n(k_\perp,k_{z0}+2\pi) |  \left[ P \prod_{j=1}^N  \mathcal{P}(k_\perp,k_{z0}+\frac{2\pi j}{N} )   \left( 1  - \frac{2\pi }{N} \partial_{k_z}|_{(k_\perp,k_{z0}+ \frac{2\pi j}{N}) }\right)   \mathcal{P}(k_\perp,k_{z0}+\frac{2\pi j}{N} ) \right] |u^m(k_\perp,k_{z0})\rangle \nonumber\\
&\approx \langle u^n(k_\perp,k_{z0})|V(2\pi\hat{z}) \hat{\Pi}(k_\perp,k_{z0} ) |u^m(k_\perp,k_{z0}) \rangle, 
\label{eq:wilsondisc}
\end{align}
where $k_\perp \equiv (k_x,k_y)$ and in the last line we have defined the ordered product of projectors,
\begin{equation} 
\hat{\Pi}(k_\perp,k_{z}) \equiv\hat{\mathcal{P}}(k_\perp, k_{z} + 2\pi )\hat{\mathcal{P}}(k_\perp, k_{z}+\frac{2\pi (N-1)}{N})\cdots \hat{\mathcal{P}}(k_\perp,k_z+\frac{2\pi }{N}).
\end{equation} 
Eq.~(\ref{eq:wilsondisc}) shows that the discretized Wilson loop can be written in the basis-invariant form, 
\begin{equation}
\mathcal{W}_{(k_\perp,k_{z0})}=\hat{V}(2\pi \hat{z}) \hat{\Pi}(k_\perp,k_{z0}).
\label{eq:wilson}
\end{equation}
By applying Eq.~(\ref{eq:projG}) to Eq.~(\ref{eq:wilson}), for a reciprocal lattice vector $\mathbf{G}$, 
\begin{equation}
\mathcal{W}_{\mathbf{k}+\mathbf{G}} = \hat{V}(\mathbf{G})^\dagger \mathcal{W}_{\mathbf{k}+\mathbf{G}}\hat{V}(\mathbf{G}),
\label{eq:wilsonG}
\end{equation}
which shows that the Wilson loop eigenvalues are invariant under shifts of $\mathbf{G}$, whether $\mathbf{G}$ is along or perpendicular to the direction of the loop.

\subsection{Effect of Time-Reversal-Like Symmetries on the Wilson Loop}
\label{sec:WilsonTR}

Here, we consider an antiunitary symmetry, $\mathcal{T}_g\equiv \mathcal{T}\hat{g}$, where $\hat{g}$ rotates $k_\perp$ but does not affect $k_z$ and $\hat{g}$ does not include a real space translation along $\hat{z}$. If we write the action of $\hat{g}$ in real space by $\mathbf{r} \rightarrow D_g\mathbf{r} + \mathbf{\delta}$, these conditions require that $D_g(k_\perp,k_z) = (D_gk_\perp, k_z)$ and $\delta=(\delta_x,\delta_y,0)$.
Using Eq.~(\ref{eq:FourierT}), these properties imply $\mathcal{T}_{g,(k_\perp,k_z)} = \mathcal{T}_{g,(k_\perp,0)}\equiv \mathcal{T}_{g,k_\perp}$.  We would like to relate $\mathcal{W}_{(-D_g k_\perp,0)}$ to $\mathcal{W}_{(k_\perp,0)}$, to which end we compute:
\begin{align}
\mathcal{T}_{g,k_\perp}^{-1}\mathcal{W}_{(-D_gk_\perp,0)} \mathcal{T}_{g,k_\perp} &=\mathcal{T}_{g,k_\perp}^{-1}\hat{V}(2\pi \hat{z})\hat{\Pi}(-D_g k_\perp,0)\mathcal{T}_{g,k_\perp}  \nonumber\\
&= \mathcal{T}_{g,k_\perp}^{-1}\hat{V}(2\pi \hat{z})\hat{\mathcal{P}}(-D_g k_\perp,2\pi )  \hat{\mathcal{P}}(-D_g k_\perp,2\pi - \frac{2\pi}{N})  \cdots \hat{\mathcal{P}}(-D_g k_\perp,\frac{2\pi}{N})  \mathcal{T}_{g,k_\perp}    \nonumber\\
&= \mathcal{T}_{g,k_\perp}^{-1}\hat{V}(2\pi \hat{z})\mathcal{T}_{g,k_\perp} \hat{\mathcal{P}}( k_\perp,-2\pi )  \hat{\mathcal{P}}( k_\perp,-2\pi + \frac{2\pi}{N})  \cdots \hat{\mathcal{P}}(k_\perp,-\frac{2\pi}{N})\mathcal{T}_{g,k_\perp}^\dagger  \mathcal{T}_{g,k_\perp}  \nonumber\\
&=\hat{V}(2\pi \hat{z})^\dagger \hat{V}(2\pi \hat{z})\hat{\mathcal{P}}( k_\perp, 0 )  \hat{\mathcal{P}}( k_\perp, \frac{2\pi}{N})  \cdots \hat{\mathcal{P}}(k_\perp,2\pi-\frac{2\pi}{N})\hat{V}(2\pi \hat{z})^\dagger\nonumber\\
&= \mathcal{W}_{(k_\perp,0)}^\dagger.
\end{align}

We have shown that, $\mathcal{T}_{g,k_\perp}^{-1}\mathcal{W}_{(-D_g k_\perp,0)}\mathcal{T}_{g,k_\perp} = \mathcal{W}_{(k_\perp,0)}^\dagger$. 
Thus, if $\psi(k_\perp)$ is an eigenvector of $\mathcal{W}_{(k_\perp,0)}$ with eigenvalue $e^{i\theta(k_\perp)}$, then $\mathcal{T}_{g,k_\perp} \psi(k_\perp)$ is an eigenvector of $\mathcal{W}_{(-D_g k_\perp,0)}$ with the same eigenvalue.  Furthermore, if for some $k_\perp$, $\mathcal{T}_g$ leaves $k_\perp$ invariant up to a reciprocal lattice vector (i.e, $-D_g (k_\perp,k_z) = (k_\perp+G_\perp, -k_z)$, where $(G_\perp,0)$ is a reciprocal lattice vector), then from Eq.~(\ref{eq:wilsonG}), $\hat{V}(G_\perp)\mathcal{T}_{g,k_\perp}\psi(k_\perp)$ is an eigenstate of $\mathcal{W}_{(k_\perp,0)}$, also with the same eigenvalue as $\psi(k_\perp)$. Since $\hat{V}(G_\perp)\mathcal{T}_{g,k_\perp}$ is an antiunitary symmetry that squares to $-1$, $\psi(k_\perp)$ and $\hat{V}(G_\perp)\mathcal{T}_{g,k_\perp}\psi(k_\perp)$ are orthogonal. Thus, at momenta whose projections onto $(k_x,k_y)$ are invariant under $\mathcal{T}_g$ up to a reciprocal lattice vector, eigenstates of $\mathcal{W}_{(k_x,k_y,0)}$ come in Kramers pairs.

\subsection{Effect of Unitary Symmetries that Leave $k_z$ Invariant}
\label{sec:WilsonUnitary}

Here we consider a unitary symmetry, $\hat{g}$, which leaves $k_z$ invariant and does not translate in the $\hat{z}$ direction, i.e., $D_g(k_\perp,k_z)=(D_g k_\perp,k_z)$ and $\delta = (\delta_x,\delta_y,0)$.  Using Eq.~(\ref{eq:Fourierg}), $g_{(k_\perp,k_z)} = g_{(k_\perp,0)} \equiv g_{k_\perp}$. 
Eq.~(\ref{eq:projsymm}) shows that $ \hat{V}(2\pi \hat{z})\hat{\Pi}(D_g k_\perp,k_z) = \hat{V}(2\pi \hat{z})\hat{g}_{k_\perp} \hat{\Pi}(k_\perp,k_z) \hat{g}_{k_\perp}^\dagger=\hat{g}_{k_\perp} \hat{V}(2\pi \hat{z}) \hat{\Pi}(k_\perp,k_z) \hat{g}_{k_\perp}^\dagger$.  Thus, by definition, 
\begin{equation}
\mathcal{W}_{(D_g k_\perp,k_z)} = \hat{g}_{k_\perp}\mathcal{W}_{(k_\perp,k_z)}\hat{g}^\dagger_{k_\perp}.
\label{eq:wilsonunitary}
\end{equation}

Now specialize to momenta invariant under $\hat{g}$ up to a reciprocal lattice vector, so that $D_g(k_\perp, k_z) =(D_gk_\perp,k_z)= (k_\perp+G_\perp, k_z)$, where $(G_\perp,0)$ is a reciprocal lattice vector.  At these momenta, $\mathcal{W}_{(D_g k_\perp,k_z)} = \mathcal{W}_{(k_\perp+G_\perp,k_z)}$, which, combined with Eqs.~(\ref{eq:wilsonunitary}) and (\ref{eq:wilsonG}), shows that at these momenta, $\left[ \mathcal{W}_{(k_\perp,k_z)}, \hat{V}(G_\perp)\hat{g}_{k_\perp} \right]=0$.  Thus, at values of $k_\perp$ that are invariant under $D_g$ up to a reciprocal lattice vector, the Wilson loop, $\mathcal{W}_{(k_\perp,k_z)}$, and the operator $\hat{V}(G_\perp)\hat{g}_{k_\perp}$ can be simultaneously diagonalized. Wilson loop bands can then be labeled by their $\hat{V}(G_\perp)\hat{g}$ eigenvalue in exactly the same way as energy bands can be labeled by their $\hat{g}_{k_\perp}$ eigenvalue. 

We now apply the results of this section and the previous section to the glide symmetries defined in the main text, $g_{x,y} \equiv \{ M_{x,y}|\frac{1}{2}\frac{1}{2}0\}$.  At $k_\perp = (\pi,\pi)$, $D_{g_x}(k_\perp,k_z) = (k_\perp,k_z)-2\pi \hat{x}$ and $D_{g_y}(k_\perp,k_z) = (k_\perp,k_z)-2\pi\hat{y}$. From the previous paragraph, the Wilson loop operator $\mathcal{W}_{(k_\perp,k_z)}$ commutes with both $\hat{V}(-2\pi\hat{x})g_{x,k_\perp}$ and $\hat{V}(-2\pi\hat{y})g_{y,k_\perp}$, while $\lbrace \hat{V}(-2\pi\hat{x})g_{x,k_\perp}, \hat{V}(-2\pi\hat{y})g_{y,k_\perp} \rbrace=0$ (the last statement follows from Eq.~(\ref{eq:Vg}) and the fact that $\{g_{x,k_\perp}, g_{y,k_\perp}\}=0$ at $k_\perp=(\pi,\pi)$).  Furthermore, from~\ref{sec:WilsonTR}, eigenstates of $\mathcal{W}_{(k_\perp,0)}$ come in Kramers pairs due to time-reversal symmetry, which takes $(k_\perp,0)$ to $(k_\perp,0) -2\pi(\hat{x}+\hat{y})$.  Thus, if $\psi(k_\perp)$ is a simultaneous eigenvector of $\mathcal{W}_{(k_\perp,0)}$ and $\hat{V}(-2\pi \hat{x})g_{x,k_\perp}$, it forms a quartet with three other states, $\hat{V}(-2\pi(\hat{x}+\hat{y}))\mathcal{T}_{k_\perp}\psi(k_\perp)$, $\hat{V}(-2\pi\hat{y})g_{y,k_\perp}\psi(k_\perp)$, and $\hat{V}(-2\pi(\hat{x}+\hat{y}))\mathcal{T}_{k_\perp}\hat{V}(-2\pi\hat{y})g_{y,k_\perp}\psi(k_\perp)$, which share the same eigenvalue of $\mathcal{W}_{(k_\perp,0)}$ but are orthonormal (the orthonormality is verified because the last two states have the opposite $\hat{V}(-2\pi \hat{x})g_{x,k_\perp}$ eigenvalues as do the first two). Thus, the Wilson loop eigenvalues are fourfold-degenerate at $k_\perp = (\pi,\pi)$.

\subsection{Effect of Unitary Symmetries that Flip the Sign of $k_z$}
\label{sec:wilsoninv}

Here we consider a unitary symmetry, $\hat{g}$, which flips the sign of $k_z$ and does not translate in the $\hat{z}$ direction, i.e. $D_g(k_\perp,k_z)=(D_gk_\perp,-k_z)$ and $\delta=(\delta_x,\delta_y,0)$.  As in the previous section, Eq.~(\ref{eq:Fourierg}) guarantees $g_{(k_\perp,k_z)} = g_{(k_\perp,0)} \equiv g_{k_\perp}$.  Using Eq.~(\ref{eq:projsymm}), 
\begin{align}
\mathcal{W}_{(D_gk_\perp,k_z)} &\equiv  \hat{V}(2\pi \hat{z})\hat{\Pi}(D_gk_\perp,k_z) \nonumber\\ 
&= \hat{V}(2\pi \hat{z}) \hat{\mathcal{P}}(D_gk_\perp, k_z+2\pi)\hat{\mathcal{P}}(D_gk_\perp, k_z + \frac{2\pi(N-1)}{N} ) \cdots \hat{\mathcal{P}}(D_gk_\perp, k_z + \frac{2\pi}{N} )\nonumber\\
&= \hat{V}(2\pi \hat{z}) \hat{g}_{k_\perp}\hat{\mathcal{P}}(k_\perp, -k_z-2\pi)\hat{\mathcal{P}}(k_\perp, -k_z -\frac{2\pi(N-1)}{N} ) \cdots \hat{\mathcal{P}}(k_\perp, -k_z - \frac{2\pi}{N} )\hat{g}_{k_\perp}^\dagger \nonumber\\
&= \hat{V}(2\pi \hat{z})  \hat{g}_{k_\perp} \hat{V}(2\pi \hat{z}) \hat{\mathcal{P}}(k_\perp, -k_z)\hat{\mathcal{P}}(k_\perp, -k_z +\frac{2\pi}{N} ) \cdots \hat{\mathcal{P}}(k_\perp, -k_z +2\pi)\hat{V}(2\pi \hat{z})^\dagger\hat{g}_{k_\perp}^\dagger \nonumber\\
&= \hat{V}(2\pi \hat{z})\hat{g}_{k_\perp}\hat{V}(2\pi\hat{z}) \hat{\Pi}(k_\perp,-k_z)^\dagger \hat{V}(2\pi\hat{z})^\dagger  \hat{g}_{k_\perp}^\dagger\nonumber\\
&=\hat{g}_{k_\perp} \hat{\Pi}(k_\perp,-k_z)^\dagger  \hat{V}(2\pi\hat{z})^\dagger \hat{g}_{k_\perp}^\dagger = \hat{g}_{k_\perp} \mathcal{W}_{(k_\perp,-k_z)}^\dagger \hat{g}_{k_\perp}^\dagger.
\end{align}
Thus, the Wilson loop eigenvalues at $k_\perp$ and $D_gk_\perp$ come in complex conjugate pairs.  Consequently, at momenta whose surface projection is invariant under this symmetry (or invariant up to a shift of a reciprocal lattice vector $(G_\perp,0)$, according to Eq.~(\ref{eq:wilsonG})), the spectrum of the phase of the Wilson loop eigenvalues is particle-hole symmetric.


\section{Topological Invariant}

\subsection{Single Glide}
\label{sec:Z4}

In~\onlinecite{Shiozaki16}, the authors introduce a $\mathbb{Z}_4$ invariant to classify strong topological phases of time-reversal-invariant systems with a single glide symmetry; in~\onlinecite{AlexandradinataPub}, the authors show how to interpret this invariant in terms of Wilson loops.  In this section, we refine this invariant for systems with two glide symmetries.

We first summarize the relevant result from~\onlinecite{Shiozaki16}. Consider a time-reversal-invariant system with a single glide symmetry, $g_y=\{m_y|\frac{1}{2},t, 0\}$, where $t$ can be a fractional or integer lattice translation.  Since $g_y^2=\{\mathcal{R}|100\}$, where $\mathcal{R}$ indicates a $2\pi$ rotation, Eq.~(\ref{eq:Fourierg}) dictates that $g_{y,(k_x,k_y,k_z)}^2 = -e^{ik_x}$, where the minus sign comes because we are considering spinful systems that acquire a minus sign upon a $2\pi$ rotation.  Hence, along the glide-invariant planes, $k_y=0$ and $k_y=\pi$, each energy or Wilson band can be labeled by its $g_{y,(k_x,k_y,k_z)}$ eigenvalue, $\pm ie^{ik_x/2}$, and we say it belongs to the $\pm$ glide sector.  The $\mathbb{Z}_4$ invariant is defined as,
\begin{align}
\chi_y &\equiv \frac{2}{\pi} \int_{-\pi}^\pi dk_z \left( {\rm tr} A_{+,z}^I(\pi,\pi,k_z)- {\rm tr} A_{+,z}^I(\pi,0,k_z)\right)\nonumber\\
& +\frac{1}{\pi} \int_0^\pi dk_x\int_{-\pi}^\pi dk_z \left( {\rm tr}F_{+,y}(k_x,\pi,k_z)- {\rm tr}F_{+,y}(k_x,0,k_z)\right)\nonumber\\
& -\frac{1}{2\pi}\int_0^\pi dk_y \int_{-\pi}^\pi dk_z {\rm tr} F_x(0,k_y,k_z) \mod 4,
\label{eq:SupChi}
\end{align}
where $A_{\pm,i} \equiv i\langle u^\pm | \partial_{k_i}|u^\pm\rangle$, $F_{\pm} \equiv \nabla\times A_\pm$, the subscript $\pm$ indicates the glide sector (when there is no subscript, there is no glide symmetry and all occupied bands are summed over), and the superscript $I$ indicates one state in a Kramers pair. It is shown in~\onlinecite{Shiozaki16} that $\chi_y$ always takes integer values.  Because it is a function of Wilson loops, it is also gauge invariant: the first line compensates for any change of gauge implemented in the last two lines.  Thus, $\chi_y$ at least partially classifies time-reversal-invariant systems with one glide. Using $K$ theory, it is claimed in~\onlinecite{Shiozaki16} that this is a complete classification of strong topological phases for these space groups.  In~\onlinecite{Alexandradinata16}, the authors show that the completeness of this classification can also be deduced from the possible windings of the glide-allowed Wilson band connectivities.  

\begin{figure}[t]
\centering
\includegraphics[width=0.3\textwidth]{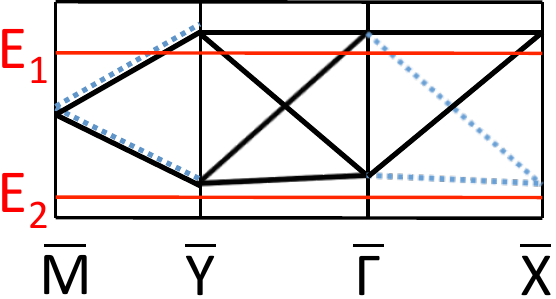}
\caption{An example to compute $\chi_y$ according to steps 1-4 in~\ref{sec:Z4}.  The $\pm$ glide sectors are identified by solid black (dashed blue) lines for the bands with $g_y$ eigenvalue $\pm ie^{ik_x/2}$ along $\bar{M}\bar{Y}$ and $\bar{\Gamma}\bar{X}$.  We now follows steps 1-4 to compute $\chi_y$: $1.$ Draw the red line labeled $E_1$.  $2.$ One positively sloped line in the $+$ sector (black solid line) crosses $E_1$ along $\bar{M}\bar{Y}$ and no negatively sloped lines cross; after multiplying by 2 the total for this step is 2.  $3.$ One positively sloped line in the $+$ sector (black solid line) crosses $E_1$ along $\bar{\Gamma}\bar{X}$ and no negatively sloped lines cross; after multiplying by 2 the total for this step is 2.  $4.$ Along $\bar{Y}\bar{\Gamma}$, one line with positive slope and one line with negative slope cross $E_1$; the total for this step is zero.  The total from steps 2, 3, and 4 is 4. Thus, $\chi_y=4\mod 4=0$ in this example.  We could have also seen that $\chi_y=0$ by choosing in step 1 the red horizontal line at energy $E_2$. Since no bands cross this line, steps 2-4 again show that $\chi_y=0$. }
\label{fig:Z4ex}
\end{figure}  

Before moving to systems with two glide symmetries, it is useful to translate Eq.~(\ref{eq:SupChi}) into a pictorial computation from a plot of the Wilson loop (defined in Eq.~(\ref{eq:wilson})) eigenvalues along the line segment $\bar{M}\bar{Y}\bar{\Gamma}\bar{X}$ (notice this is a bent line~\cite{ Alexandradinata16b} consisting of three segments, not a complete loop):
\begin{enumerate}
\item Draw a horizontal line across the plot.
\item Count the number of times a positively sloped band in the $+$ sector crosses the line along the $\bar{M}\bar{Y}$ segment and subtract from it the number of times a negatively sloped band in the $+$ sector crosses the line along the same segment. Multiply the total by 2.
\item Repeat along the $\bar{\Gamma}\bar{X}$ segment. 
\item Along the $\bar{Y}\bar{\Gamma}$ segment, count the number of times any positively sloped band crosses the line and subtract from it the number of times a negatively sloped band crosses the line (since bands along this line are not eigenstates of $g_{y}$, all bands contribute).
\item Add the numbers from the previous three steps together; taken mod 4, this sum is $\chi_y$.
\end{enumerate}
An example is shown in Fig.~\ref{fig:Z4ex}.

\subsection{Two Glides}
\label{sec:doubleZ4}

\begin{figure}[t]
\centering
\includegraphics[width=0.35\textwidth]{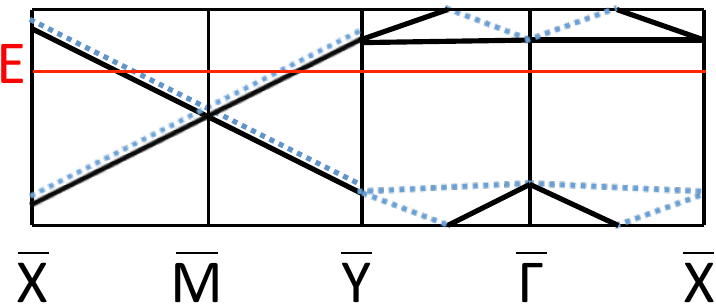}
\caption{An example to compute $(\chi_x,\chi_y)$.  The $\pm$ glide sectors are identified by solid black (dashed blue) lines for the bands with $g_y$ eigenvalue $\pm ie^{ik_x/2}$ along $\bar{M}\bar{Y}$ and $\bar{\Gamma}\bar{X}$.  Similarly, the $\pm$ glide sectors are identified by solid black (dashed blue) lines for the bands with $g_x$ eigenvalue $\pm ie^{ik_x/2}$ along $\bar{M}\bar{X}$ and $\bar{\Gamma}\bar{Y}$.  Following steps 1-4 in~\ref{sec:Z4}, we compute $\chi_y=2$.  To compute $\chi_x$, we follow the same steps but with $\bar{Y}$ exchanged with $\bar{X}$ and find that $\chi_x = 2$.
} 
\label{fig:Z4ex2}
\end{figure}

We now consider systems with two glide symmetries, $g_x$ and $g_y$, which satisfy $g_{x,(k_x,k_y,k_z)}^2=-e^{ik_y},\ g_{y,(k_x,k_y,k_z)}^2=-e^{ik_x}$.
Such a system can be described by a pair of invariants $(\chi_x,\chi_y)$, where $\chi_x$ is defined by exchanging $k_x$ and $k_y$ in Eq.~(\ref{eq:SupChi}).  $\chi_x$ can be easily computed from a plot of the Wilson loop by following steps 1-4 in the previous section after interchanging $\bar{X}$ and $\bar{Y}$. An example is shown in Fig.~\ref{fig:Z4ex2}.
 
\begin{figure}[t]
\centering
\includegraphics[width=0.35\textwidth]{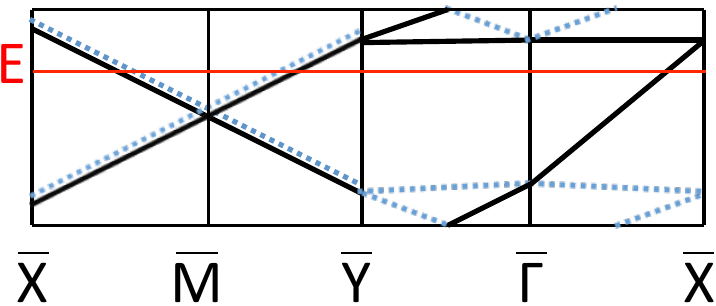}
\quad\quad\quad\quad\quad
\includegraphics[width=0.1\textwidth]{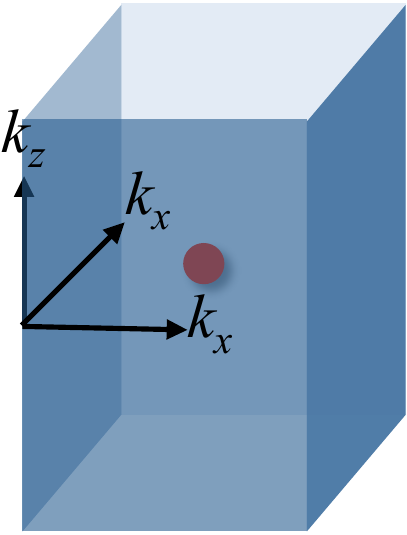}
\caption{The Wilson loop configuration on the left obeys the requirements imposed by symmetry but is forbidden from occurring in a bulk-gapped system because it has a net winding number and thus violates the constraint $\chi_x + \chi_y = 0 \mod 2$; specifically, $\chi_x = 1, \chi_y=0$. The consequence of the winding number is a bulk Weyl point enclosed by the planes $(0\leq k_x\leq \pi,0,k_z), (\pi,0\leq k_y\leq \pi,k_z), (0\leq k_x\leq \pi, \pi,k_z), (0,0\leq k_y\leq \pi,k_z)$, which are shown in blue in the right figure. The red circle represents such a bulk Weyl point.}
\label{fig:Weyl}
\end{figure}

However, not all pairs $(\chi_x,\chi_y)$ are compatible with our assumption of an insulating bulk: we now show that a bulk band insulator only permits $\chi_x+\chi_y=0\mod 2$.  The total number of Wilson bands that cross the reference line around the closed loop $\bar{\Gamma}\bar{X}\bar{M}\bar{Y}\bar{\Gamma}$ must be even because the system is gapped and, consequently, the integral of the Berry curvature over any closed surface must be zero (else there would be a Weyl point contained in the bulk region enclosed by the planes $(0\leq k_x\leq \pi,0,k_z), (\pi,0\leq k_y\leq \pi,k_z), (0\leq k_x\leq \pi, \pi,k_z), (0,0\leq k_y\leq \pi,k_z)$, as shown in Fig.~\ref{fig:Weyl}).  Since the bands along the segment $\bar{Y}\bar{M}\bar{X}$ are doubly degenerate (shown in~\ref{sec:WilsonTR}), the parity of the number of bands that cross the reference line around the closed loop $\bar{\Gamma}\bar{X}\bar{M}\bar{Y}\bar{\Gamma}$ is equal to the parity of the number of bands that cross the reference line along the segment $\bar{X}\bar{\Gamma}\bar{Y}$.  Since, in the computation of $\chi_{x,y}$, the results of steps 2 and 3 are necessarily even, the parity of the number of bands that cross along $\bar{Y}\bar{\Gamma}$ is exactly $\chi_y \mod 2$; similarly, the parity of the number of bands that cross along $\bar{X}\bar{\Gamma}$ is exactly $\chi_x \mod 2$.  Thus if we disallow Weyl points by mandating a gapped bulk,
\begin{equation}\chi_x + \chi_y = 0 \mod 2.
\label{eq:sameparity}
\end{equation}
Consequently, there are eight topologically distinct surface phases that describe a gapped system with two glide symmetries. The eight possible Wilson loops corresponding to the pair of invariants are shown in Fig.~\ref{fig:SurfaceBands}.

Because $\chi_x+\chi_y=0\mod 2$, we can rewrite each pair of invariants as $(\chi_x,\chi_y )= (\chi_x, \chi_x + 1-(-1)^{(\chi_x-\chi_y)/2} )$, which shows that the eight topological phases are classified by a $\mathbb{Z}_4\times\mathbb{Z}_2$ index: the $\mathbb{Z}_4$ index is given by the $\mathbb{Z}_4$ index of a single $g_x$ glide, $\chi_x$, and the $\mathbb{Z}_2$ index is $\eta_{\chi_x,\chi_y} \equiv \frac{1}{2}\left( 1- (-1)^{(\chi_x-\chi_y)/2}\right)$.
It is straightforward to check that the $\mathbb{Z}_2$ index satisfies the desired group addition:
\begin{align}
\eta_{\chi_x,\chi_y}  + \eta_{\chi_x',\chi_y'} &\equiv 
\frac{1}{2}\left( 1- (-1)^{(\chi_x-\chi_y)/2}\right) + \frac{1}{2}\left( 1- (-1)^{(\chi_x'-\chi_y')/2}\right)\nonumber\\
&= 
\frac{1}{2}\left( 1 - (-1)^{(\chi_x + \chi_x'-\chi_y - \chi_y')/2}  + \left(1 - (-1)^{(\chi_x-\chi_y)/2}\right)\left(1 - (-1)^{(\chi_x'-\chi_y')/2}\right) \right) \nonumber\\
&= \frac{1}{2}\left( 1 - (-1)^{(\chi_x + \chi_x'-\chi_y - \chi_y')/2}\right) \text{ mod } 2 \equiv \eta_{\chi_x+\chi_x', \chi_y + \chi_y'}.
\end{align}

We now consider what would happen if instead of computing $\chi_{x,y}$ in the first Brillouin zone, we looked at an adjacent Brillouin zone, shifted by $2\pi$ along the $k_x$ axis. In this case, the $g_y$ eigenvalues change sign, $\pm i e^{ik_x/2} \rightarrow \mp i e^{i(k_x+2\pi)/2}$, while the $g_x$ eigenvalues remain invariant.  Consequently, $\chi_y \rightarrow -\chi_y$, while $\chi_x$ remains unchanged.  Similarly, if we moved to a Brillouin zone shifted by $2\pi$ in the $k_y$ direction, $\chi_y$ would be unchanged, but $\chi_x \rightarrow -\chi_x$.  Thus, one might worry that the $\mathbb{Z}_4$ invariant is not a robust characterization of the phase.  However, the characterization remains robust: while the labels depend on our choice of Brillouin zone, once that choice is made, there are always eight distinct topological phases.  Within a given choice of BZ labeling, transitioning between $\chi_{x,y}=1$ and $\chi_{x,y}=3$ requires closing a bulk gap.  Because the phases with odd $\chi_{x,y}$ have topological indices that depend on choice of BZ, and Bloch's theorem requires that all choices of BZ be equivalent under reciprocal lattice translation, physically distinguishing between $\chi_{x,y}=1,3$ for a single, isolated sample requires a glide-polarized measurement scheme.  However, one can compare the relative values of $\chi_{x,y}$ for two samples stacked in the $z$-direction by measuring the presence or absence of a boundary mode, which depends on the \emph{difference} between $\chi_{x,y}$ in each sample.  For example, placing two samples with $\chi_{x}=\chi_{y}=1$ next to each other would yield a trivial boundary, whereas placing such a sample next to one with $\chi_{x}=\chi_{y}=3$ would yield a $(2,2)$ topological Dirac point on the interface.  

This effect is similar to what occurs in time-dependent adiabatic pumping cycles of topological superconducting Josephson junctions, for which there are two distinct QSH-like phases that in practice are only distinguishable when coupled to other similar systems~\cite{Fan14}.



\subsection{$\mathbb{Z}_2$ Topological Invariant in the Presence of Inversion Symmetry}
\label{sec:z4inv}

In the presence of inversion symmetry, $\mathcal{I}$, the $\mathbb{Z}_2$ strong topological invariant, $\nu$, is given by,
\begin{equation} 
\nu = \prod_{\mathbf{k}_{\text{inv}}} \xi_{\mathbf{k}_{\text{inv}}},
\label{eq:topinv}
\end{equation}
where the product is over the eight inversion-symmetric points, $\mathbf{k}_\text{inv}$, and $\xi_{\mathbf{k}_{\text{inv}}}$ is the product of the inversion eigenvalues of the occupied bands at $\mathbf{k}_\text{inv}$ that are not time-reversal partners (since time-reversal partners have the same inversion eigenvalues, there is no ambiguity in choosing just one of the partners).  In this section, we show that in the presence of the two glides, $g_x,\ g_y$, and inversion symmetry, Eq.~(\ref{eq:topinv}) can be simplified to only involve two points:
\begin{equation} \nu = \xi_{(0,0,0)}\xi_{(0,0,\pi)}.
\label{eq:topinv2}
\end{equation}

Without loss of generality, in this section we choose the crystal origin so that the inversion operator involves no translation. The two glides can be expressed as $g_x=\lbrace m_x| t_x, \frac{1}{2}, 0\rbrace,\ g_y=\lbrace m_y|\frac{1}{2}, t_y ,0\rbrace$, where $t_{x,y}=0$ or $\frac{1}{2}$, depending on the space group. We do not consider translations in the $\hat{z}$ direction since these symmetries would not be preserved by a surface parallel to the $xy$-plane. The operators obey the following commutation relation:
\begin{equation} 
\mathcal{I}g_{x} = g_{x} \mathcal{I} t_{-2t_x\hat{x}+\hat{y}},
\label{eq:invcomm}
\end{equation}
as do the operators under the transformation $x\leftrightarrow y$; $t_{\mathbf{v}}$ indicates a translation by $\mathbf{v}$. 

At the two points $(\pi,\pi,0(\pi))$, which project in the surface Brillouin zone to the $\bar{M}$ point, where a Dirac node is located, filled bands come in groups of four that are eigenstates of $g_y$: $\psi$, $g_x\psi$ and their time-reversed partners, $\mathcal{T}\psi, \mathcal{T}g_x\psi$; notice $\psi$ and $g_x\psi$ are linearly independent because they have different $g_y$ eigenvalues (assume $g_y\psi = \pm \psi$; then because $\{ g_x,g_y\}=0$ at the $\bar{M}$ point, $g_y(g_x\psi) = -g_xg_y\psi = \mp (g_x\psi)$); they are not time-reversed partners because their distinct $g_y$ eigenvalues are real.  At $(\pi,\pi,0)$, for each eigenstate, $\psi$, with inversion eigenvalue $\lambda=\pm 1$, the state $g_x\psi$ has inversion eigenvalue $\lambda e^{(-2t_x + 1)\pi i}$, following Eq.~(\ref{eq:invcomm}). Thus, $\xi_{(\pi,\pi,0)}=\left( -e^{2\pi i t_x}\right)^{n_{\rm occ}/4}$. Since the same logic holds at the $(\pi,\pi,\pi)$ point, $\xi_{(\pi,\pi,\pi)}=\left( -e^{2\pi i t_x}\right)^{n_{\rm occ}/4}$, as well. Since $t_x=0$ or $\frac{1}{2}$, $\xi_{(\pi,\pi,0)}\xi_{(\pi,\pi,\pi)}=\left(e^{4\pi i t_x}\right)^{n_{\rm occ}/4}=1$.

We now assume the presence of $C_{4z}$ to show that the points $(0,\pi,0(\pi))$ and $(\pi,0,0(\pi))$ also contribute a factor of $+1$ to $\nu$.  Since $\left[ C_{4z}, \mathcal{I} \right] =0$, an eigenstate at $(0,\pi, 0(\pi))$ has a $C_{4z}$ partner at $(\pi,0,0(\pi))$ with the same inversion eigenvalues; hence the product of the inversion eigenvalues at these two points is $+1$.  This proves Eq.~(\ref{eq:topinv2}) in the presence of $C_{4z}$ symmetry.

We now prove Eq.~(\ref{eq:topinv2}) without $C_{4z}$ symmetry using the $\mathbb{Z}_4$ invariants, $\chi_{x,y}$.  We already proved in Eq.~(\ref{eq:sameparity}) and the surrounding text that the two $\mathbb{Z}_4$ invariants have the same parity: $(-1)^{\chi_x}=(-1)^{\chi_y}$.  It is proved in~\onlinecite{Shiozaki16} that the parity of the $\mathbb{Z}_4$ invariant, $\chi_{x,y}$, is exactly the $\mathbb{Z}_2$ invariant of the $k_{y,x}=0$ plane.  Writing the $\mathbb{Z}_2$ invariant of the $k_{y,x}=0$ plane in terms of its inversion eigenvalues and equating the parity of $\chi_x$ with $\chi_y$ yields  $\xi_{(0,0,0)}\xi_{(0,0,\pi)}\xi_{(\pi,0,0)}\xi_{(\pi,0,\pi)} = (-1)^{\chi_x} = (-1)^{\chi_y}=\xi_{(0,0,0)}\xi_{(0,0,\pi)}\xi_{(0,\pi,0)}\xi_{(0,\pi,\pi)}$. By equating the left-most and right-most expressions, we find that $\xi_{(\pi,0,0)}\xi_{(\pi,0,\pi)} = \xi_{(0,\pi,0)}\xi_{(0,\pi,\pi)}$. Hence, the inversion eigenvalues at these points contribute a (trivial) $+1$ to Eq.~(\ref{eq:topinv}). We already showed above that the two points $(\pi,\pi,0(\pi))$ also contribute a factor of $+1$. Together, this proves Eq.~(\ref{eq:topinv2}).


\subsection{Mirror Chern Number in Wallpaper Group $p4g$}
\label{sec:mirrorcherncalc}

The wallpaper group $p4g$ has $C_{4z}$ symmetry, in addition to the two glides, $g_{x,y}$. Consequently, it has the two mirror symmetries $m_{1\bar{1}}\equiv \{m_{1\bar 10}|\frac{1}{2}\bar{\frac{1}{2}}0\}$ and $m_{110} \equiv \{m_{110}|\frac{1}{2}\frac{1}{2}0\}$, pictured Fig.~\ref{fig:combinedWall}.  Though these symmetries contain translations, they are symmorphic mirrors, and not glides, because their associated translations are along the same axes as their reflections. Equivalently, for a different choice of origin, these symmetries could therefore be written without an accompanying translation. One can define mirror Chern numbers~\cite{Teo2008}, $n_{1\bar{1}0}$ and $n_{110}$, associated with $m_{1\bar{1}0}$ and $m_{110}$, respectively.

We now show that $(-1)^{n_{1\bar{1}0}} = (-1)^{n_{110}} = (-1)^{\chi_x} = (-1)^{\chi_y}$ (the last equality was proved above Eq.~(\ref{eq:sameparity})). We focus on the mirror Chern number $n_{1\bar{1}0}$, associated with $m_{1\bar{1}0}$, which leaves the line $(k,k,k_z)$ invariant; an identical argument holds for $n_{110}$. As shown in~\onlinecite{Alexandradinata14a},
\begin{equation}
n_{1\bar{1}0} = \frac{1}{2\pi}\int_0^{\pi} dk_{11} \int_0^{2\pi} dk_z {\rm tr}\left[ F_{+,1\bar{1}}- F_{-,1\bar{1}} \right],
\label{eq:mirrorcherndef}
\end{equation}
where $k_{11(1\bar{1})} \equiv \frac{1}{\sqrt{2}}(k_x \pm k_y)$ and here, $F_{\pm,1\bar{1}} \equiv \partial_{k_{11}} A_{\pm,z} - \partial_{k_z} A_{\pm,11} $, where $\pm$ indicates that the trace is over bands with $m_{1\bar{1}0}$ eigenvalue $\pm i$.  This quantity is gauge invariant, but we can evaluate it in our gauge choice of Eq.~(\ref{eq:fourier}).  In particular, $A_{\pm,11}(k,k,2\pi) \equiv \langle u^i(k,k,2\pi) | \partial_{k_{11}} | u^j(k,k,2\pi)\rangle = \langle u^i(k,k,0) |V(2\pi\hat{z}) \partial_{k_{11}} \left( V(2\pi \hat{z} )^{-1} | u^j(k,k,0)\rangle\right) = \langle u^i(k,k,0) |\partial_{k_{11}} | u^j(k,k,0)\rangle \equiv A_{\pm,11}(k,k,0)$, using Eq.~(\ref{eq:vecaddBZ}) and the fact that $\partial_{k_{11}}V(\mathbf{G})=0$. Thus, $\int_0^{2\pi} \partial_{k_z} A_{\pm,11} = 0$, and 
\begin{equation} 
n_{1\bar{1}0}= \frac{1}{2\pi} \int_0^\pi dk_{11}  \partial_{k_{11}} {\rm Tr}\left[ \int_0^{2\pi} dk_z \left( A_{+,z}- A_{-,z}\right) \right]= -\frac{i}{2\pi} \int_0^\pi dk\partial_{k_{11}} {\rm Tr}\left[ \ln \mathcal{W}_{(k,k,0)}^+ - \ln \mathcal{W}_{(k,k,0)}^- \right],
\end{equation}
where $\mathcal{W}^\pm$ is defined as the Wilson loop evaluated on the bands with $m_{1\bar{1}0}$ eigenvalue $\pm i$.  Thus, $n_{1\bar{1}0}$ can be evaluated from a plot of the phases of the eigenvalues of the Wilson loop $\mathcal{W}_{(k,k,0)}$ along the line $\bar{\Gamma}\bar{M}$. in a similar manner to the $\mathbb{Z}_{4}$ invariant calculation in~\ref{sec:Z4}:
\begin{enumerate}
\item Draw a horizontal reference line across the plot.
\item Count the number of times a positively sloped line in the $+$ sector crosses the horizontal reference line along $\bar{\Gamma}\bar{M}$ and subtract from that the number of times a negatively sloped line in the $+$ sector crosses the horizontal reference line; this gives the $F_{+,1\bar{1}}$ part of Eq.~(\ref{eq:mirrorcherndef}), up to edge contributions.
\item Repeat for the $-$ sector.
\item $n_{1\bar{1}0}$ is equal to the result from step 2 minus the result from step 3.
\end{enumerate}

\begin{figure}[t]
\centering
\includegraphics[width=0.33\textwidth]{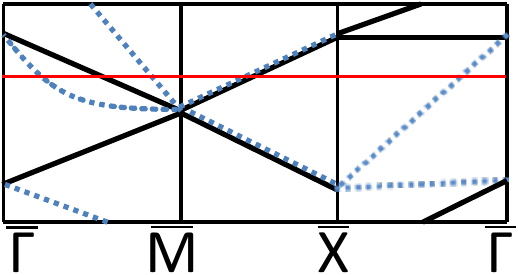}
\caption{An example to compute $n_{110}$ according to steps 1-4 in~\ref{sec:mirrorcherncalc}.  Along $\bar{\Gamma}\bar{M}$ the solid black (dashed blue) lines indicate the bands with $m_{1\bar{1}0}$ eigenvalues $\pm i$. Along $\bar{M}\bar{X}\ (\bar{X}\bar{\Gamma})$ the solid black (dashed blue) lines indicate bands with $g_x\ (g_y)$ eigenvalue $\pm i e^{ik_y/2}\ (\pm ie^{ik_x/2})$.  To compute steps 1-4, we need only examine the segment $\bar{\Gamma}\bar{M}$, along which one negatively sloped line in the $+$ sector and two negatively sloped lines in the $-$ sector cross the red horizontal reference line. Thus, the result from step 2 is -1, the result from step 3 is -2, and $n_{1\bar{1}0}=1.$ Evaluating $\chi_x$ according to~\ref{sec:Z4} shows that $\chi_x=1\mod 2$, exemplifying the proof that $\chi_x = n_{1\bar{1}0}\mod 2$.}
\label{fig:mirrorchern}
\end{figure}

This is illustrated in~\onlinecite{Alexandradinata14a} and in Fig.~\ref{fig:mirrorchern}.  From steps 1-4, it is evident that the parity of $n_{1\bar{1}0}$ is equal to the parity of the number of lines crossing the horizontal reference line drawn in step 1, along the segment $\bar{\Gamma}\bar{M}$ (the negative/positive slope no longer matters, since we are considering parity).  Now consider the closed loop in the surface Brillouin zone, $\bar{\Gamma}\bar{M}\bar{X}\bar{\Gamma}$.  Because the system is insulating, when the Wilson loop, $\mathcal{W}_{(k_x,k_y,0)}$, is plotted along this path, an even number of bands must cross the reference line (the assumption that the system is insulating actually places a stronger constraint -- that there must be the same number of bands crossing the reference line with positive as with negative slope -- but again, since we are considering parity, we only need this constraint mod 2).  Since the bands come in pairs along the line $\bar{M}\bar{X}$, the parity of the number of bands crossing the reference line along $\bar{\Gamma}\bar{M}$ is equal to the parity of the number of bands crossing the reference line along $\bar{X}\bar{\Gamma}$; the latter is equal to the parity of $\chi_y$, as derived at the end of~\ref{sec:Z4}. It follows that $n_{1\bar{1}0} = \chi_y\mod 2$, completing the proof.

Since $(-1)^{\chi_x} = (-1)^{\chi_y}$ gives us the strong $\mathbb{Z}_2$ index, as can be seen from the examples in Fig.~\ref{fig:SurfaceBands}, which show that the parity of $\chi_{x,y}$ is exactly the parity of the number of linearly dispersing surface states (the Dirac cone at $\bar{M}$ counts twice towards this parity since it comes from four intersecting bands), we conclude that $(-1)^{n_{1\bar{1}0}}$ also provides the strong $\mathbb{Z}_2$ index, i.e., if $n_{1\bar{1}0}$ is odd, the occupied bands constitute a strong topological insulating phase. 



\section{Density Functional Theory Methods and Additional Double-Glide Topological Materials}
\label{sec:DFTstuff}

In this section, we provide additional details on the density functional theory (DFT) calculations in the main text and present supplemental materials results.  First, we detail our methodology for generating the bulk electronic band structures for both the materials in the main text and for those analyzed in this supplement.  We then provide material-specific descriptions of the $(001)$-directed Wilson loop calculations used to confirm the bulk topology.  Finally, we present the details of the semi-infinite surface Green's function calculations used to predict topological surface features for the materials in the main text.  All of the materials candidates presented in this work were previously synthesized, and all were sufficiently stable as to have been powderized and examined under X-ray diffraction in an argon environment~\cite{Merlo84p78,Bruzzone81p155, Chai11pi53,Druska96p401,Gumiski05p81,ZJGermanBA}.  

To identify our topological materials candidates, we first searched the previously calculated electronic structures tabulated in the open online database~\textsc{Materials Project}~\cite{MatProject} for double-glide materials in space groups 32, 50, 54, 55, 100, 106, 117, 125, and 127 with small or negligible listed band gaps, zero net magnetic moment, and fewer than 50 atoms per unit cell.  Using the most promising candidates from this search, as well as additional materials listed in the inorganic crystal structure database (ICSD)~\cite{ICSD} with fewer than 50 atoms per unit cell, we performed DFT examinations of $\sim 100$ materials.  In Tables~\ref{tb:SG55},~\ref{tb:SG125}, and~\ref{tb:SG127}, we list the chemical formulas and ICSD numbers for the materials which our DFT and Wilson loop calculations determined to be gapless at the Fermi energy or gapped with topologically trivial $\mathbb{Z}_{4}\times\mathbb{Z}_{2}$ glide indices.  

\begin{table}[H]
\begin{center}
\begin{tabular}{ |c|c|c|c|c|c|c|c| } 
\hline
\multicolumn{8}{|c|}{Glide-Trivial Tested Materials in SG 55 (22 Materials)} \\
 \hline
Chemical Formula & ICSD Number & \ \ \ \ \  & Chemical Formula & ICSD Number & \ \ \ \ \  & Chemical Formula & ICSD Number \\
\hline
Sc$_2$Pt$_3$Si$_2$ & 247425 & & Y$_5$Pd$_2$In$_4$ & 165133 & & Y$_2$RuB$_6$ & 427249, 603493,\\
& & & & & & & 615404\\
\hline
Na$_3$Cu$_4$S$_4$ & 10004 & & NaIn$_3$S$_5$ & 426706 & & AgHg$_2$PO$_4$ & 2208 \\
\hline
Al$_3$Pt$_5$ & 55579, 58135, & & CuHgSeCl & 16450 & & La$_2$InSi$_2$ & 193223 \\
 & 656681 & & & & & & \\
\hline
La$_2$SnS$_5$ &2313, 641853 & & Ca$_2$PbO$_4$ & 36629 & & Ca$_2$SnO$_4$ & 9011, 173626, \\
& & & & & & & 187742\\
\hline
Cd$_2$SnO$_4$ & 69296, 9010 & & Cs$_2$Te$_2$ & 83351 & & Mg$_2$Ru$_5$B$_4$ & 61039 \\
\hline
Sr$_5$Sn$_2$P$_6$ & 63593 & & Ta$_2$Pd$_3$Se$_8$ & 73318 & & Ca$_5$Sn$_2$As$_6$ & 61037 \\
\hline
Rb$_2$Te$_2$ & 83350 & & Ta$_4$SiTe$_4$ & 40207, 659266 & & Tl$_2$PdSe$_2$ & 79601 \\
\hline
YCrB$_4$ & 16171 & & & & & & \\
\hline
\end{tabular}
\caption{Materials in SG 55 identified by DFT to be gapless at the Fermi energy or topologically trivial by glide indices.}
\label{tb:SG55}
\end{center}
\end{table}

\begin{table}[H]
\begin{center}
\begin{tabular}{ |c|c|c|c|c|c|c|c| } 
\hline
\multicolumn{8}{|c|}{Glide-Trivial Tested Materials in SG 125 (18 Materials)} \\
 \hline
Chemical Formula & ICSD Number & \ \ \ \ \  & Chemical Formula & ICSD Number & \ \ \ \ \  & Chemical Formula & ICSD Number \\
\hline
SrNa$_2$P$_4$O$_{12}$ & 37171 & & La$_2$NiGa$_{12}$ & 161765 & & K$_3$AgSn$_3$Se$_8$ & 416330 \\
\hline
K$_3$NaSn$_3$Se$_8$ & 280286 & & Ge$_4$N$_6$Sr$_{11}$ & 170982 & & La$_2$PdGa$_{12}$ & 171486, 183717\\
\hline
K$_2$Sr(VO$_3$)$_4$ & 155420, 250103 & & BaAg$_2$Hg$_2$O$_4$ & 40835 & & BaAl$_2$Te$_4$ & 41165 \\
\hline
CeMn$_2$Ge$_4$O$_{12}$ & 50695 & & Cs$_2$Sr(VO$_3$)$_4$ & 250105 & & KCeSe$_4$ & 67656 \\
\hline
Ag$_2$Ca(VO$_3$)$_4$ & 161369 & & Ag$_2$Sr(VO$_3$)$_4$ & 161371 & & Ce$_2$CuGa$_{12}$ & 161767 \\
\hline
La$_2$NiGa$_{12}$ & 161765 & & Na$_2$Sr(VO$_3$)$_4$ & 155419, 161370, & & RbAg$_5$Se$_3$ & 50738 \\
 & & & &  250102 & & & \\
\hline
\end{tabular}
\caption{Materials in SG 125 identified by DFT to be gapless at the Fermi energy or topologically trivial by glide indices.}
\label{tb:SG125}
\end{center}
\end{table}

\begin{table}[H]
\begin{center}
\begin{tabular}{ |c|c|c|c|c|c|c|c| } 
\hline
\multicolumn{8}{|c|}{Glide-Trivial Tested Materials in SG 127 (57 Materials)} \\
 \hline
Chemical Formula & ICSD Number & \ \ \ \ \  & Chemical Formula & ICSD Number & \ \ \ \ \  & Chemical Formula & ICSD Number \\
\hline
CaCu$_9$Cd$_2$ & 424134 & & YB$_2$C$_2$ & 427155 & & Sc$_2$MgGa$_2$ & 260213 \\
\hline
NaNbO$_3$ & 192406, 280100, & & NaMgF$_3$ & 193088 & & MgY$_2$Ge$_2$ & 423457 \\
 & 23563, 236892 & & & & & & \\
\hline
BaPtLa$_2$O$_5$ & 68794 & & InNi$_2$Sc$_2$ & 107333 & & CsSnI$_3$ & 69995, 262925 \\
\hline
CdY$_2$Ge$_2$ & 414169 & & B$_2$Nb$_3$Ru$_5$ & 423469 & & AlPt$_3$ & 107439, 609126, \\
& & & & & & & 609153, 656679 \\
\hline
Y$_2$Cu$_2$Mg & 411711, 419472 & & Sc$_2$Ni$_2$Sn & 54348 & & Pb$_2$Br$_2$CO$_3$ & 250396, 29114 \\
\hline
Na$_2$Bi$_5$AuO$_{11}$ & 74365, 164986 & & Mn$_2$Ga$_5$ & 249632, 634613, & & La$_7$Ni$_2$Zn & 159116 \\
 & & & &  634639 & & & \\
\hline
La$_2$Ni$_2$Mg & 107327 & & La$_2$Cu$_2$In & 411708, 628002 & & K$_3$Li$_2$(NbO$_{3}$)$_5$ & 164890 \\
\hline
K$_2$LaTa$_5$O$_{15}$ & 421750 & & Co$_2$Zr$_2$In & 107331 & & CeB$_4$ & 417745, 24682, \\
 & & & & & & & 612705 \\
\hline
CoIn$_3$ & 102501, 623922 & &  Ba$_4$In$_2$Te$_2$Se$_5$ & 425588 & & BaHg$_2$O$_2$Cl$_2$ & 77509 \\
\hline
Ba$_3$Nb$_5$O$_{15}$ & 69993 & & Ba$_3$Ta$_5$O$_{15}$ & 79810 & & CeB$_2$C$_2$ & 88560, 94036 \\
 & & & & & & & 280180, 88857 \\
\hline
Ca$_2$Au$_2$Pb & 409531 & & Cs$_3$GeF$_7$ & 202917 & & Hg$_2$PbI$_2$S$_2$ & 59204 \\
\hline
KAlF$_4$ & 16413, 60524, & & K$_2$CsPdF$_5$ & 72301 & & KCuF$_3$ & 21110 \\ 
 & 201947 & & & & & & \\
\hline
KMo$_4$O$_6$ & 68533 & & LaB$_2$C$_2$ & 94035 & & La$_2$InGe$_2$ & 87511 \\ 
\hline
Li$_2$Sn$_5$ & 26200 & & Mg$_2$SiIr$_5$B$_2$ & 69487 & & NaMo$_4$O$_6$ & 40962 \\ 
\hline
SnMo$_4$O$_6$ & 92839 & & V$_3$B$_2$ & 88317, 107318, & & La$_2$Pd$_2$Pb & 99190 \\
 & & & & 615662 & & & \\ 
\hline
Y$_2$Pd$_2$Pb & 99189 & & Y$_2$BaPdO$_5$ & 202819 & & NaTaO$_3$ & 23322, 88377 \\
 & & & & & & & 239692, 280101 \\ 
\hline
La$_2$Ni$_2$Mg & 107327 & & La$_2$Cu$_2$Mg & 411709 & & La$_2$Ni$_5$C$_3$ & 62277, 67376 \\ 
\hline
Ta$_4$ & 54204 & & Ti$_2$In$_5$ & 401730 & & Au$_2$InY$_2$ & 658835 \\ 
\hline
Au$_2$SnTb$_2$ & 658834 & & Ta$_3$Ga$_2$ & 107309, 635467, & & Tl$_2$GeTe$_5$ & 69035 \\
 & & & & 635490 & & &  \\
\hline
\end{tabular}
\caption{Materials in SG 127 identified by DFT to be gapless at the Fermi energy or topologically trivial by glide indices.}
\label{tb:SG127}
\end{center}
\end{table}

Though our investigations only yielded four materials with band gaps at the Fermi energy and nontrivial glide indices, we note that expanding consideration to materials with greater unit cell complexity, the ICSD reports over 2000 previously synthesized materials in SGs 55 and 127 alone.  By examining these more complicated materials, or by performing strain or symmetry-preserving chemical substitution, improved topological double-glide materials candidates should be readily discoverable.

Of the two materials presented in the main text, we first detail our analysis of Ba$_5$In$_2$Sb$_6$ in SG 55~\cite{ZJGermanBA}.  We then consider members of the the Si$_2$U$_3$-like A$_2$B$_3$ family of materials in SG 127, which include Sr$_2$Pb$_3$, highlighted in the main text, as well as Au$_2$Y$_3$ and Hg$_2$Sr$_3$.  All of the numerical data for the first-principles calculations and figures in this work are freely accessible at~\url{https://dataverse.harvard.edu/dataset.xhtml?persistentId=doi:10.7910/DVN/EUGQDU}.

\subsection{Ba$_5$In$_2$Sb$_6$ in SG 55}
\label{sec:ZJmatdeets}

To explore the electronic properties of Ba$_5$In$_2$Sb$_6$ (ICSD No. 62305), we performed first-principles calculations with the projector-augmented wave (PAW) potential method~\cite{PAW} as implemented in the Vienna ab initio simulation package software (\textsc{VASP})~\cite{VASP}, using the Perdew-Burke-Ernzerhof (PBE)-type generalized gradient approximation (GGA)~\cite{Perdew96p3865} for the exchange-correlation functional.  The cutoff energy for the wave-function expansion was set to 400 eV, and the $k$-point sampling grid was set at $6\times 4 \times 12$.  The experimental lattice parameters~\cite{ZJGermanBA} were used for Ba$_5$In$_2$Sb$_6$ without structural relaxation, as the band structure near the Fermi level and (thus potentially the bulk topology) was found to be sensitive to the structural parameters, possibly due to the material's relatively small band gap of 5 meV (indirect) and 17 meV (direct).  We additionally confirmed convergency using the above settings.  

To calculate the Wilson loop spectrum, we used the following subroutine in the~\textsc{mlwf.f90} module in VASP to calculate the overlap of wavefunctions:
\\ \\
\textsc{CALL PEAD\_CALC\_OVERLAP(W,KI,KJ,ISP,P,CQIJ,LATT\_CUR,T\_INFO,S,LQIJB$=$.TRUE.)} \\
\\
which we then inputted into Eq.~(\ref{eq:SupWilsoncont}).  The Wilson loop spectrum was calculated for the highest 60 valence bands.  To separate the Wilson spectrum into glide sectors, we determined the position-space form of the glide matrices $\hat{g}_{x,y}$ in the plane wave basis:
\begin{equation}
\psi_{\bf k}^{n}({\bf r}) = \langle {\bf r }|\psi^{n}_{\bf{k}}\rangle\equiv\sum_{\bf{G}}C^{n}_{\bf{k}+\bf{G}}e^{i(\bf{k}+\bf{G})\cdot\bf{r}},
\end{equation}
where, along the glide-invariant lines:
\begin{equation}
\hat{g}_{x,y}\psi_{\bf k}^{n}({\bf r}) = \pm i e^{ik_{y,x}}\psi_{\bf k}^{n}({\bf r}).
\end{equation}
We then calculated the Wilson loop individually for the energy eigenstates within each sector of $\hat{g}_{x,y}$.  We determined the bulk topology of Ba$_5$In$_2$Sb$_6$ by utilizing the rules in~\ref{sec:Z4}.  In Fig.~\ref{Fig_BABand}(d) we draw a dashed green line across the Wilson spectrum and count the glide-polarized Wilson bands that cross it.  Calculating the eigenvalues of $g_{y}$, where valid, along the path $\bar{M}\bar{Y}\bar{\Gamma}\bar{X}$, we count no bands in the $(+)$ sector along $\bar{M}\bar{Y}$, one positively and one negatively sloped band along $\bar{Y}\bar{\Gamma}$, and no bands in the $(+)$ sector along $\bar{\Gamma}\bar{X}$.  This results in $\chi_{y}= 0 + 1 -1 + 0 \mod 4 = 0$.  To obtain $\chi_{x}$, we perform a similar calculation using the Wilson spectrum along $\bar{M}\bar{X}\bar{\Gamma}\bar{Y}$ labeled by the eigenvalues of $g_{x}$ along the two lines where that symmetry is valid.  We count no bands in the $(+)$ sector along $\bar{M}\bar{X}$, no bands along $\bar{X}\bar{\Gamma}$, and one negatively sloped band in the $(+)$ sector along $\bar{\Gamma}\bar{Y}$ (the slope here is taken to be negative because we are moving right to left along this line in Fig.~\ref{Fig_BABand}(d)).  This results in $\chi_{x} = 0 + 0 + (0-1)\times 2 \mod 4 = 2$, giving an overall bulk topology of $(\chi_{x},\chi_{y})=(2,0)$.  The projected surface states were obtained from the surface Green's function of a semi-infinite system. For this purpose, we constructed maximally localized Wannier functions for the $s$ orbitals of Ba and the $p$ orbitals of Sb from first-principles calculations using \textsc{Wannier90}~\cite{ZJWannier, Marzari97p12847, Mostofi08p685, Souza01p035109}.

\subsection{Sr$_2$Pb$_3$, Au$_2$Y$_3$, and Hg$_2$Sr$_3$ in SG 127}
\label{sec:YKmatdeets}

We find that three previously-synthesized members of the Si$_2$U$_3$ family of materials in SG 127, Sr$_2$Pb$_3$ (ICSD No. 105627)~\cite{Merlo84p78,Bruzzone81p155}, Au$_2$Y$_3$ (ICSD No. 262043)~\cite{Chai11pi53}, and Hg$_2$Sr$_3$ (ICSD Nos. 107371, 247135)~\cite{Druska96p401,Gumiski05p81}, are capable of hosting double-glide topological crystalline phases.  To explore the electronic properties of these materials, we employed first-principles calculations based on DFT. The PBE-type GGA~\cite{Perdew96p3865} was used to describe exchange correlation as implemented in the \textsc{Quantum Espresso} package~\cite{Giannozzi09p395502}. Core electrons were treated by norm-conserving,  optimized, designed nonlocal pseudopotentials, generated using the \textsc{Opium} package~\cite{Rappe90p1227,Ramer99p12471}.  We used the energy threshold of 680 eV for the plane wave basis. Relativistic effects of spin-orbit interaction were fully described using a non--collinear scheme. The atomic structures were initially obtained from the ICSD~\cite{ICSD}, and then fully relaxed to achieve consistency between the electronic and structural states using a force threshold of 0.005 eV/\AA.  The lattice constants of Sr$_2$Pb$_3$ changed negligibly after the full structural relaxation from a (c) = 8.367 (4.883) \AA\ to 8.383 (4.944) \AA.   We calculated the Wilson bands of Sr$_2$Pb$_3$ within each glide sector following the methodology detailed in~\ref{sec:ZJmatdeets}.  We found that the bulk topology was unaffected by this structural relaxation.  The Wilson bands of Sr$_2$Pb$_3$ were calculated using 20 bands from the $N-19$-th to the $N$-th bulk bands, where N is the number of the valence electrons per unit cell.  We determined the bulk topology of Sr$_2$Pb$_3$ by utilizing the rules in~\ref{sec:Z4}.  In Fig.~\ref{Fig_SRBand}(d) we draw a dashed green line across the Wilson spectrum and count the glide-polarized Wilson bands that cross it.  Calculating the eigenvalues of $g_{y}$, where valid, along the path $\bar{M}\bar{X}'\bar{\Gamma}\bar{X}$, we count one positively sloped $(+)$ band along $\bar{M}\bar{X}'$, one positively and one negatively sloped band along $\bar{X}'\bar{\Gamma}$, and no bands in the $(+)$ sector along $\bar{\Gamma}\bar{X}$.  This results in $\chi_{y}=(1\times 2) + 1 -1 + 0 \mod 4 = 2$.  By the $C_{4z}$ symmetry of SG 127, $\chi_{x}=\chi_{y}$, giving an overall bulk topology of $(\chi_{x},\chi_{y})=(2,2)$, that of a nonsymmorphic Dirac insulator.  The surface states of Sr$_2$Pb$_3$ were obtained by calculating the surface Green's function for a semi-infinite system~\cite{Sancho84p1205,Sancho85p851} based on the maximally localized Wannier functions for the $d$ orbitals of Sr and the $p$ orbitals of Pb using \textsc{Wannier90}~\cite{ZJWannier, Marzari97p12847, Mostofi08p685, Souza01p035109}.

\begin{figure*}[tp!]
\includegraphics[width=0.9\textwidth]{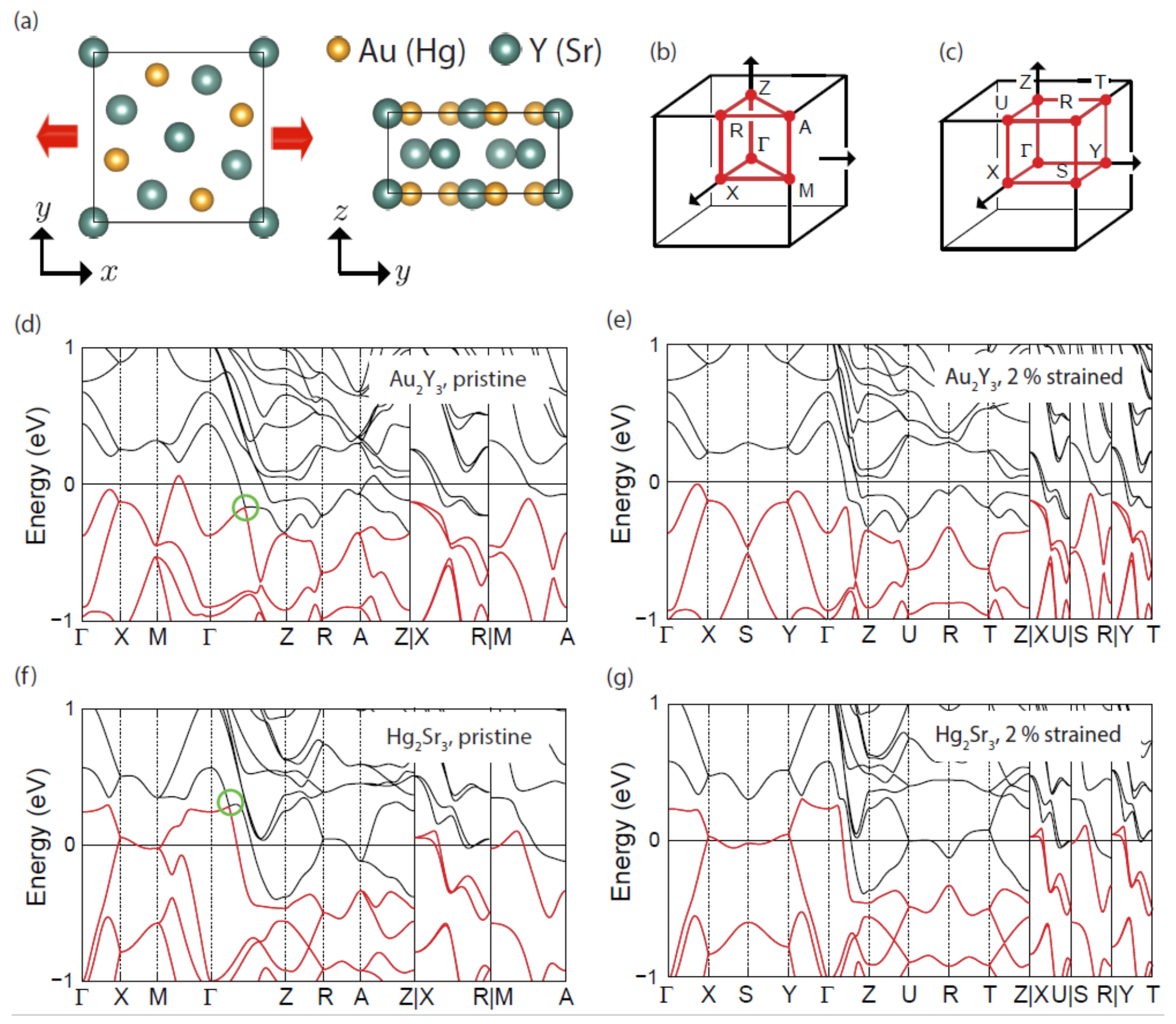}
\caption{The crystal and electronic band structures of Au$_2$Y$_3$ and Hg$_2$Sr$_3$.  In their natural forms in SG 127, they are gapless, with $C_{4z}$-protected Dirac points near the Fermi energy, indicated here with green circles.  We therefore apply $(100)$-direction strain to open up a gap and allow for the possibility of topological crystalline insulating bulk topologies protected by the remaining $(001)$-surface wallpaper group symmetries ($pgg$).  (a) The unit cell of Au$_2$Y$_3$ (Hg$_2$Sr$_3$). The red arrows illustrate the direction of strain. (b) The Brillouin zone of tetragonal SG 127. (c) The Brillouin zone of orthorhombic SG 55, the space group which results when $C_{4z}$ in SG 127 is broken while preserving the two glide symmetries. The electronic band structures of (d) pristine and (e) 2 \% strained Au$_2$Y$_3$. The electronic band structures of (f) pristine and (g) 2 \% strained Hg$_2$Sr$_3$. The bands with indices lower than $N-1$ for Au$_2$Y$_3$ ($N$ for Hg$_2$Sr$_3$) are highlighted in red.}
\label{ay_sr}
\end{figure*}

As detailed in the main text and in Fig.~\ref{ay_sr}, all three of these materials share the same unit cell structure, with two orthogonal glide mirrors in the $x$ and $y$ directions related by $C_{4z}$ symmetry and a mirror in the $z$ direction that relates the upper and lower layers of the unit cell.   We find that Sr$_2$Pb$_3$ posses a gap at the Fermi energy at each crystal momentum but that, as shown in Fig.~\ref{ay_sr}(d,f), pristine Au$_2$Y$_3$ and Hg$_2$Sr$_3$ are gapless, with $C_{4z}$-protected Dirac points near the Fermi energy.  Applying strain in the $(100)$-direction ($x$) breaks $C_{4z}$ while still preserving the two glides that project to form wallpaper group $pgg$ on the $(001)$-surface.  As this lower-symmetry bulk system still possess inversion and time-reversal symmetries, Weyl points are forbidden~\cite{Vafek14}, and therefore under $C_{4z}$-breaking strain the Dirac points in Au$_2$Y$_3$ and Hg$_2$Sr$_3$ become fully gapped.  We applied incremental strain and subsequent relaxation of the internal atomic coordinates and evaluated the symmetries of the strained systems using \textsc{FINDSYM}~\cite{findsym} with a tolerance of 0.001 \AA.  We find that Au$_2$Y$_3$ develops a gap for up to 5\% strain, whereas  Hg$_2$Sr$_3$ develops a gap for up to around 4\% strain, after which it becomes unstable and undergoes a structural transition to the symmorphic SG 75. 

\begin{figure*}[tp!]
\includegraphics[width=0.7\textwidth]{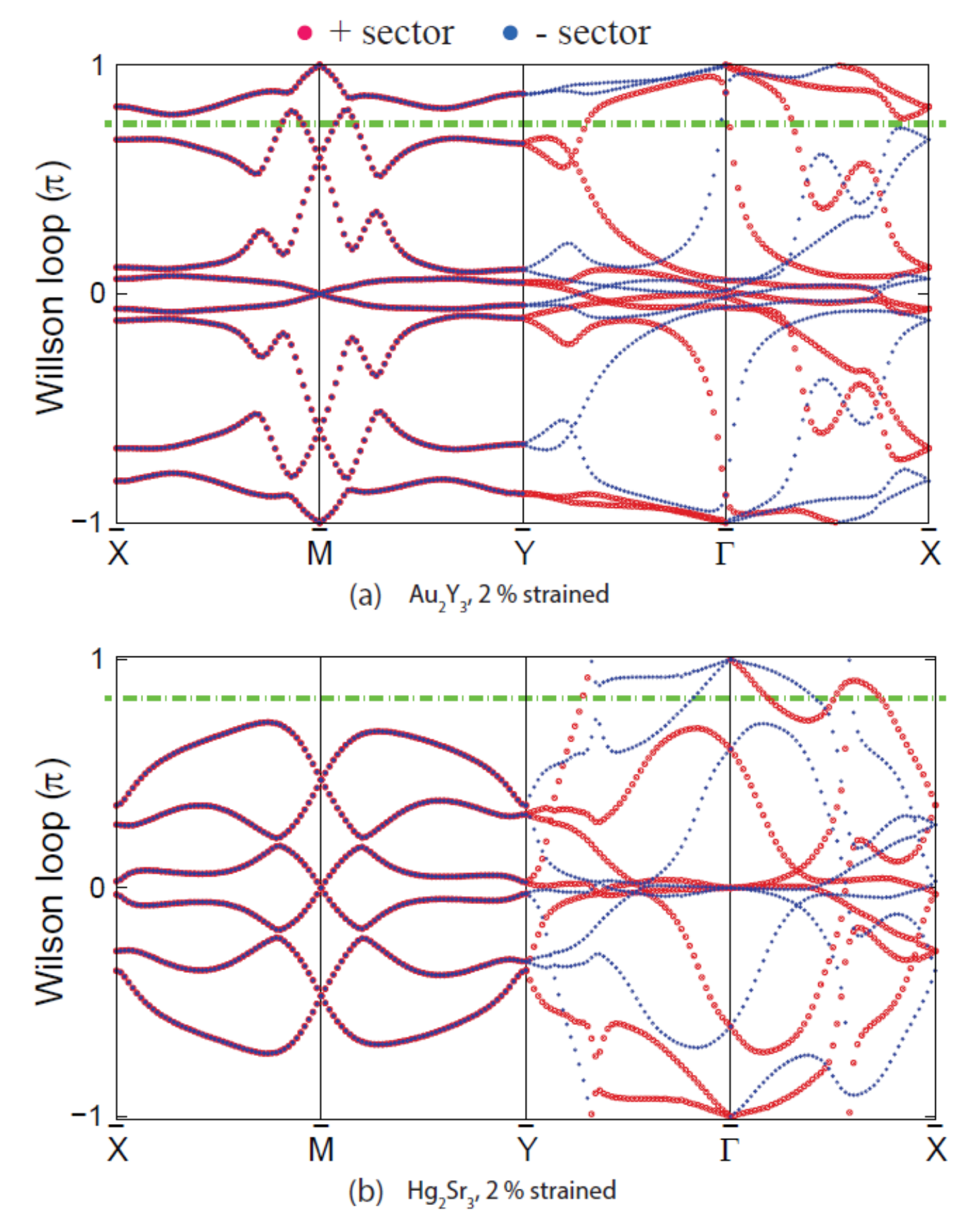}
\caption{The glide-resolved Wilson bands of (a) 2 \%-strained Au$_2$Y$_3$ and (b) 2 \%-strained Hg$_2$Sr$_3$, with red (blue) points indicating Wilson bands with glide eigenvalues $\lambda^{+ (-)}_{x,y}$.  Using the rules detailed in~\ref{sec:Z4} for evaluating the $\mathbb{Z}_{4}$ invariants on the dotted green line in each plot, we find that both strained systems host the $(\chi_x,\chi_y) = (0,2)$ double-glide topological hourglass connectivity. }
\label{ay_wl} 
\end{figure*}

For each of these materials, we therefore evaluate the $(001)$-directed Wilson loop under 2\% strain, shown in Fig.~\ref{ay_wl}, where the Wilson bands within each glide sector are calculated following the methodology detailed in~\ref{sec:ZJmatdeets}.  The Wilson bands for Au$_2$Y$_3$ are calculated using the 16 bands from the $N-17$-th to the $N-2$-th bands, valid as the band gap between the $N-2$-th and the $N-1$-th bands widely overlaps with the Fermi level.  For Hg$_2$Sr$_3$, we use the 12 bands from band indices $N-11$ to $N$.  Though the Wilson spectra for these materials appear to be more complicated than those of Ba$_5$In$_2$Sb$_6$ and Sr$_2$Pb$_3$, we can still use the simple rules developed in \ref{sec:Z4} to evaluate the $\mathbb{Z}_{4}$ indices.   

We determine the bulk topology of these two strained materials by following the rules detailed in~\ref{sec:Z4} and used previously in~\ref{sec:ZJmatdeets} to evaluate the topology of Ba$_5$In$_2$Sb$_6$ in SG 55.  For 2\%-strained Au$_2$Y$_3$, we draw a dashed green line across the Wilson spectrum in Fig.~\ref{ay_wl}(a) and count the glide-polarized Wilson bands that cross it.  Calculating the eigenvalues of $g_{y}$, where valid, along the path $\bar{M}\bar{Y}\bar{\Gamma}\bar{X}$, we count one positively and one negatively sloped band in the $(+)$ sector along $\bar{M}\bar{Y}$, two positively sloped bands along $\bar{Y}\bar{\Gamma}$, and two negatively sloped bands in the $(+)$ sector along $\bar{\Gamma}\bar{X}$.  This results in $\chi_{y}= (1-1)\times 2 + 2 + (0-2)\times 2 \mod 4 = 2$.  To obtain $\chi_{x}$, we perform a similar calculation using the Wilson spectrum along $\bar{M}\bar{X}\bar{\Gamma}\bar{Y}$ labeled by the eigenvalues of $g_{x}$ where applicable.  We note that as we are now taking a right-to-left path in Fig.~\ref{ay_wl}(a), the positive or negative Wilson band slope labeling will here be the \emph{opposite} of what it was in the calculation of $\chi_{y}$.  We count one positively and one negatively sloped band in the $(+)$ sector along $\bar{M}\bar{X}$, two positively sloped bands along $\bar{X}\bar{\Gamma}$, and one negatively sloped band in the $(+)$ sector along $\bar{\Gamma}\bar{Y}$.  This results in $\chi_{x} = (1 -1)\times 2 + 2 + (0-1)\times 2 \mod 4 = 0$, giving an overall bulk topology of $(\chi_{x},\chi_{y})=(0,2)$.

For 2\%-strained Hg$_2$Sr$_3$, we draw a dashed green line across the Wilson spectrum in Fig.~\ref{ay_wl}(b) and count the glide-polarized Wilson bands that cross it.  Calculating the eigenvalues of $g_{y}$, where valid, along the path $\bar{M}\bar{Y}\bar{\Gamma}\bar{X}$, we count no bands in the $(+)$ sector along $\bar{M}\bar{Y}$, two positively sloped bands along $\bar{Y}\bar{\Gamma}$, and two positively and two negatively sloped bands in the $(+)$ sector along $\bar{\Gamma}\bar{X}$ (one of the $(+)$ bands has a very sharp slope, and its continuity across the green line can be inferred by tracing where it appears again at $\theta=-\pi$).  This results in $\chi_{y}= 0 + 2 + (2-2)\times 2 \mod 4 = 2$.  To obtain $\chi_{x}$, we perform a similar calculation using the Wilson spectrum along $\bar{M}\bar{X}\bar{\Gamma}\bar{Y}$ labeled by the eigenvalues of $g_{x}$ where applicable.  We note that as we are now taking a right-to-left path in Fig.~\ref{ay_wl}(b), the positive or negative Wilson band slope labeling will here be the \emph{opposite} of what it was in the calculation of $\chi_{y}$.  We count no bands in the $(+)$ sector along $\bar{M}\bar{X}$, four positively and two negatively sloped bands along $\bar{X}\bar{\Gamma}$, and one negatively sloped band in the $(+)$ sector along $\bar{\Gamma}\bar{Y}$.  This results in $\chi_{x} = 0 + 4 - 2 + (0-1)\times 2 \mod 4 = 0$, giving an overall bulk topology of $(\chi_{x},\chi_{y})=(0,2)$.

To summarize, we find that both 2\% strained Au$_2$Y$_3$ and 2\% strained Hg$_2$Sr$_3$ host the $(\chi_x,\chi_y) = (0,2)$ double-glide topological hourglass connectivity. 

\section{Tight-Binding Model and the SSH Limit}
\label{sec:myTB}

In this section, we present a simplified tight-binding model that can realize all of the $\mathbb{Z}_{4}\times\mathbb{Z}_{2}$ insulating phases allowed for wallpaper groups $pgg$ and $p4g$, which have perpendicular glides, $g_{x,y}$, in the $x$ and $y$ directions.  For simplicity, we further specialize to systems with inversion symmetry, $\mathcal{I}$; as shown in~\ref{sec:z4inv}, this simplifies the computation of the $z$-projected Wilson loop.  Inversion symmetry also implies the presence of a mirror in the $z$ direction: $\mathcal{I}=g_{x}g_{y}m_{z}$.  We find a fine-tuned limit in which the $\mathbb{Z}_4$ topological invariants defined in~\ref{sec:Z4} can be computed by comparing the relative values of three Su-Schrieffer-Heeger (SSH) $\mathbb{Z}_2$ invariants~\cite{SSH} defined by effective 1D chains at the corners of the BZ which project to $\bar{X}$, $\bar{Y}$, and $\bar{M}$ on the $z$-normal surface.  This limit provides an alternative, intuitive way to understand the four topological phases where $\chi_x = \chi_y= 0\mod 2$.  We note that the Wilson spectrum pinning in this tight-binding model can alternatively be understood in terms of the bulk inversion eigenvalues as detailed in~\onlinecite{Alexandradinata14}.

\subsection{Tight-Binding Model for Space Groups 55 and 127}
\label{sec:TBwC4}

As a starting point, consider a single layer of the two-site unit cell shown in Fig.~\ref{fig:combinedWall}.  The sites, designated $A$ and $B$, are related to each other by glide reflections, $g_{x,y}=\{m_{x,y}|\frac{1}{2}\frac{1}{2}0\}$; the origin is defined to sit at an $A$ site.  An orthorhombic stack of this single layer, with no other symmetries, is in space group 32, $Pba2$, and its $z$-normal $(001)$ surface is characterized by the two-dimensional wallpaper group, $pgg$.

\begin{figure}[t]
 \centering
\includegraphics[width=0.85\textwidth]{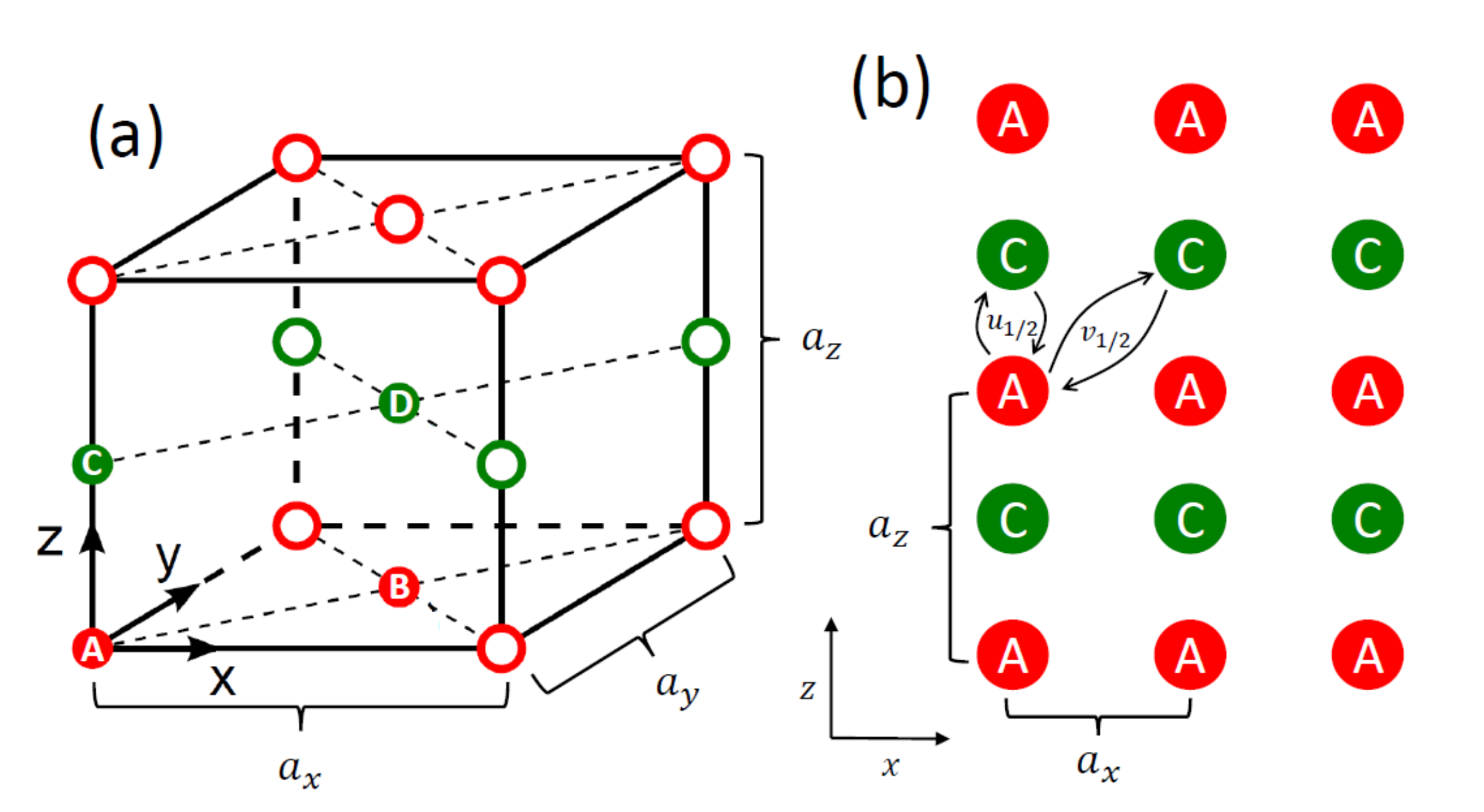}
\caption{The unit cell (a) for our tight-binding model.  When $a_{x}=a_{y}$, the bulk is in space group 127 and has a $z$-normal wallpaper group $p4g$; when $a_{x}\neq a_{y}$, the bulk is in space group 55 and has a $z$-normal wallpaper group $pgg$.  The layers (labeled in alternating red and green) each contain two sites related by the bulk glide reflections $g_{x,y}$.  The red and green layers are related to each other by mirrors $m_{z}$ that sit at $z=1/4,3/4$, in units of the lattice constant $a_{z}$.  We first determine in Eqs.~(\ref{eq:Hxy55}) and (\ref{c4map}) all of the symmetry-allowed in-plane hopping terms up to second-nearest neighbor.  By artificially turning off some of the SOC terms, we reach a limit where, if we then couple the layers, this system can effectively be described by two independent copies of the Su-Schrieffer-Heeger (SSH) model~\cite{SSH} living at the BZ boundary.  By tuning the relative values of two terms which dimerize the layers (b), we can control the relative polarization invariants of the two SSH chains and therefore the overall bulk topology.  Panel (b) shows the $y=0$ plane; there are also equivalent hopping terms $u_{1,2}$ and $v_{1,2}$ between the $B$ and $D$ sublattices on the $y=1/2$ plane.}
\label{fig:tbLattice}
\end{figure}

We enforce an extra mirror symmetry, $m_{z}=\{m_{z}|00\frac{1}{2}\}$, by adding two more sublattices, $C$ and $D$, sitting $t_{z/2}$ above the $A$ and $B$ sites respectively (Fig.~\ref{fig:tbLattice}(a)); this mirror also adds an inversion symmetry $\mathcal{I}=g_{x}g_{y}m_{z}$.  The resulting system is in space group 55, $Pbam$.  In this space group, when the two layers are decoupled, each sheet is equivalent to a single layer of the two-dimensional Dirac semimetal model in~\onlinecite{Steve2D}, which possesses fourfold degeneracies at $X$, $Y$, and $M$ due to the algebraic relations between $\mathcal{I}$, $g_{x}$, and $g_{y}$ at those TRIMs.  For our 3D orthorhombic system, at all six bulk TRIMs for which $k_{x}$ or $k_{y}$ is equal to $\pi$, states are fourfold-degenerate by the algebra from~\ref{sec:Diracdouble} and Eq.~(\ref{someDiracAlgebra}).  At the four TRIMs for which $(k_{x},k_{y})=(0,\pi)$ or $(\pi,0)$, at least one of the glides anticommutes with inversion, which, combined with $\mathcal{I}^{2}=-\mathcal{T}^{2}=+1$, requires states to be fourfold-degenerate.  At the two TRIMs for which $(k_{x},k_{y})=(\pi,\pi)$, the two glides anticommute and square to $+1$, which, combined with $\mathcal{T}^{2}=-1$, also enforces a fourfold degeneracy.  

The following Hamiltonian includes all allowed in-plane hopping terms up to second-nearest-neighbor:
\begin{eqnarray}
\mathcal{H}_{xy}(\vec{k}) &=& \cos\left(\frac{k_{x}}{2}\right)\cos\left(\frac{k_{y}}{2}\right)\left[t_{1}\tau^{x} + v_{r1}\tau^{y}\sigma^{z}\right] \nonumber \\
&+& \sin\left(\frac{k_{x}}{2}\right)\cos\left(\frac{k_{y}}{2}\right)\left[v_{s1}\tau^{x}\mu^{z}\sigma^{y}\right] \nonumber \\
&+& \cos\left(\frac{k_{x}}{2}\right)\sin\left(\frac{k_{y}}{2}\right)\left[v_{s1}'\tau^{x}\mu^{z}\sigma^{x}\right] \nonumber \\
&+& \cos\left(k_{x}\right)t_{2x} + \cos\left(k_{y}\right)t_{2y} \nonumber \\
&+& \sin\left(k_{x}\right)\left[v_{s2}\tau^{z}\mu^{z}\sigma^{x} + v_{s2}'\mu^{z}\sigma^{y}\right] \nonumber \\
&+& \sin\left(k_{y}\right)\left[v_{s2}''\mu^{z}\sigma^{x} + v_{s2}'''\tau^{z}\mu^{z}\sigma^{y}\right],
\label{eq:Hxy55}
\end{eqnarray}
where $\tau^x$ corresponds to hopping between the $A$ and $B$ (or $C$ and $D$) orbitals, $\mu^x$ corresponds to hopping between the $A$ and $C$ (or $B$ and $D$), orbitals and the $\sigma$ Pauli matrices correspond to an on-site spin.

To further simplify, we impose $C_{4z}$ symmetry, which is implemented by the operator, $C_{4z}=\sqrt{i\sigma^{z}}f_{4z}(\vec{k})$, where $f_{4z}(\vec{k})$ acts on the crystal momenta by enforcing the cyclical mapping:
\begin{eqnarray}
\sigma^{x}\rightarrow\sigma^{y},\ \sigma^{y}\rightarrow-\sigma^{x},\ \sigma^{z}\rightarrow\sigma^{z} \nonumber \\
k_{x}\rightarrow k_{y},\ k_{y}\rightarrow -k_{x},\ k_{z}\rightarrow k_{z}.
\label{c4map}
\end{eqnarray}
With this additional symmetry, $v_{s1}=-v_{s1}',\ v_{s2}=v_{s2}''',\ v_{s2}'=-v_{s2}''$ in Eq.~(\ref{eq:Hxy55}) and the system is now in the higher-symmetry space group 127, $P4/mbm$, with an in-plane Hamiltonian up to second-nearest neighbor hopping given by Eq.~(\ref{eq:Hxy55}) with these restrictions.

We now introduce hopping in the $z$-direction to couple the layers.  We begin in a relatively artificial limit by only turning on specific SOC-free dimerizing terms, (Fig.~\ref{fig:tbLattice}(b)):
\begin{eqnarray}
V_{z}(k_{z}) &=& \cos\left(\frac{k_{z}}{2}\right)u_{1}\mu^{x} + \sin\left(\frac{k_{z}}{2}\right)u_{2}\mu^{y}  \nonumber \\
&+& \cos\left(\frac{k_{z}}{2}\right)\left[\cos\left(k_{x}\right) + \cos\left(k_{y}\right)\right]v_{1}\mu^{x} \nonumber \\
&+& \sin\left(\frac{k_{z}}{2}\right)\left[\cos\left(k_{x}\right) + \cos\left(k_{y}\right)\right]v_{2}\mu^{y}
\end{eqnarray}
and
\begin{equation}
\mathcal{H}_{127}(\vec{k}) = \mathcal{H}_{xy}(\vec{k}) + V_{z}(k_{z}).
\label{H127eq}
\end{equation}
The terms proportional to $u_{1,2}$ correspond to hopping between nearest-neighbor $A$ ($B$) and $C$ ($D$) sites.  Terms proportional to $v_{1,2}$ originate from longer-range versions of the same type of hopping, and connect $A$ ($B$) and $C$ ($D$) sites separated by $\vec{d}=\{1 0 \frac{1}{2}\}$ and $\vec{d}=\{0 1 \frac{1}{2}\}$.  

 \begin{figure}[t]
 \centering
\includegraphics[width=0.8\textwidth]{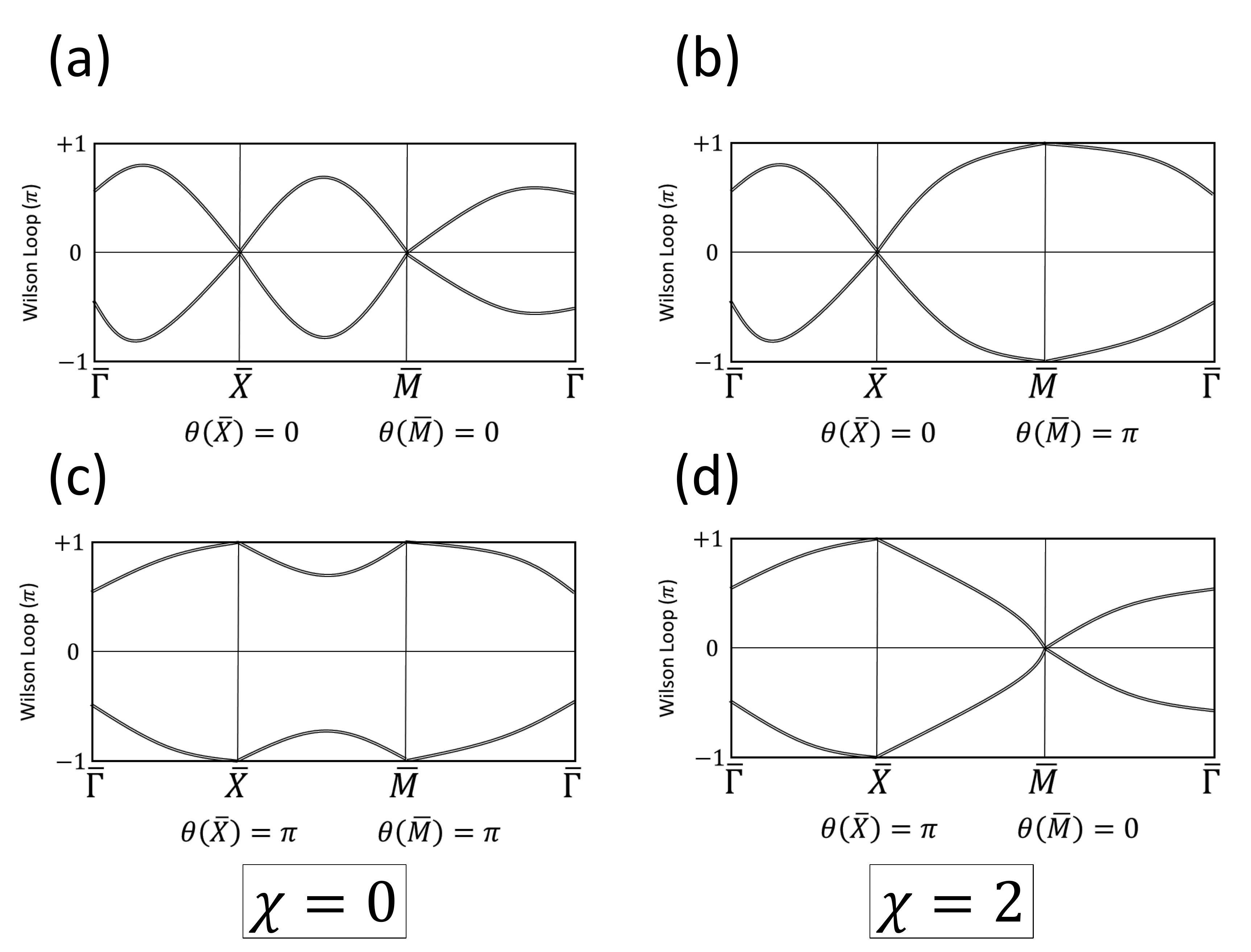}
\caption{The $\mathbb{Z}_{2}\times\mathbb{Z}_{2}$ QSH-trivial Wilson loops (a-d) for our model of SG 127 (Eq.~(\ref{H127eq})) in the SSH limit, sorted by the values of the SSH polarization invariants $\theta(\bar{X}/\bar{M})$.  Even though there are 4 possible Wilson spectra, there are only 2 topologically distinct connectivities, characterized by the $\mathbb{Z}_{2}$ invariant $\chi$ defined in Eq.~(\ref{SSHchi}) by the \emph{relative} values of $\theta(\bar{X}/\bar{M})$.}
\label{fig:SSHcomposite}
\end{figure}

\begin{figure}[t]
 \centering
\includegraphics[width=1.0\textwidth]{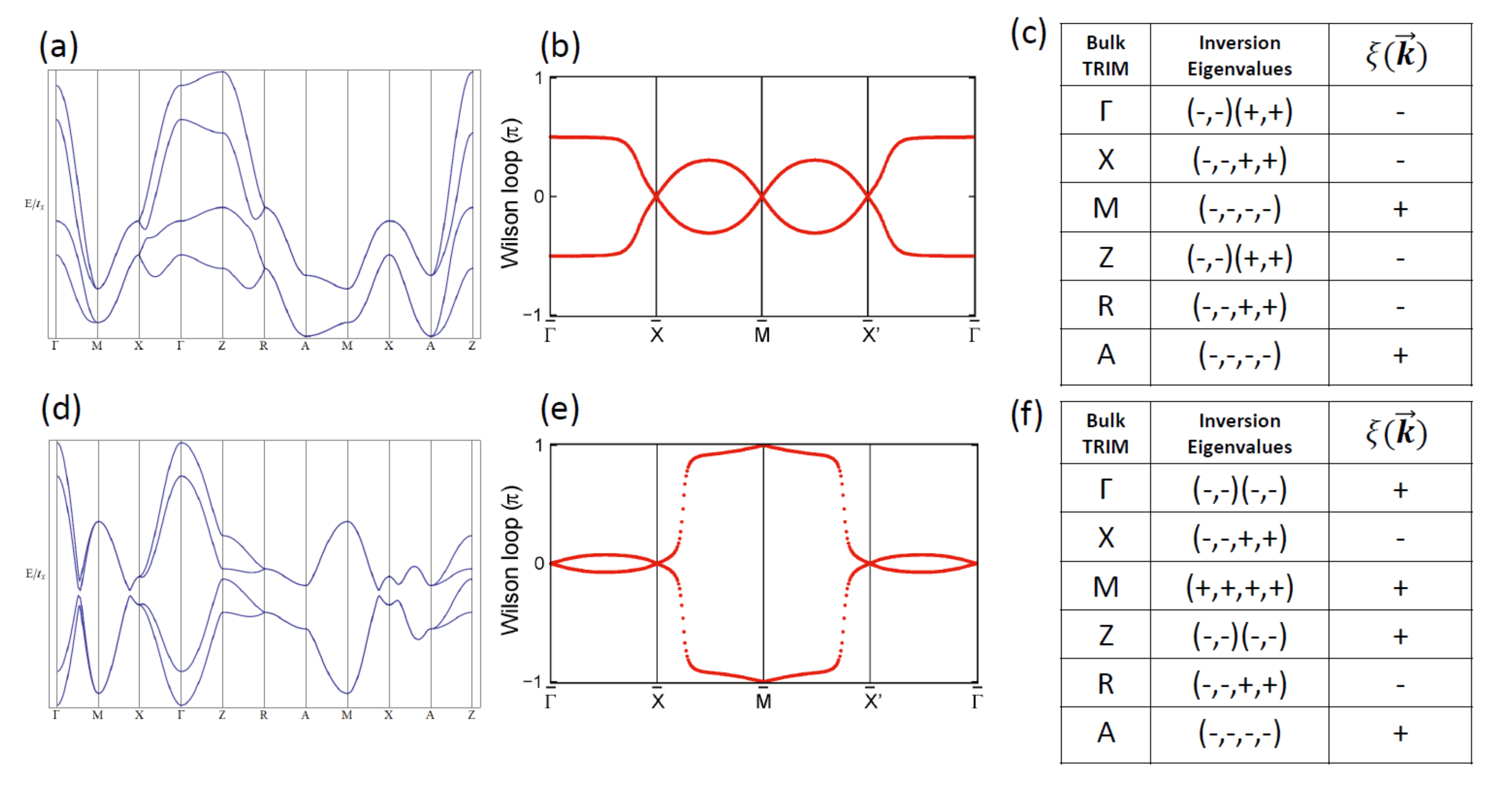}
\caption{Energy bands and $z$-projection Wilson bands for the SSH limit of the tight-binding model for SG 127 in Eq.~(\ref{H127eq}), with the filling chosen such that the bottom four bands are occupied.  Bands (a) and Wilson bands (b) display the trivial connectivity $\chi=0$ when the polarization invariants at $\bar{X}$ and $\bar{M}$ are the same (obtained using $t_{1}=1,t_{2}=0.5,v_{s2}=-0.2,v_{s2}'=0.15,u_{1}=0.25,u_{2}=0.45,v_{r1}=v_{s1}=v_{1}=v_{2}=0$).  When the polarization invariants at $\bar{X}$ and $\bar{M}$ differ, which can be induced by a band inversion about an $\bar{M}$-projecting TRIM (d), $\chi=2$ and the Wilson spectrum is nontrivially connected (e), demonstrating the SSH limit of the nonsymmorphic Dirac insulating phase in SG 127 (obtained using Eq.~(\ref{H127eq}) with $t_{1}=1,t_{2}=0.5,v_{s2}=-0.2,v_{s2}'=0.15,u_{1}=0.85,u_{2}=1.3,v_{1}=3,v_{r1}=v_{s1}=v_{2}=0$).  Bulk inversion eigenvalues and their products for the lower four bands of the trivial (c) and nonsymmorphic Dirac (f) insulating phases, using the conventional TRIM labeling for the tetragonal BZ (Fig.~\ref{ay_sr}(b)).  Inversion eigenvalues are grouped by parentheses according to their pairing at two- or fourfold bulk degeneracies.  The product of the inversion eigenvalues $\xi(\vec{k})$ at a TRIM with momentum $\vec{k}$ is defined using \emph{only one} inversion eigenvalue per Kramers pair, and the product of $\xi(\vec{k})$ over all 8 bulk TRIMs is the $\mathbb{Z}_{2}$ strong QSH invariant (discussed in further detail in~\ref{sec:z4inv}).  For this realization of the nonsymmorphic Dirac insulating phase, all four of the inversion eigenvalues are the same at the $\bar{\Gamma}$-projecting bulk TRIMs ($\Gamma$ and $Z$), resulting in an additional fourfold degeneracy in the Wilson spectrum at $\bar{\Gamma}$ as detailed in~\onlinecite{Alexandradinata14}.}
\label{fig:SSH}
\end{figure}

We now consider the fine-tuned limit where $v_{r1}=v_{s1}=0$, or one in which there is no spin-orbit coupling at $(k_{x},k_{y})=(\pi,0)$ and $(0,\pi)$.  In this limit, we can write down the bulk Hamiltonian on the line that projects to $\bar{X}$ on the $z$-normal surface:
\begin{equation}
\mathcal{H}_{\bar{X}}^{SSH}= \cos\left(\frac{k_{z}}{2}\right)u_{1}\mu^{x} + \sin\left(\frac{k_{z}}{2}\right)u_{2}\mu^{y},
\end{equation}
where we have shifted the energy to make $t_{2}=0$ without loss of generality.  We note that the $\tau$ sublattice and $\sigma$ spin degrees of freedom no longer play a role, they merely impose additional degeneracies.  

This limit places strong constraints on the Wilson loop bands: since the spins are decoupled, we can consider the spinless time-reversal operator, $\tilde{\mathcal{T}}$, which satisfies $\tilde{\mathcal{T}}^2=+1$ and the spinless glide, $\tilde{g}_y$, which, at the $\bar{X}$ point, satisfies $\tilde{g}_y^2=-1$. Thus, at $\bar{X}$, $\tilde{\mathcal{T}}\tilde{g}_y$ is an antiunitary operator that squares to $-1$, enforcing that all eigenstates are doubly degenerate, within each spin sector.  Since there is no SOC, the two spin sectors are also degenerate, resulting in a fourfold degeneracy at the $\bar{X}$ point, though one which is broken into twofold degeneracies when SOC is realistically reintroduced.  In addition, inversion requires that the Wilson bands are particle-hole symmetric~\cite{Alexandradinata16}.  Thus, our four-band model has all four Wilson bands degenerate at the $\bar{X}$ point and, because of inversion symmetry, they are pinned to either $0$ or $\pi$.

It was shown in~\ref{sec:WilsonUnitary} that all four Wilson bands are degenerate at $\bar{M}$, as well, and by inversion are also pinned to either $0$ or $\pi$.  Since the Wilson bands must continuously connect the bands at $\bar{X}$ to the bands at $\bar{M}$, there is a $\mathbb{Z}_2$ invariant that characterizes the possible connectivities for fixed band inversion at $\bar{\Gamma}$ (Fig.~\ref{fig:SSHcomposite}):
\begin{equation}
\chi=2\left(\left[\frac{1}{\pi}\left(\theta(\bar{M}) - \theta(\bar{X})\right)\right]\mod 2\right).
\label{SSHchi}
\end{equation}
One of these connectivities is a trivial phase and the other is the nonsymmorphic Dirac insulating phase.  In Fig.~\ref{fig:SSH}, we plot the bulk and Wilson bands for our model in this limit and demonstrate both trivial and nonsymmorphic Dirac insulating phases.  As long as the system is QSH-trivial and diagonal mirror Chern-trivial, then the Wilson loop eigenvalues are generically unpinned at $\bar{\Gamma}$ and along all low symmetry lines (though, as seen in Fig.~\ref{fig:SSH}(e), in the non-generic case that all of bulk inversion eigenvalues are the same at the $\bar{\Gamma}$-projecting TRIMs, the Wilson loop will \emph{additionally} be pinned at $\bar{\Gamma}$ by the mechanism detailed in~\onlinecite{Alexandradinata14}).  Therefore, in this particular model of SG 127, all of the possible crystalline connectivities for the path $\bar{\Gamma}\bar{X}\bar{M}\bar{X}'$ are entirely determined by how the eigenvalues are pinned at $\bar{X}$ and $\bar{M}$.  

This limit can also be described by two, spin-degenerate Su-Schrieffer-Heeger (SSH) chains:  the layer degree of freedom $\mu$ acts like the sublattice degree of freedom in the original, two-band SSH model, and each chain in this limit has an additional degeneracy in the spin subspace $\sigma$.  The edge state of one chain is the projection of the Hamiltonian onto $\bar{X}$, and the edge state of the other is the projection onto $\bar{M}$.  The relative values of $u_{1,2}$ and $v_{1,2}$, which parameterize hopping in the $z$-direction, correspond to the two choices of dimerization for each SSH chain.
For each of these two chains, there is a $\mathbb{Z}_{2}$ polarization, $\theta=0,\pi$, which directly corresponds to the Wilson phases at those surface TRIMs, and the overall bulk topology for these crystalline phases is given by the $\mathbb{Z}_{2}$ relative polarization between the two SSH models.  

For the chain which projects to $\bar{M}$, the fourfold surface degeneracy prevents the two Hamiltonians from coupling.  However, for the chain projecting to $\bar{X}$, the two, spin-degenerate SSH Hamiltonians are only prevented from coupling in the limit that there is no SOC at any bulk $k$-point projecting to $\bar{X}$.  As symmetry-allowed SOC terms are turned on, a real system will escape this limit, and the two copies of the SSH Hamiltonian at $\bar{X}$ will couple and their surface states will gap into pairs.  Nevertheless, as discussed in further detail in the next section, if introducing SOC does not result in a bulk gap closure, then the value of $\chi$ cannot change, and the bulk topology will remain the same as it was in the SSH limit.

\begin{figure}[b]
 \centering
\includegraphics[width=1.0\textwidth]{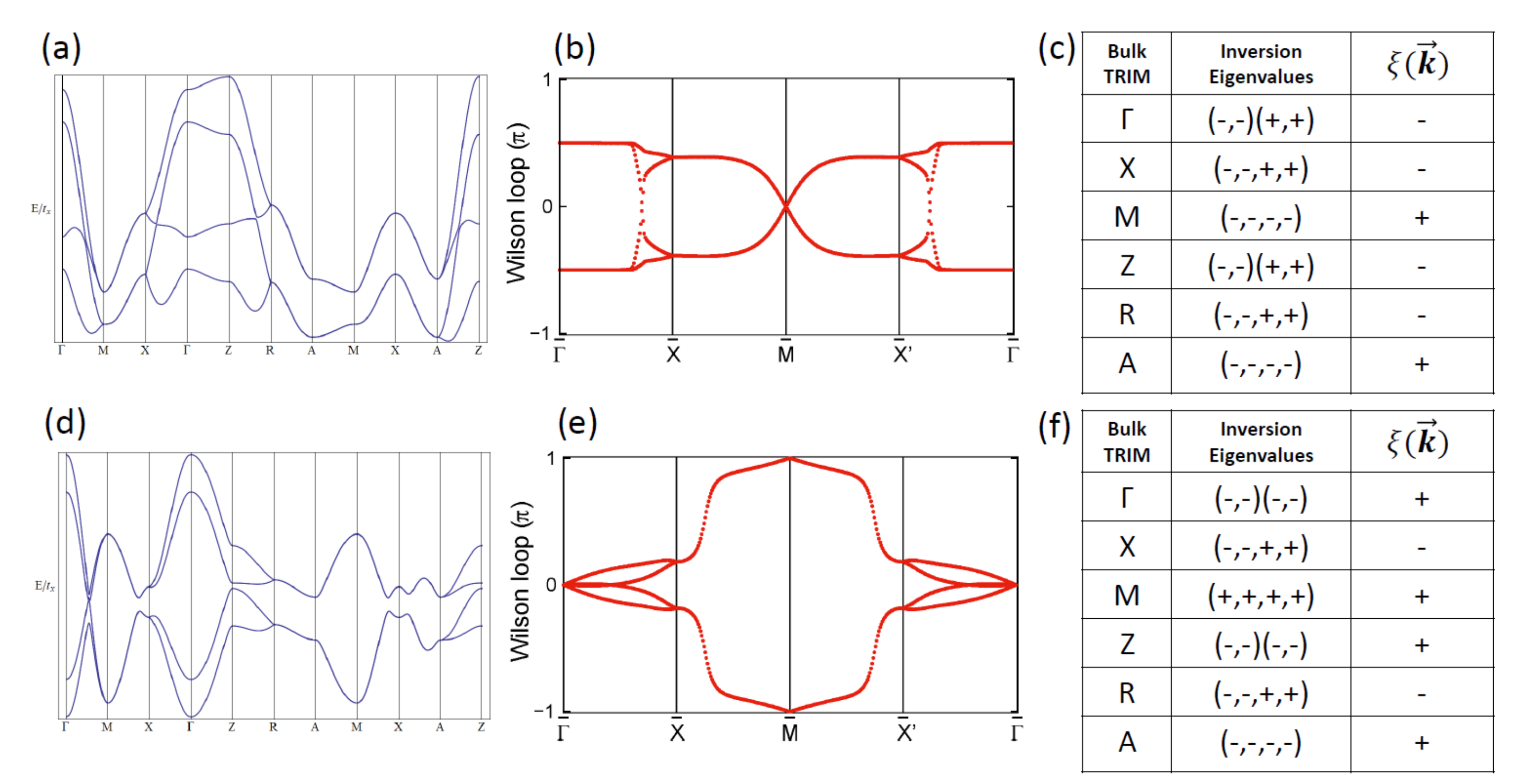}
\caption{Bulk bands (a,d) and Wilson bands (b,e) for the tight-binding model (Eq.~(\ref{H127eq})) away from the SSH limit.  The bands along $\bar{\Gamma}\bar{X}$ open up into hourglasses and the SSH edge states at $\bar{X}$ couple and gap out.  For the trivial phase in panel (a), the bottom four bands approach the top four very closely in a few places, but there remains at each $k$ point a gap such that the four-band Wilson matrix is well-defined for the whole $z$-surface BZ.  These figures were obtained by tuning $v_{r1}\rightarrow0.55,v_{s1}\rightarrow0.4$ from the values used in Fig.~\ref{fig:SSH} for each phase.  Bulk inversion eigenvalues and their products are shown for the lower four bands of the trivial (c) and nonsymmorphic Dirac (f) insulating phases, using the conventional TRIM labeling for the tetragonal BZ (Fig.~\ref{ay_sr}(b)).   Inversion eigenvalues are grouped by parentheses according to their pairing at two- or fourfold bulk degeneracies.  The product of the inversion eigenvalues $\xi(\vec{k})$ at a TRIM with momentum $\vec{k}$ is defined using \emph{only one} inversion eigenvalue per Kramers pair, and the product of $\xi(\vec{k})$ over all 8 bulk TRIMs is the $\mathbb{Z}_{2}$ strong QSH invariant (discussed in further detail in~\ref{sec:z4inv}).  For this realization of the nonsymmorphic Dirac insulating phase, all four of the inversion eigenvalues are the same at the $\bar{\Gamma}$-projecting bulk TRIMs ($\Gamma$ and $Z$), resulting in an additional fourfold degeneracy in the Wilson spectrum at $\bar{\Gamma}$ as detailed in~\onlinecite{Alexandradinata14}.}
\label{fig:noSSH}
\end{figure}

\subsection{Beyond the SSH Limit}

Generically, all symmetry-allowed hopping terms will be present in a real material, including $v_{r1}$, $v_{s1}$, and the other $z$-direction hopping terms.  In this section, we examine how this affects the SSH-model definition of crystalline invariants. 

The line that projects to $\bar{M}$ continues to be described by a doubly degenerate SSH chain even after generic terms are added, as it still hosts a fourfold-degenerate surface state with a $\mathbb{Z}_{2}$ polarization.  As states at $\bar{M}$ are fourfold-degenerate due to surface wallpaper symmetries, the Wilson phases $\theta(\bar{M})$ must still be either $0$ or $\pi$ as long as there is bulk inversion symmetry.  Furthermore, in the limit of weak spin-orbit interaction, a band inversion at an $\bar{M}$-projecting TRIM will result in the presence of bulk Dirac point or line nodes.  However, the properties of these line nodes and their relationship to nonsymmorphic symmetries are space-group dependent, and in general go beyond the focus of this manuscript.

At the TRIMs which project onto $\bar{X}$, the immediate consequences of allowing nonzero values of $v_{r1}$ and $v_{s1}$ are to couple the two copies of the SSH Hamiltonians under $\bar{X}$ and gap out their surface states.  As states at $\bar{X}$ are only generically twofold-degenerate, the presence of additional SOC terms breaks the artificial symmetries $\tilde{g}_{y}$ and $\tilde{\mathcal{T}}$ and gaps the states into the two ends of an hourglass.  Nevertheless, as pictured in Fig.~\ref{fig:noSSH}, this hourglass along $\bar{\Gamma}\bar{X}$ is still characterized by a $\mathbb{Z}_{2}$ invariant that characterizes whether it is centered about a Wilson phase of $0$ or $\pi$.  Therefore, as long as there are no additional band inversions, then the Wilson connectivity remains unchanged and, away from the SSH limit there still is an overall $\mathbb{Z}_{2}$ quantity that characterizes the topological crystalline phases.  However, as the phase of the hourglass center can be moved by any band inversion \emph{on the plane} which projects to $\bar{\Gamma}\bar{X}$, then we find that this $\mathbb{Z}_{2}$ invariant is no longer just a property of the TRIMs.  Therefore, for a real double-glide system, the Wilson loop remains the only generic method for evaluating the bulk topology.  

\begin{figure}[b]
 \centering
\includegraphics[width=0.8\textwidth]{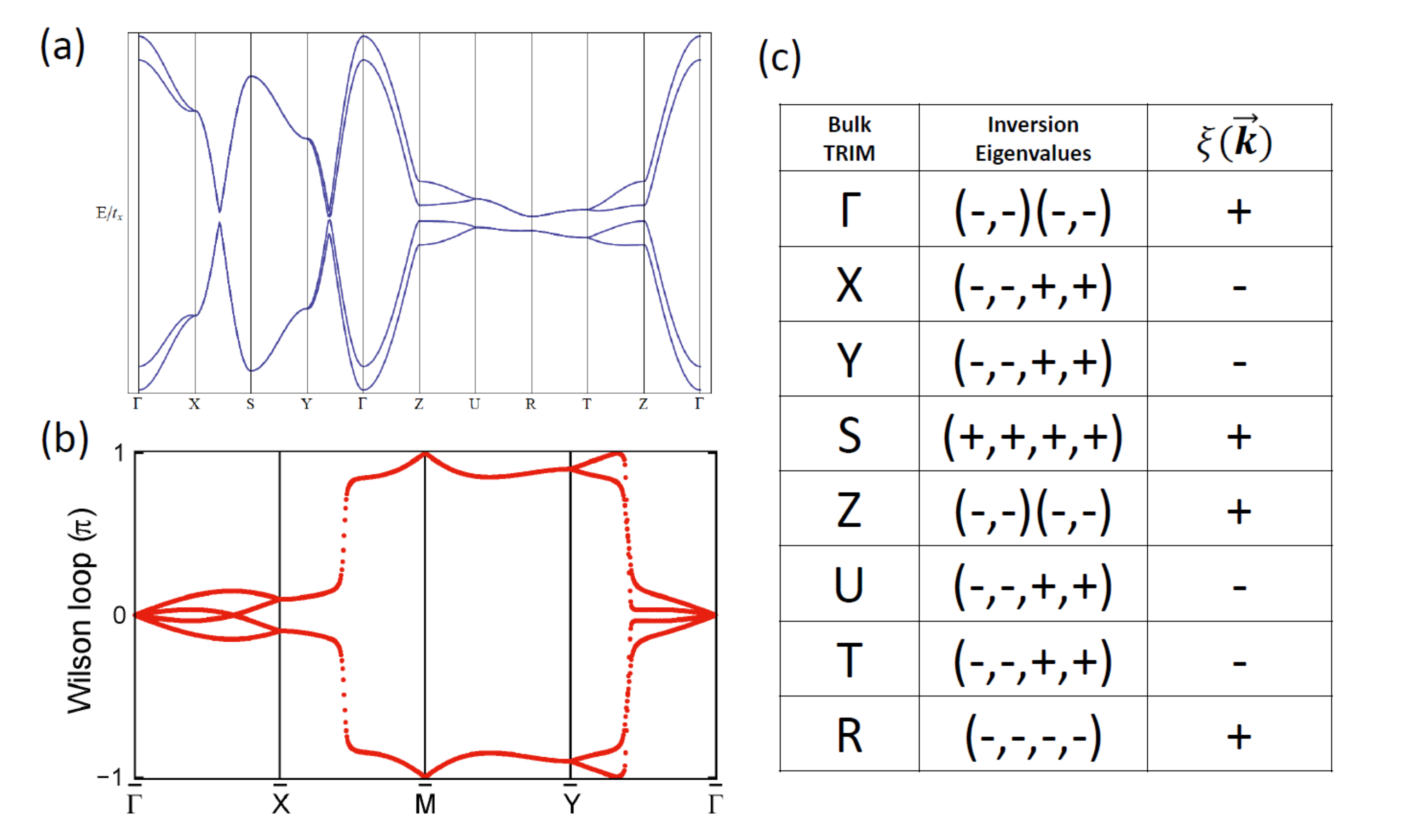}
\caption{Bulk bands (a) and Wilson bands (b) for the broken-$C_{4z}$ crystalline phase labeled by $(\chi_{x},\chi_{y})=(0,2)$ (obtained by using Eq.~(\ref{H127eq}) and the additional term in Eq.~(\ref{eqC4}) with $t_{1}=1,v_{r1}=0.55,v_{s1}=0.4,t_{2}=0.5,v_{s2}=-0.2,v_{s2}'=0.35,u_{1}=0.85,u_{2}=1.3,v_{1}=3,v_{2}=0.3,$ and $v_{C4}=12$).  In this phase, the hourglass along $\bar{Y}\bar{\Gamma}$ is sharply distorted and centered around $\pi$.  Bulk inversion eigenvalues and their products are shown for the lower four bands (c), using the conventional TRIM labeling for the orthorhombic BZ (Fig.~\ref{ay_sr}(c)).  Inversion eigenvalues are grouped by parentheses according to their pairing at two- or fourfold bulk degeneracies.  The product of the inversion eigenvalues $\xi(\vec{k})$ at a TRIM with momentum $\vec{k}$ is defined using \emph{only one} inversion eigenvalue per Kramers pair, and the product of $\xi(\vec{k})$ over all 8 bulk TRIMs is the $\mathbb{Z}_{2}$ strong QSH invariant (discussed in further detail in~\ref{sec:z4inv}).  For this realization of the topological double-glide hourglass phase, all four of the inversion eigenvalues are the same at the $\bar{\Gamma}$-projecting bulk TRIMs ($\Gamma$ and $Z$), resulting in an additional fourfold degeneracy in the Wilson spectrum at $\bar{\Gamma}$ as detailed in~\onlinecite{Alexandradinata14}.}
\label{fig:C4}
\end{figure}

\subsection{Broken $C_{4z}$ Phases in and beyond the SSH Limit}

\begin{figure}[b]
 \centering
\includegraphics[width=0.90\textwidth]{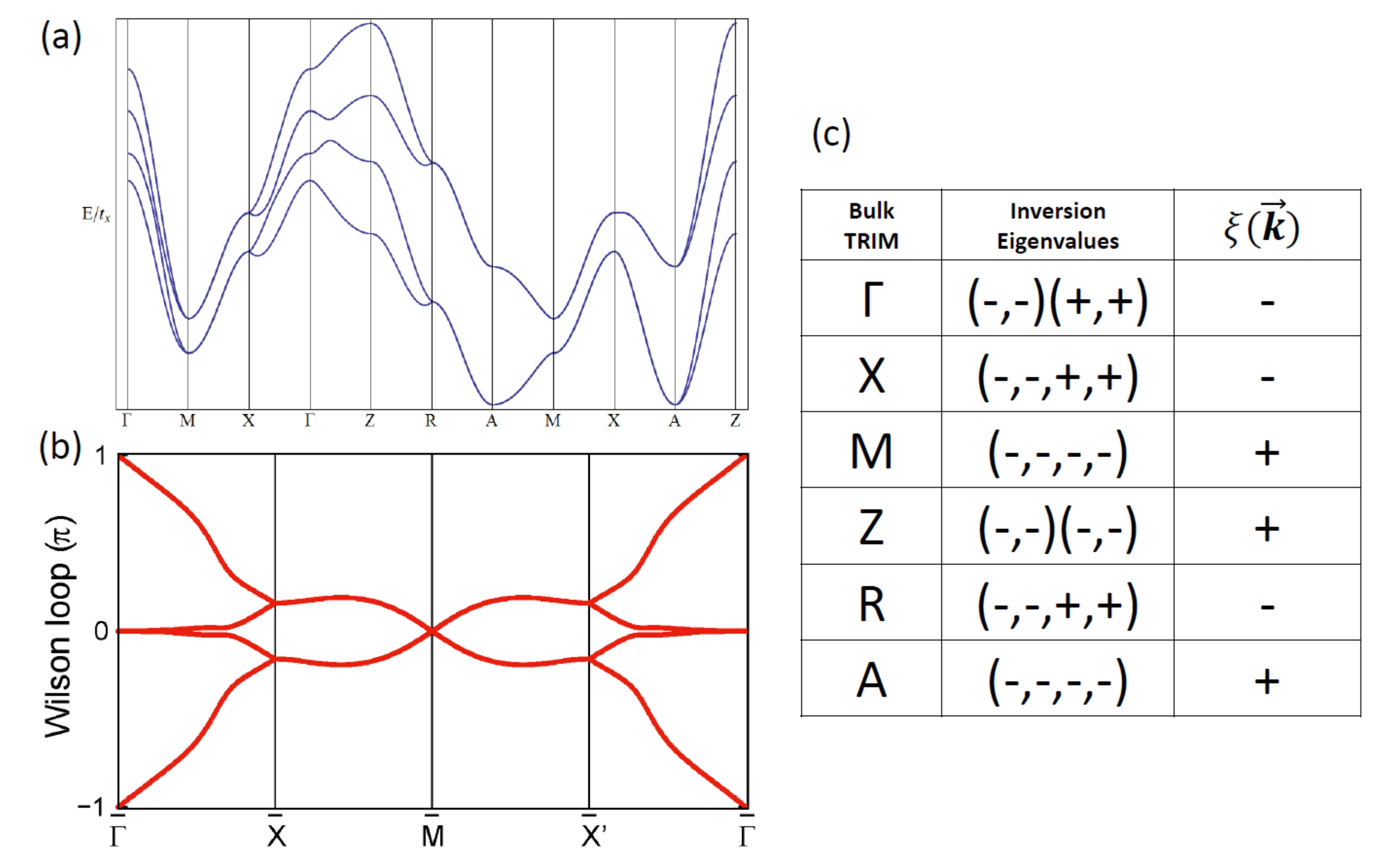}
\caption{Bulk bands (a) and Wilson bands (b) for one of the $C_{4z}$-symmetric double-glide QSH phases.  This phase can be obtained by adding the term $V_{TI}$ in Eq.~(\ref{TITermeq}) to the Hamiltonian in Eq.~(\ref{H127eq}) away from the SSH limit.  Bands for this figure were generated using $t_{1}=1,v_{r1}=0.3,v_{s1}=0.25,t_{2}=1.5,v_{s2}=-0.2,v_{s2}'=0.15,u_{1}=0.5,u_{2}=2,v_{1}=v_{2}=0,$ and $v_{TI}=0.4$.  Bulk inversion eigenvalues and their products are shown for the lower four bands (c), using the conventional TRIM labeling for the tetragonal BZ (Fig.~\ref{ay_sr}(b)).  Inversion eigenvalues are grouped by parentheses according to their pairing at two- or fourfold bulk degeneracies.  The product of the inversion eigenvalues $\xi(\vec{k})$ at a TRIM with momentum $\vec{k}$ is defined using \emph{only one} inversion eigenvalue per Kramers pair, and the product of $\xi(\vec{k})$ over all 8 bulk TRIMs is the $\mathbb{Z}_{2}$ strong QSH invariant (discussed in further detail in~\ref{sec:z4inv}).   Grouping this product of $\xi(\vec{k})$ by the contributions from pairs of bulk TRIMs that project to a given $z$-normal surface TRIM, we find that  $\xi(\bar{\Gamma})=-1$ and $\xi(\bar{X})=\xi(\bar{X}')=\xi(\bar{M})=+1$, confirming that this system is a strong topological insulator.}
\label{fig:TI}
\end{figure}

Additional Wilson band topologies are possible in systems without $C_{4z}$ symmetry, whose surfaces are described by the wallpaper group $pgg$.  As described in the main text and in~\ref{sec:DFTstuff}, Ba$_5$In$_2$Sb$_6$, as well as strained Au$_2$Y$_3$ and Hg$_2$Sr$_3$, are expected to exhibit such phases.  We break $C_{4z}$ symmetry in the tight-binding model by adding another interlayer hopping term:
\begin{equation}
V_{C4} = v_{C4}\mu^{x}\cos\left(k_{y}\right)\cos\left(\frac{k_{z}}{2}\right).
\label{eqC4}
\end{equation}
The resulting system is now in SG 55.  Without $C_{4z}$ symmetry, the SSH polarizations at $\bar{X}$ and $\bar{Y}$ can now differ, leading us to define a second crystalline invariant for the independent $y$ direction:
\begin{equation}
\chi_{x,y}=2\left(\left[\frac{1}{\pi}\left(\theta(\bar{M}) - \theta(\bar{Y},\bar{X})\right)\right]\mod 2\right).
\end{equation}
The reason that $\chi_{x,y}$ is determined by $\theta(\bar{Y},\bar{X})$, as opposed to $\theta(\bar{X},\bar{Y})$, is because $g_y$ enforces a fourfold degeneracy at $\bar{X}$, as explained in~\ref{sec:TBwC4}.  Thus, $\chi_{x,y}$ is determined by $g_{x,y}$, consistent with the notation in~\ref{sec:doubleZ4}.

If the polarization at $\bar{X}$ differs from that at $\bar{Y}$ and $\bar{M}$, the system can display a single fourfold point at $\bar{X}$.  As more symmetry-allowed SOC terms are added, the states at $\bar{X}$ will couple and the four-band crossing will open up into an hourglass along $\bar{\Gamma}\bar{X}$.  Nevertheless, as long as the bulk symmetries remain unchanged and the introduction of SOC preserves the bulk gap and band ordering, the Wilson connectivity will be preserved and the $C_{4z}$-broken SSH phase will evolve into a crystalline hourglass phase.  An observation of the allowed Wilson band connectivities with broken $C_{4z}$ symmetry confirms in fact that these two phases (hourglasses along $\bar{\Gamma}\bar{X}$ or along $\bar{\Gamma}\bar{Y}$) are the only two allowed topological crystalline connectivities which fundamentally violate $C_{4z}$.  The $(\chi_{x},\chi_{y})=(0,2)$ phase is demonstrated in our tight-binding model in Figure~\ref{fig:C4}. 

If symmetry-allowed terms are made larger, other band inversions may occur and $\chi_{x/y}$ may eventually change.  However, the resulting phases, if QSH-trivial, will still ultimately be tunable back to an SSH limit.  Therefore the SSH limit can be considered as a parent phase to all four possible $\chi_{x,y}=0,2$ crystalline insulating phases.

\subsection{Strong Topological Insulating Phases}

Away from the SSH limit, but preserving $C_{4z}$ symmetry such that the system is in SG 127, as shown in~\ref{sec:z4inv}, a band inversion about a TRIM which projects to $\bar{\Gamma}$ can flip the product of parity eigenvalues and induce a strong topological insulating phase.  In practice, this band inversion can be accomplished in our tight-binding model by adding the term:
\begin{equation}
V_{TI} = v_{TI}\tau^{y}\mu^{x}\sigma^{z}\cos\left(\frac{k_{x}}{2}\right)\cos\left(\frac{k_{y}}{2}\right)\cos\left(\frac{k_{z}}{2}\right)
\label{TITermeq}
\end{equation}
to the Hamiltonian in Eq.~(\ref{H127eq}).  Counting the product of the parity eigenvalues at the bulk TRIMs (Fig.~\ref{fig:TI}(c)) and grouping them into their projections to $z$-normal TRIMs:
\begin{equation}
\xi(\bar{\Gamma})=\xi(\Gamma)\xi(Z),\ \xi(\bar{X})=\xi(X)\xi(R),\ \xi(\bar{X}')=\xi(X')\xi(R'),\ \xi(\bar{M})=\xi(M)\xi(A),
\end{equation}
we confirm that this system is a strong topological insulator with $\xi(\bar{\Gamma})=-1$ and $\xi(\bar{X})=\xi(\bar{X}')=\xi(\bar{M})=+1$.  The  Wilson loop in Fig.~\ref{fig:TI}(b) further shows that this system is in a double-glide spin Hall phase characterized by $\chi_{x}=\chi_{y}=1,3$, depending on the choice of an odd- or even-numbered surface BZ.
\end{appendix}

\bibliography{finalBibsMain}

\begin{thebibliography}{106}
\expandafter\ifx\csname natexlab\endcsname\relax\def\natexlab#1{#1}\fi
\expandafter\ifx\csname bibnamefont\endcsname\relax
  \def\bibnamefont#1{#1}\fi
\expandafter\ifx\csname bibfnamefont\endcsname\relax
  \def\bibfnamefont#1{#1}\fi
\expandafter\ifx\csname citenamefont\endcsname\relax
  \def\citenamefont#1{#1}\fi
\expandafter\ifx\csname url\endcsname\relax
  \def\url#1{\texttt{#1}}\fi
\expandafter\ifx\csname urlprefix\endcsname\relax\def\urlprefix{URL }\fi
\providecommand{\bibinfo}[2]{#2}
\providecommand{\eprint}[2][]{\url{#2}}

\bibitem[{\citenamefont{Teo et~al.}(2008)\citenamefont{Teo, Fu, and
  Kane}}]{Teo2008}
\bibinfo{author}{\bibfnamefont{J.~C.~Y.} \bibnamefont{Teo}},
  \bibinfo{author}{\bibfnamefont{L.}~\bibnamefont{Fu}}, \bibnamefont{and}
  \bibinfo{author}{\bibfnamefont{C.~L.} \bibnamefont{Kane}},
  \bibinfo{journal}{Phys. Rev. B} \textbf{\bibinfo{volume}{78}},
  \bibinfo{pages}{045426} (\bibinfo{year}{2008}).

\bibitem[{\citenamefont{Fu}(2011)}]{Fu2011}
\bibinfo{author}{\bibfnamefont{L.}~\bibnamefont{Fu}}, \bibinfo{journal}{Phys.
  Rev. Lett.} \textbf{\bibinfo{volume}{106}}, \bibinfo{pages}{106802}
  (\bibinfo{year}{2011}),
  \urlprefix\url{https://link.aps.org/doi/10.1103/PhysRevLett.106.106802}.

\bibitem[{\citenamefont{Kim et~al.}(2015)\citenamefont{Kim, Kane, Mele, and
  Rappe}}]{Kin15p086802}
\bibinfo{author}{\bibfnamefont{Y.}~\bibnamefont{Kim}},
  \bibinfo{author}{\bibfnamefont{C.~L.} \bibnamefont{Kane}},
  \bibinfo{author}{\bibfnamefont{E.~J.} \bibnamefont{Mele}}, \bibnamefont{and}
  \bibinfo{author}{\bibfnamefont{A.~M.} \bibnamefont{Rappe}},
  \bibinfo{journal}{Phys. Rev. Lett.} \textbf{\bibinfo{volume}{115}},
  \bibinfo{pages}{086802} (\bibinfo{year}{2015}),
  \urlprefix\url{http://link.aps.org/doi/10.1103/PhysRevLett.115.086802}.

\bibitem[{\citenamefont{Hsieh et~al.}(2012)\citenamefont{Hsieh, Lin, Liu, Duan,
  Bansil, and Fu}}]{Hsieh2012}
\bibinfo{author}{\bibfnamefont{T.~H.} \bibnamefont{Hsieh}},
  \bibinfo{author}{\bibfnamefont{H.}~\bibnamefont{Lin}},
  \bibinfo{author}{\bibfnamefont{J.}~\bibnamefont{Liu}},
  \bibinfo{author}{\bibfnamefont{W.}~\bibnamefont{Duan}},
  \bibinfo{author}{\bibfnamefont{A.}~\bibnamefont{Bansil}}, \bibnamefont{and}
  \bibinfo{author}{\bibfnamefont{L.}~\bibnamefont{Fu}},
  \bibinfo{journal}{Nature Commun.} \textbf{\bibinfo{volume}{3}},
  \bibinfo{pages}{982} (\bibinfo{year}{2012}).

\bibitem[{\citenamefont{Tanaka et~al.}(2012)\citenamefont{Tanaka, Ren, Sato,
  Nakayama, Souma, Takahashi, and Segawa}}]{Tanaka2012}
\bibinfo{author}{\bibfnamefont{Y.}~\bibnamefont{Tanaka}},
  \bibinfo{author}{\bibfnamefont{Z.}~\bibnamefont{Ren}},
  \bibinfo{author}{\bibfnamefont{T.}~\bibnamefont{Sato}},
  \bibinfo{author}{\bibfnamefont{K.}~\bibnamefont{Nakayama}},
  \bibinfo{author}{\bibfnamefont{S.}~\bibnamefont{Souma}},
  \bibinfo{author}{\bibfnamefont{T.}~\bibnamefont{Takahashi}},
  \bibnamefont{and} \bibinfo{author}{\bibfnamefont{K.}~\bibnamefont{Segawa}},
  \bibinfo{journal}{Nat. Phys.} \textbf{\bibinfo{volume}{8}},
  \bibinfo{pages}{800} (\bibinfo{year}{2012}).

\bibitem[{\citenamefont{Dziawa et~al.}(2012)\citenamefont{Dziawa, Kowalski,
  Dybko, Buczko, Szczerbakow, Szot, {\L}usakowska, Balasubramanian, Wojek,
  Berntsen et~al.}}]{Dziawa2012}
\bibinfo{author}{\bibfnamefont{P.}~\bibnamefont{Dziawa}},
  \bibinfo{author}{\bibfnamefont{B.~J.} \bibnamefont{Kowalski}},
  \bibinfo{author}{\bibfnamefont{K.}~\bibnamefont{Dybko}},
  \bibinfo{author}{\bibfnamefont{R.}~\bibnamefont{Buczko}},
  \bibinfo{author}{\bibfnamefont{A.}~\bibnamefont{Szczerbakow}},
  \bibinfo{author}{\bibfnamefont{M.}~\bibnamefont{Szot}},
  \bibinfo{author}{\bibfnamefont{E.}~\bibnamefont{{\L}usakowska}},
  \bibinfo{author}{\bibfnamefont{T.}~\bibnamefont{Balasubramanian}},
  \bibinfo{author}{\bibfnamefont{B.~M.} \bibnamefont{Wojek}},
  \bibinfo{author}{\bibfnamefont{M.~H.} \bibnamefont{Berntsen}},
  \bibnamefont{et~al.}, \bibinfo{journal}{Nat. Mater.}
  \textbf{\bibinfo{volume}{11}}, \bibinfo{pages}{1023} (\bibinfo{year}{2012}).

\bibitem[{\citenamefont{Xu et~al.}(2012)\citenamefont{Xu, Liu, Alidoust,
  Neupane, Qian, Belopolski, Denlinger, Wang, Lin, Wray et~al.}}]{Xu2012}
\bibinfo{author}{\bibfnamefont{S.-Y.} \bibnamefont{Xu}},
  \bibinfo{author}{\bibfnamefont{C.}~\bibnamefont{Liu}},
  \bibinfo{author}{\bibfnamefont{N.}~\bibnamefont{Alidoust}},
  \bibinfo{author}{\bibfnamefont{M.}~\bibnamefont{Neupane}},
  \bibinfo{author}{\bibfnamefont{D.}~\bibnamefont{Qian}},
  \bibinfo{author}{\bibfnamefont{I.}~\bibnamefont{Belopolski}},
  \bibinfo{author}{\bibfnamefont{J.~D.} \bibnamefont{Denlinger}},
  \bibinfo{author}{\bibfnamefont{Y.~J.} \bibnamefont{Wang}},
  \bibinfo{author}{\bibfnamefont{H.}~\bibnamefont{Lin}},
  \bibinfo{author}{\bibfnamefont{L.~A.} \bibnamefont{Wray}},
  \bibnamefont{et~al.}, \bibinfo{journal}{Nat Commun}
  \textbf{\bibinfo{volume}{3}}, \bibinfo{pages}{1192} (\bibinfo{year}{2012}).

\bibitem[{\citenamefont{Liu et~al.}(2014{\natexlab{a}})\citenamefont{Liu,
  Zhang, and VanLeeuwen}}]{Liu2014}
\bibinfo{author}{\bibfnamefont{C.-X.} \bibnamefont{Liu}},
  \bibinfo{author}{\bibfnamefont{R.-X.} \bibnamefont{Zhang}}, \bibnamefont{and}
  \bibinfo{author}{\bibfnamefont{B.~K.} \bibnamefont{VanLeeuwen}},
  \bibinfo{journal}{Phys. Rev. B} \textbf{\bibinfo{volume}{90}},
  \bibinfo{pages}{085304} (\bibinfo{year}{2014}{\natexlab{a}}),
  \urlprefix\url{http://link.aps.org/doi/10.1103/PhysRevB.90.085304}.

\bibitem[{\citenamefont{Wang et~al.}(2016)\citenamefont{Wang, Alexandradinata,
  Cava, and Bernevig}}]{Wang16}
\bibinfo{author}{\bibfnamefont{Z.}~\bibnamefont{Wang}},
  \bibinfo{author}{\bibfnamefont{A.}~\bibnamefont{Alexandradinata}},
  \bibinfo{author}{\bibfnamefont{R.~J.} \bibnamefont{Cava}}, \bibnamefont{and}
  \bibinfo{author}{\bibfnamefont{B.~A.} \bibnamefont{Bernevig}},
  \bibinfo{journal}{Nature} \textbf{\bibinfo{volume}{532}},
  \bibinfo{pages}{189} (\bibinfo{year}{2016}).

\bibitem[{\citenamefont{Alexandradinata
  et~al.}(2016)\citenamefont{Alexandradinata, Wang, and
  Bernevig}}]{Alexandradinata16}
\bibinfo{author}{\bibfnamefont{A.}~\bibnamefont{Alexandradinata}},
  \bibinfo{author}{\bibfnamefont{Z.}~\bibnamefont{Wang}}, \bibnamefont{and}
  \bibinfo{author}{\bibfnamefont{B.~A.} \bibnamefont{Bernevig}},
  \bibinfo{journal}{Physical Review X} \textbf{\bibinfo{volume}{6}},
  \bibinfo{pages}{021008} (\bibinfo{year}{2016}).

\bibitem[{\citenamefont{Shiozaki et~al.}(2016)\citenamefont{Shiozaki, Sato, and
  Gomi}}]{Shiozaki16}
\bibinfo{author}{\bibfnamefont{K.}~\bibnamefont{Shiozaki}},
  \bibinfo{author}{\bibfnamefont{M.}~\bibnamefont{Sato}}, \bibnamefont{and}
  \bibinfo{author}{\bibfnamefont{K.}~\bibnamefont{Gomi}},
  \bibinfo{journal}{Phys. Rev. B} \textbf{\bibinfo{volume}{93}},
  \bibinfo{pages}{195413} (\bibinfo{year}{2016}),
  \urlprefix\url{http://link.aps.org/doi/10.1103/PhysRevB.93.195413}.

\bibitem[{\citenamefont{Chang et~al.}(2017{\natexlab{a}})\citenamefont{Chang,
  Erten, and Coleman}}]{Chang2016}
\bibinfo{author}{\bibfnamefont{P.-Y.} \bibnamefont{Chang}},
  \bibinfo{author}{\bibfnamefont{O.}~\bibnamefont{Erten}}, \bibnamefont{and}
  \bibinfo{author}{\bibfnamefont{P.}~\bibnamefont{Coleman}},
  \bibinfo{journal}{Nat Phys} \textbf{\bibinfo{volume}{13}},
  \bibinfo{pages}{794} (\bibinfo{year}{2017}{\natexlab{a}}), ISSN
  \bibinfo{issn}{1745-2473}, \bibinfo{note}{article},
  \urlprefix\url{http://dx.doi.org/10.1038/nphys4092}.

\bibitem[{\citenamefont{Ma et~al.}(2017)\citenamefont{Ma, Yi, Lv, Wang, Nie,
  Wang, Kong, Huang, Richard, Zhang et~al.}}]{HourglassObserve}
\bibinfo{author}{\bibfnamefont{J.}~\bibnamefont{Ma}},
  \bibinfo{author}{\bibfnamefont{C.}~\bibnamefont{Yi}},
  \bibinfo{author}{\bibfnamefont{B.}~\bibnamefont{Lv}},
  \bibinfo{author}{\bibfnamefont{Z.}~\bibnamefont{Wang}},
  \bibinfo{author}{\bibfnamefont{S.}~\bibnamefont{Nie}},
  \bibinfo{author}{\bibfnamefont{L.}~\bibnamefont{Wang}},
  \bibinfo{author}{\bibfnamefont{L.}~\bibnamefont{Kong}},
  \bibinfo{author}{\bibfnamefont{Y.}~\bibnamefont{Huang}},
  \bibinfo{author}{\bibfnamefont{P.}~\bibnamefont{Richard}},
  \bibinfo{author}{\bibfnamefont{P.}~\bibnamefont{Zhang}},
  \bibnamefont{et~al.}, \bibinfo{journal}{Science Advances}
  \textbf{\bibinfo{volume}{3}} (\bibinfo{year}{2017}),
  \urlprefix\url{http://advances.sciencemag.org/content/3/5/e1602415}.

\bibitem[{\citenamefont{Haldane}(2004)}]{Haldane2004}
\bibinfo{author}{\bibfnamefont{F.~D.~M.} \bibnamefont{Haldane}},
  \bibinfo{journal}{Phys. Rev. Lett.} \textbf{\bibinfo{volume}{93}},
  \bibinfo{pages}{206602} (\bibinfo{year}{2004}).

\bibitem[{\citenamefont{Fu and Kane}(2007)}]{Fu07}
\bibinfo{author}{\bibfnamefont{L.}~\bibnamefont{Fu}} \bibnamefont{and}
  \bibinfo{author}{\bibfnamefont{C.~L.} \bibnamefont{Kane}},
  \bibinfo{journal}{Phys. Rev. B} \textbf{\bibinfo{volume}{76}},
  \bibinfo{pages}{045302} (\bibinfo{year}{2007}),
  \urlprefix\url{http://link.aps.org/doi/10.1103/PhysRevB.76.045302}.

\bibitem[{\citenamefont{Moore and Balents}(2007)}]{Moore07}
\bibinfo{author}{\bibfnamefont{J.~E.} \bibnamefont{Moore}} \bibnamefont{and}
  \bibinfo{author}{\bibfnamefont{L.}~\bibnamefont{Balents}},
  \bibinfo{journal}{Phys. Rev. B} \textbf{\bibinfo{volume}{75}},
  \bibinfo{pages}{121306} (\bibinfo{year}{2007}),
  \urlprefix\url{http://link.aps.org/doi/10.1103/PhysRevB.75.121306}.

\bibitem[{\citenamefont{Young et~al.}(2012)\citenamefont{Young, Zaheer, Teo,
  Kane, Mele, and Rappe}}]{SteveDirac}
\bibinfo{author}{\bibfnamefont{S.~M.} \bibnamefont{Young}},
  \bibinfo{author}{\bibfnamefont{S.}~\bibnamefont{Zaheer}},
  \bibinfo{author}{\bibfnamefont{J.~C.~Y.} \bibnamefont{Teo}},
  \bibinfo{author}{\bibfnamefont{C.~L.} \bibnamefont{Kane}},
  \bibinfo{author}{\bibfnamefont{E.~J.} \bibnamefont{Mele}}, \bibnamefont{and}
  \bibinfo{author}{\bibfnamefont{A.~M.} \bibnamefont{Rappe}},
  \bibinfo{journal}{Phys. Rev. Lett.} \textbf{\bibinfo{volume}{108}},
  \bibinfo{pages}{140405} (\bibinfo{year}{2012}),
  \urlprefix\url{http://link.aps.org/doi/10.1103/PhysRevLett.108.140405}.

\bibitem[{\citenamefont{Steinberg et~al.}(2014)\citenamefont{Steinberg, Young,
  Zaheer, Kane, Mele, and Rappe}}]{JuliaDirac}
\bibinfo{author}{\bibfnamefont{J.~A.} \bibnamefont{Steinberg}},
  \bibinfo{author}{\bibfnamefont{S.~M.} \bibnamefont{Young}},
  \bibinfo{author}{\bibfnamefont{S.}~\bibnamefont{Zaheer}},
  \bibinfo{author}{\bibfnamefont{C.~L.} \bibnamefont{Kane}},
  \bibinfo{author}{\bibfnamefont{E.~J.} \bibnamefont{Mele}}, \bibnamefont{and}
  \bibinfo{author}{\bibfnamefont{A.~M.} \bibnamefont{Rappe}},
  \bibinfo{journal}{Phys. Rev. Lett.} \textbf{\bibinfo{volume}{112}},
  \bibinfo{pages}{036403} (\bibinfo{year}{2014}),
  \urlprefix\url{http://link.aps.org/doi/10.1103/PhysRevLett.112.036403}.

\bibitem[{\citenamefont{Wieder et~al.}(2016)\citenamefont{Wieder, Kim, Rappe,
  and Kane}}]{DDP}
\bibinfo{author}{\bibfnamefont{B.~J.} \bibnamefont{Wieder}},
  \bibinfo{author}{\bibfnamefont{Y.}~\bibnamefont{Kim}},
  \bibinfo{author}{\bibfnamefont{A.~M.} \bibnamefont{Rappe}}, \bibnamefont{and}
  \bibinfo{author}{\bibfnamefont{C.~L.} \bibnamefont{Kane}},
  \bibinfo{journal}{Phys. Rev. Lett.} \textbf{\bibinfo{volume}{116}},
  \bibinfo{pages}{186402} (\bibinfo{year}{2016}),
  \urlprefix\url{http://link.aps.org/doi/10.1103/PhysRevLett.116.186402}.

\bibitem[{\citenamefont{Bradlyn et~al.}(2016)\citenamefont{Bradlyn, Cano, Wang,
  Vergniory, Felser, Cava, and Bernevig}}]{NewFermions}
\bibinfo{author}{\bibfnamefont{B.}~\bibnamefont{Bradlyn}},
  \bibinfo{author}{\bibfnamefont{J.}~\bibnamefont{Cano}},
  \bibinfo{author}{\bibfnamefont{Z.}~\bibnamefont{Wang}},
  \bibinfo{author}{\bibfnamefont{M.~G.} \bibnamefont{Vergniory}},
  \bibinfo{author}{\bibfnamefont{C.}~\bibnamefont{Felser}},
  \bibinfo{author}{\bibfnamefont{R.~J.} \bibnamefont{Cava}}, \bibnamefont{and}
  \bibinfo{author}{\bibfnamefont{B.~A.} \bibnamefont{Bernevig}},
  \bibinfo{journal}{Science} \textbf{\bibinfo{volume}{353}},
  \bibinfo{pages}{aaf5037} (\bibinfo{year}{2016}), ISSN
  \bibinfo{issn}{0036-8075},
  \urlprefix\url{http://science.sciencemag.org/content/353/6299/aaf5037}.

\bibitem[{\citenamefont{Young and Kane}(2015)}]{Steve2D}
\bibinfo{author}{\bibfnamefont{S.~M.} \bibnamefont{Young}} \bibnamefont{and}
  \bibinfo{author}{\bibfnamefont{C.~L.} \bibnamefont{Kane}},
  \bibinfo{journal}{Phys. Rev. Lett.} \textbf{\bibinfo{volume}{115}},
  \bibinfo{pages}{126803} (\bibinfo{year}{2015}),
  \urlprefix\url{http://link.aps.org/doi/10.1103/PhysRevLett.115.126803}.

\bibitem[{\citenamefont{Conway et~al.}(2008)\citenamefont{Conway, Burgiel, and
  Goodman-Strauss}}]{ConwayWallpaper}
\bibinfo{author}{\bibfnamefont{J.~H.} \bibnamefont{Conway}},
  \bibinfo{author}{\bibfnamefont{H.}~\bibnamefont{Burgiel}}, \bibnamefont{and}
  \bibinfo{author}{\bibfnamefont{C.}~\bibnamefont{Goodman-Strauss}},
  \emph{\bibinfo{title}{The Symmetries of Things}} (\bibinfo{publisher}{A K
  Peters/CRC Press}, \bibinfo{address}{Worcester, MA}, \bibinfo{year}{2008}),
  ISBN \bibinfo{isbn}{1568812205}.

\bibitem[{\citenamefont{Fu and Kane}(2006)}]{Fu06}
\bibinfo{author}{\bibfnamefont{L.}~\bibnamefont{Fu}} \bibnamefont{and}
  \bibinfo{author}{\bibfnamefont{C.~L.} \bibnamefont{Kane}},
  \bibinfo{journal}{Phys. Rev. B} \textbf{\bibinfo{volume}{74}},
  \bibinfo{pages}{195312} (\bibinfo{year}{2006}).

\bibitem[{\citenamefont{Ryu et~al.}(2010)\citenamefont{Ryu, Mudry, Obuse, and
  Furusaki}}]{Ryu10}
\bibinfo{author}{\bibfnamefont{S.}~\bibnamefont{Ryu}},
  \bibinfo{author}{\bibfnamefont{C.}~\bibnamefont{Mudry}},
  \bibinfo{author}{\bibfnamefont{H.}~\bibnamefont{Obuse}}, \bibnamefont{and}
  \bibinfo{author}{\bibfnamefont{A.}~\bibnamefont{Furusaki}},
  \bibinfo{journal}{New Journal of Physics} \textbf{\bibinfo{volume}{12}},
  \bibinfo{pages}{065005} (\bibinfo{year}{2010}).

\bibitem[{\citenamefont{Soluyanov and Vanderbilt}(2011)}]{Soluyanov11}
\bibinfo{author}{\bibfnamefont{A.~A.} \bibnamefont{Soluyanov}}
  \bibnamefont{and}
  \bibinfo{author}{\bibfnamefont{D.}~\bibnamefont{Vanderbilt}},
  \bibinfo{journal}{Physical Review B} \textbf{\bibinfo{volume}{83}},
  \bibinfo{pages}{235401} (\bibinfo{year}{2011}).

\bibitem[{\citenamefont{Yu et~al.}(2011)\citenamefont{Yu, Qi, Bernevig, Fang,
  and Dai}}]{Yu11}
\bibinfo{author}{\bibfnamefont{R.}~\bibnamefont{Yu}},
  \bibinfo{author}{\bibfnamefont{X.~L.} \bibnamefont{Qi}},
  \bibinfo{author}{\bibfnamefont{A.}~\bibnamefont{Bernevig}},
  \bibinfo{author}{\bibfnamefont{Z.}~\bibnamefont{Fang}}, \bibnamefont{and}
  \bibinfo{author}{\bibfnamefont{X.}~\bibnamefont{Dai}},
  \bibinfo{journal}{Phys. Rev. B} \textbf{\bibinfo{volume}{84}},
  \bibinfo{pages}{075119} (\bibinfo{year}{2011}).

\bibitem[{\citenamefont{Taherinejad et~al.}(2014)\citenamefont{Taherinejad,
  Garrity, and Vanderbilt}}]{Taherinejad14}
\bibinfo{author}{\bibfnamefont{M.}~\bibnamefont{Taherinejad}},
  \bibinfo{author}{\bibfnamefont{K.~F.} \bibnamefont{Garrity}},
  \bibnamefont{and}
  \bibinfo{author}{\bibfnamefont{D.}~\bibnamefont{Vanderbilt}},
  \bibinfo{journal}{Physical Review B} \textbf{\bibinfo{volume}{89}},
  \bibinfo{pages}{115102} (\bibinfo{year}{2014}).

\bibitem[{\citenamefont{Alexandradinata
  et~al.}(2014{\natexlab{a}})\citenamefont{Alexandradinata, Dai, and
  Bernevig}}]{Alexandradinata14}
\bibinfo{author}{\bibfnamefont{A.}~\bibnamefont{Alexandradinata}},
  \bibinfo{author}{\bibfnamefont{X.}~\bibnamefont{Dai}}, \bibnamefont{and}
  \bibinfo{author}{\bibfnamefont{B.~A.} \bibnamefont{Bernevig}},
  \bibinfo{journal}{Physical Review B} \textbf{\bibinfo{volume}{89}},
  \bibinfo{pages}{155114} (\bibinfo{year}{2014}{\natexlab{a}}).

\bibitem[{\citenamefont{Su et~al.}(1979)\citenamefont{Su, Schrieffer, and
  Heeger}}]{SSH}
\bibinfo{author}{\bibfnamefont{W.~P.} \bibnamefont{Su}},
  \bibinfo{author}{\bibfnamefont{J.~R.} \bibnamefont{Schrieffer}},
  \bibnamefont{and} \bibinfo{author}{\bibfnamefont{A.~J.}
  \bibnamefont{Heeger}}, \bibinfo{journal}{Phys. Rev. Lett.}
  \textbf{\bibinfo{volume}{42}}, \bibinfo{pages}{1698} (\bibinfo{year}{1979}),
  \urlprefix\url{https://link.aps.org/doi/10.1103/PhysRevLett.42.1698}.

\bibitem[{\citenamefont{Dong and Liu}(2016)}]{PSUTCI}
\bibinfo{author}{\bibfnamefont{X.-Y.} \bibnamefont{Dong}} \bibnamefont{and}
  \bibinfo{author}{\bibfnamefont{C.-X.} \bibnamefont{Liu}},
  \bibinfo{journal}{Phys. Rev. B} \textbf{\bibinfo{volume}{93}},
  \bibinfo{pages}{045429} (\bibinfo{year}{2016}),
  \urlprefix\url{https://link.aps.org/doi/10.1103/PhysRevB.93.045429}.

\bibitem[{\citenamefont{Bradlyn et~al.}(2017)\citenamefont{Bradlyn, Elcoro,
  Cano, Vergniory, Wang, Felser, Aroyo, and Bernevig}}]{QuantumChemistry}
\bibinfo{author}{\bibfnamefont{B.}~\bibnamefont{Bradlyn}},
  \bibinfo{author}{\bibfnamefont{L.}~\bibnamefont{Elcoro}},
  \bibinfo{author}{\bibfnamefont{J.}~\bibnamefont{Cano}},
  \bibinfo{author}{\bibfnamefont{M.~G.} \bibnamefont{Vergniory}},
  \bibinfo{author}{\bibfnamefont{Z.}~\bibnamefont{Wang}},
  \bibinfo{author}{\bibfnamefont{C.}~\bibnamefont{Felser}},
  \bibinfo{author}{\bibfnamefont{M.~I.} \bibnamefont{Aroyo}}, \bibnamefont{and}
  \bibinfo{author}{\bibfnamefont{B.~A.} \bibnamefont{Bernevig}},
  \bibinfo{journal}{Nature} \textbf{\bibinfo{volume}{547}},
  \bibinfo{pages}{298} (\bibinfo{year}{2017}), ISSN \bibinfo{issn}{0028-0836},
  \bibinfo{note}{article},
  \urlprefix\url{http://dx.doi.org/10.1038/nature23268}.

\bibitem[{\citenamefont{Fang and Fu}(2015)}]{FangFuNSandC2}
\bibinfo{author}{\bibfnamefont{C.}~\bibnamefont{Fang}} \bibnamefont{and}
  \bibinfo{author}{\bibfnamefont{L.}~\bibnamefont{Fu}}, \bibinfo{journal}{Phys.
  Rev. B} \textbf{\bibinfo{volume}{91}}, \bibinfo{pages}{161105}
  (\bibinfo{year}{2015}),
  \urlprefix\url{https://link.aps.org/doi/10.1103/PhysRevB.91.161105}.

\bibitem[{\citenamefont{Shiozaki et~al.}(2017)\citenamefont{Shiozaki, Sato, and
  Gomi}}]{ShiozakiWallpaper}
\bibinfo{author}{\bibfnamefont{K.}~\bibnamefont{Shiozaki}},
  \bibinfo{author}{\bibfnamefont{M.}~\bibnamefont{Sato}}, \bibnamefont{and}
  \bibinfo{author}{\bibfnamefont{K.}~\bibnamefont{Gomi}},
  \bibinfo{journal}{Phys. Rev. B} \textbf{\bibinfo{volume}{95}},
  \bibinfo{pages}{235425} (\bibinfo{year}{2017}),
  \urlprefix\url{https://link.aps.org/doi/10.1103/PhysRevB.95.235425}.

\bibitem[{\citenamefont{Song et~al.}(2017{\natexlab{a}})\citenamefont{Song,
  Huang, Fu, and Hermele}}]{LiangPointGroup}
\bibinfo{author}{\bibfnamefont{H.}~\bibnamefont{Song}},
  \bibinfo{author}{\bibfnamefont{S.-J.} \bibnamefont{Huang}},
  \bibinfo{author}{\bibfnamefont{L.}~\bibnamefont{Fu}}, \bibnamefont{and}
  \bibinfo{author}{\bibfnamefont{M.}~\bibnamefont{Hermele}},
  \bibinfo{journal}{Phys. Rev. X} \textbf{\bibinfo{volume}{7}},
  \bibinfo{pages}{011020} (\bibinfo{year}{2017}{\natexlab{a}}),
  \urlprefix\url{https://link.aps.org/doi/10.1103/PhysRevX.7.011020}.

\bibitem[{\citenamefont{Watanabe et~al.}(2015)\citenamefont{Watanabe, Po,
  Vishwanath, and Zaletel}}]{WPVZ}
\bibinfo{author}{\bibfnamefont{H.}~\bibnamefont{Watanabe}},
  \bibinfo{author}{\bibfnamefont{H.~C.} \bibnamefont{Po}},
  \bibinfo{author}{\bibfnamefont{A.}~\bibnamefont{Vishwanath}},
  \bibnamefont{and} \bibinfo{author}{\bibfnamefont{M.}~\bibnamefont{Zaletel}},
  \bibinfo{journal}{Proceedings of the National Academy of Sciences}
  \textbf{\bibinfo{volume}{112}}, \bibinfo{pages}{14551}
  (\bibinfo{year}{2015}),
  \eprint{http://www.pnas.org/content/112/47/14551.full.pdf},
  \urlprefix\url{http://www.pnas.org/content/112/47/14551.abstract}.

\bibitem[{\citenamefont{Wieder and Kane}(2016)}]{WiederLayers}
\bibinfo{author}{\bibfnamefont{B.~J.} \bibnamefont{Wieder}} \bibnamefont{and}
  \bibinfo{author}{\bibfnamefont{C.~L.} \bibnamefont{Kane}},
  \bibinfo{journal}{Phys. Rev. B} \textbf{\bibinfo{volume}{94}},
  \bibinfo{pages}{155108} (\bibinfo{year}{2016}),
  \urlprefix\url{http://link.aps.org/doi/10.1103/PhysRevB.94.155108}.

\bibitem[{\citenamefont{Shiozaki et~al.}(2015)\citenamefont{Shiozaki, Sato, and
  Gomi}}]{SSGMobius}
\bibinfo{author}{\bibfnamefont{K.}~\bibnamefont{Shiozaki}},
  \bibinfo{author}{\bibfnamefont{M.}~\bibnamefont{Sato}}, \bibnamefont{and}
  \bibinfo{author}{\bibfnamefont{K.}~\bibnamefont{Gomi}},
  \bibinfo{journal}{Phys. Rev. B} \textbf{\bibinfo{volume}{91}},
  \bibinfo{pages}{155120} (\bibinfo{year}{2015}),
  \urlprefix\url{https://link.aps.org/doi/10.1103/PhysRevB.91.155120}.

\bibitem[{\citenamefont{Fidkowski et~al.}(2011)\citenamefont{Fidkowski,
  Jackson, and Klich}}]{Fidkowski2011}
\bibinfo{author}{\bibfnamefont{L.}~\bibnamefont{Fidkowski}},
  \bibinfo{author}{\bibfnamefont{T.~S.} \bibnamefont{Jackson}},
  \bibnamefont{and} \bibinfo{author}{\bibfnamefont{I.}~\bibnamefont{Klich}},
  \bibinfo{journal}{Phys. Rev. Lett.} \textbf{\bibinfo{volume}{107}},
  \bibinfo{pages}{036601} (\bibinfo{year}{2011}).

\bibitem[{\citenamefont{Alexandradinata}()}]{AlexandradinataConvo}
\bibinfo{author}{\bibfnamefont{A.}~\bibnamefont{Alexandradinata}},
  \bibinfo{howpublished}{Private communication}.

\bibitem[{\citenamefont{Xiong and Alexandradinata}(2018)}]{AlexandradinataPub}
\bibinfo{author}{\bibfnamefont{C.~Z.} \bibnamefont{Xiong}} \bibnamefont{and}
  \bibinfo{author}{\bibfnamefont{A.}~\bibnamefont{Alexandradinata}},
  \emph{\bibinfo{title}{Organizing symmetry-protected topological phases by
  layering and symmetry reduction: A minimalist perspective}}
  (\bibinfo{year}{2018}),
  \urlprefix\url{https://link.aps.org/doi/10.1103/PhysRevB.97.115153}.

\bibitem[{\citenamefont{Fang et~al.}(2016)\citenamefont{Fang, Lu, Liu, and
  Fu}}]{Fang15}
\bibinfo{author}{\bibfnamefont{C.}~\bibnamefont{Fang}},
  \bibinfo{author}{\bibfnamefont{L.}~\bibnamefont{Lu}},
  \bibinfo{author}{\bibfnamefont{J.}~\bibnamefont{Liu}}, \bibnamefont{and}
  \bibinfo{author}{\bibfnamefont{L.}~\bibnamefont{Fu}}, \bibinfo{journal}{Nat
  Phys} \textbf{\bibinfo{volume}{12}}, \bibinfo{pages}{936}
  (\bibinfo{year}{2016}), ISSN \bibinfo{issn}{1745-2473},
  \bibinfo{note}{article}, \urlprefix\url{http://dx.doi.org/10.1038/nphys3782}.

\bibitem[{\citenamefont{Alexandradinata and
  Bernevig}(2016)}]{Alexandradinata16b}
\bibinfo{author}{\bibfnamefont{A.}~\bibnamefont{Alexandradinata}}
  \bibnamefont{and} \bibinfo{author}{\bibfnamefont{B.~A.}
  \bibnamefont{Bernevig}}, \bibinfo{journal}{Phys. Rev. B}
  \textbf{\bibinfo{volume}{93}}, \bibinfo{pages}{205104}
  (\bibinfo{year}{2016}),
  \urlprefix\url{http://link.aps.org/doi/10.1103/PhysRevB.93.205104}.

\bibitem[{\citenamefont{Merlo}(1984)}]{Merlo84p78}
\bibinfo{author}{\bibfnamefont{F.}~\bibnamefont{Merlo}},
  \bibinfo{journal}{Revue de chimie min{\'e}rale}
  \textbf{\bibinfo{volume}{21}}, \bibinfo{pages}{78} (\bibinfo{year}{1984}).

\bibitem[{\citenamefont{Bruzzone et~al.}(1981)\citenamefont{Bruzzone,
  Franceschi, and Merlo}}]{Bruzzone81p155}
\bibinfo{author}{\bibfnamefont{G.}~\bibnamefont{Bruzzone}},
  \bibinfo{author}{\bibfnamefont{E.}~\bibnamefont{Franceschi}},
  \bibnamefont{and} \bibinfo{author}{\bibfnamefont{F.}~\bibnamefont{Merlo}},
  \bibinfo{journal}{Journal of the Less Common Metals}
  \textbf{\bibinfo{volume}{81}}, \bibinfo{pages}{155} (\bibinfo{year}{1981}).

\bibitem[{\citenamefont{Chai and Corbett}(2011)}]{Chai11pi53}
\bibinfo{author}{\bibfnamefont{P.}~\bibnamefont{Chai}} \bibnamefont{and}
  \bibinfo{author}{\bibfnamefont{J.~D.} \bibnamefont{Corbett}},
  \bibinfo{journal}{Acta Crystallographica Section C}
  \textbf{\bibinfo{volume}{67}}, \bibinfo{pages}{i53} (\bibinfo{year}{2011}),
  \urlprefix\url{https://doi.org/10.1107/S010827011103589X}.

\bibitem[{\citenamefont{Druska et~al.}(1996)\citenamefont{Druska, Doert, and
  Böttcher}}]{Druska96p401}
\bibinfo{author}{\bibfnamefont{C.}~\bibnamefont{Druska}},
  \bibinfo{author}{\bibfnamefont{T.}~\bibnamefont{Doert}}, \bibnamefont{and}
  \bibinfo{author}{\bibfnamefont{P.}~\bibnamefont{Böttcher}},
  \bibinfo{journal}{Zeitschrift für anorganische und allgemeine Chemie}
  \textbf{\bibinfo{volume}{622}}, \bibinfo{pages}{401} (\bibinfo{year}{1996}),
  ISSN \bibinfo{issn}{1521-3749},
  \urlprefix\url{http://dx.doi.org/10.1002/zaac.19966220304}.

\bibitem[{\citenamefont{Gumi{\'{n}}ski}(2005)}]{Gumiski05p81}
\bibinfo{author}{\bibfnamefont{C.}~\bibnamefont{Gumi{\'{n}}ski}},
  \bibinfo{journal}{Journal of Phase Equilibria and Diffusion}
  \textbf{\bibinfo{volume}{26}}, \bibinfo{pages}{81} (\bibinfo{year}{2005}),
  ISSN \bibinfo{issn}{1863-7345},
  \urlprefix\url{http://dx.doi.org/10.1007/s11669-005-0070-z}.

\bibitem[{\citenamefont{Wang et~al.}(2013)\citenamefont{Wang, Weng, Wu, Dai,
  and Fang}}]{Wang13}
\bibinfo{author}{\bibfnamefont{Z.}~\bibnamefont{Wang}},
  \bibinfo{author}{\bibfnamefont{H.}~\bibnamefont{Weng}},
  \bibinfo{author}{\bibfnamefont{Q.}~\bibnamefont{Wu}},
  \bibinfo{author}{\bibfnamefont{X.}~\bibnamefont{Dai}}, \bibnamefont{and}
  \bibinfo{author}{\bibfnamefont{Z.}~\bibnamefont{Fang}},
  \bibinfo{journal}{Phys. Rev. B} \textbf{\bibinfo{volume}{88}},
  \bibinfo{pages}{125427} (\bibinfo{year}{2013}),
  \urlprefix\url{http://link.aps.org/doi/10.1103/PhysRevB.88.125427}.

\bibitem[{\citenamefont{Cordier and Steher}(1988)}]{ZJGermanBA}
\bibinfo{author}{\bibfnamefont{G.}~\bibnamefont{Cordier}} \bibnamefont{and}
  \bibinfo{author}{\bibfnamefont{M.}~\bibnamefont{Steher}},
  \bibinfo{journal}{Zeitschrift f{\"u}r Naturforschung B}
  \textbf{\bibinfo{volume}{43}}, \bibinfo{pages}{463} (\bibinfo{year}{1988}),
  ISSN \bibinfo{issn}{0932-0776}.

\bibitem[{\citenamefont{Jackiw and Rebbi}(1976)}]{JackiwRebbi}
\bibinfo{author}{\bibfnamefont{R.}~\bibnamefont{Jackiw}} \bibnamefont{and}
  \bibinfo{author}{\bibfnamefont{C.}~\bibnamefont{Rebbi}},
  \bibinfo{journal}{Phys. Rev. D} \textbf{\bibinfo{volume}{13}},
  \bibinfo{pages}{3398} (\bibinfo{year}{1976}),
  \urlprefix\url{http://link.aps.org/doi/10.1103/PhysRevD.13.3398}.

\bibitem[{\citenamefont{Tinkham}(1964)}]{TinkhamBook}
\bibinfo{author}{\bibfnamefont{M.}~\bibnamefont{Tinkham}},
  \emph{\bibinfo{title}{Group Theory and Quantum Mechanics}}
  (\bibinfo{publisher}{McGraw-Hill Book Company}, \bibinfo{address}{New York
  City, New York}, \bibinfo{year}{1964}), ISBN \bibinfo{isbn}{0486432475}.

\bibitem[{\citenamefont{Zhang et~al.}(2011)\citenamefont{Zhang, Jung, Fiete,
  Niu, and MacDonald}}]{BLG3}
\bibinfo{author}{\bibfnamefont{F.}~\bibnamefont{Zhang}},
  \bibinfo{author}{\bibfnamefont{J.}~\bibnamefont{Jung}},
  \bibinfo{author}{\bibfnamefont{G.~A.} \bibnamefont{Fiete}},
  \bibinfo{author}{\bibfnamefont{Q.}~\bibnamefont{Niu}}, \bibnamefont{and}
  \bibinfo{author}{\bibfnamefont{A.~H.} \bibnamefont{MacDonald}},
  \bibinfo{journal}{Phys. Rev. Lett.} \textbf{\bibinfo{volume}{106}},
  \bibinfo{pages}{156801} (\bibinfo{year}{2011}),
  \urlprefix\url{http://link.aps.org/doi/10.1103/PhysRevLett.106.156801}.

\bibitem[{\citenamefont{Ohta et~al.}(2006)\citenamefont{Ohta, Bostwick,
  Seyller, Horn, and Rotenberg}}]{BLG4}
\bibinfo{author}{\bibfnamefont{T.}~\bibnamefont{Ohta}},
  \bibinfo{author}{\bibfnamefont{A.}~\bibnamefont{Bostwick}},
  \bibinfo{author}{\bibfnamefont{T.}~\bibnamefont{Seyller}},
  \bibinfo{author}{\bibfnamefont{K.}~\bibnamefont{Horn}}, \bibnamefont{and}
  \bibinfo{author}{\bibfnamefont{E.}~\bibnamefont{Rotenberg}},
  \bibinfo{journal}{Science} \textbf{\bibinfo{volume}{313}},
  \bibinfo{pages}{951} (\bibinfo{year}{2006}), ISSN \bibinfo{issn}{0036-8075},
  \urlprefix\url{http://science.sciencemag.org/content/313/5789/951}.

\bibitem[{\citenamefont{Oostinga et~al.}(2007)\citenamefont{Oostinga, Heersche,
  Liu, Morpurgo, and Vandersypen}}]{BLG5}
\bibinfo{author}{\bibfnamefont{J.~B.} \bibnamefont{Oostinga}},
  \bibinfo{author}{\bibfnamefont{H.~B.} \bibnamefont{Heersche}},
  \bibinfo{author}{\bibfnamefont{X.}~\bibnamefont{Liu}},
  \bibinfo{author}{\bibfnamefont{A.~F.} \bibnamefont{Morpurgo}},
  \bibnamefont{and} \bibinfo{author}{\bibfnamefont{L.~M.~K.}
  \bibnamefont{Vandersypen}}, \bibinfo{journal}{Nature Materials}
  \textbf{\bibinfo{volume}{7}}, \bibinfo{pages}{151} (\bibinfo{year}{2007}),
  ISSN \bibinfo{issn}{1476--1122},
  \urlprefix\url{http://www.nature.com/nmat/journal/v7/n2/suppinfo/nmat2082_S1.html}.

\bibitem[{\citenamefont{Schindler et~al.}(2018)\citenamefont{Schindler, Cook,
  Vergniory, Wang, Parkin, Bernevig, and Neupert}}]{HigherOrderTIBernevig}
\bibinfo{author}{\bibfnamefont{F.}~\bibnamefont{Schindler}},
  \bibinfo{author}{\bibfnamefont{A.~M.} \bibnamefont{Cook}},
  \bibinfo{author}{\bibfnamefont{M.~G.} \bibnamefont{Vergniory}},
  \bibinfo{author}{\bibfnamefont{Z.}~\bibnamefont{Wang}},
  \bibinfo{author}{\bibfnamefont{S.~S.~P.} \bibnamefont{Parkin}},
  \bibinfo{author}{\bibfnamefont{B.~A.} \bibnamefont{Bernevig}},
  \bibnamefont{and} \bibinfo{author}{\bibfnamefont{T.}~\bibnamefont{Neupert}},
  \bibinfo{journal}{Science Advances} \textbf{\bibinfo{volume}{4}}
  (\bibinfo{year}{2018}),
  \urlprefix\url{http://advances.sciencemag.org/content/4/6/eaat0346}.

\bibitem[{\citenamefont{Song et~al.}(2017{\natexlab{b}})\citenamefont{Song,
  Fang, and Fang}}]{HigherOrderTIChen}
\bibinfo{author}{\bibfnamefont{Z.}~\bibnamefont{Song}},
  \bibinfo{author}{\bibfnamefont{Z.}~\bibnamefont{Fang}}, \bibnamefont{and}
  \bibinfo{author}{\bibfnamefont{C.}~\bibnamefont{Fang}},
  \bibinfo{journal}{Phys. Rev. Lett.} \textbf{\bibinfo{volume}{119}},
  \bibinfo{pages}{246402} (\bibinfo{year}{2017}{\natexlab{b}}),
  \urlprefix\url{https://link.aps.org/doi/10.1103/PhysRevLett.119.246402}.

\bibitem[{\citenamefont{{Schindler} et~al.}(2018)\citenamefont{{Schindler},
  {Wang}, {Vergniory}, {Cook}, {Murani}, {Sengupta}, {Kasumov}, {Deblock},
  {Jeon}, {Drozdov} et~al.}}]{HOTIBismuth}
\bibinfo{author}{\bibfnamefont{F.}~\bibnamefont{{Schindler}}},
  \bibinfo{author}{\bibfnamefont{Z.}~\bibnamefont{{Wang}}},
  \bibinfo{author}{\bibfnamefont{M.~G.} \bibnamefont{{Vergniory}}},
  \bibinfo{author}{\bibfnamefont{A.~M.} \bibnamefont{{Cook}}},
  \bibinfo{author}{\bibfnamefont{A.}~\bibnamefont{{Murani}}},
  \bibinfo{author}{\bibfnamefont{S.}~\bibnamefont{{Sengupta}}},
  \bibinfo{author}{\bibfnamefont{A.~Y.} \bibnamefont{{Kasumov}}},
  \bibinfo{author}{\bibfnamefont{R.}~\bibnamefont{{Deblock}}},
  \bibinfo{author}{\bibfnamefont{S.}~\bibnamefont{{Jeon}}},
  \bibinfo{author}{\bibfnamefont{I.}~\bibnamefont{{Drozdov}}},
  \bibnamefont{et~al.}, \bibinfo{journal}{ArXiv e-prints}
  (\bibinfo{year}{2018}), \eprint{1802.02585}.

\bibitem[{\citenamefont{Wu et~al.}(2006)\citenamefont{Wu, Bernevig, and
  Zhang}}]{BernevigLuttinger}
\bibinfo{author}{\bibfnamefont{C.}~\bibnamefont{Wu}},
  \bibinfo{author}{\bibfnamefont{B.~A.} \bibnamefont{Bernevig}},
  \bibnamefont{and} \bibinfo{author}{\bibfnamefont{S.-C.} \bibnamefont{Zhang}},
  \bibinfo{journal}{Phys. Rev. Lett.} \textbf{\bibinfo{volume}{96}},
  \bibinfo{pages}{106401} (\bibinfo{year}{2006}),
  \urlprefix\url{https://link.aps.org/doi/10.1103/PhysRevLett.96.106401}.

\bibitem[{\citenamefont{Semenoff}(1984)}]{Semenoff}
\bibinfo{author}{\bibfnamefont{G.~W.} \bibnamefont{Semenoff}},
  \bibinfo{journal}{Phys. Rev. Lett.} \textbf{\bibinfo{volume}{53}},
  \bibinfo{pages}{2449} (\bibinfo{year}{1984}),
  \urlprefix\url{https://link.aps.org/doi/10.1103/PhysRevLett.53.2449}.

\bibitem[{\citenamefont{DiVincenzo and Mele}(1984)}]{MeleDirac}
\bibinfo{author}{\bibfnamefont{D.~P.} \bibnamefont{DiVincenzo}}
  \bibnamefont{and} \bibinfo{author}{\bibfnamefont{E.~J.} \bibnamefont{Mele}},
  \bibinfo{journal}{Phys. Rev. B} \textbf{\bibinfo{volume}{29}},
  \bibinfo{pages}{1685} (\bibinfo{year}{1984}),
  \urlprefix\url{https://link.aps.org/doi/10.1103/PhysRevB.29.1685}.

\bibitem[{\citenamefont{Hsieh et~al.}(2008)\citenamefont{Hsieh, Qian, Wray,
  Xia, Hor, Cava, and Hasan}}]{Hsieh2008}
\bibinfo{author}{\bibfnamefont{D.}~\bibnamefont{Hsieh}},
  \bibinfo{author}{\bibfnamefont{D.}~\bibnamefont{Qian}},
  \bibinfo{author}{\bibfnamefont{L.}~\bibnamefont{Wray}},
  \bibinfo{author}{\bibfnamefont{Y.}~\bibnamefont{Xia}},
  \bibinfo{author}{\bibfnamefont{Y.~S.} \bibnamefont{Hor}},
  \bibinfo{author}{\bibfnamefont{R.~J.} \bibnamefont{Cava}}, \bibnamefont{and}
  \bibinfo{author}{\bibfnamefont{M.~Z.} \bibnamefont{Hasan}},
  \bibinfo{journal}{Nature} \textbf{\bibinfo{volume}{452}},
  \bibinfo{pages}{970} (\bibinfo{year}{2008}).

\bibitem[{\citenamefont{Liu et~al.}(2014{\natexlab{b}})\citenamefont{Liu, Zhou,
  Zhang, Wang, Weng, Prabhakaran, Mo, Shen, Fang, Dai et~al.}}]{Dirac2}
\bibinfo{author}{\bibfnamefont{Z.~K.} \bibnamefont{Liu}},
  \bibinfo{author}{\bibfnamefont{B.}~\bibnamefont{Zhou}},
  \bibinfo{author}{\bibfnamefont{Y.}~\bibnamefont{Zhang}},
  \bibinfo{author}{\bibfnamefont{Z.~J.} \bibnamefont{Wang}},
  \bibinfo{author}{\bibfnamefont{H.~M.} \bibnamefont{Weng}},
  \bibinfo{author}{\bibfnamefont{D.}~\bibnamefont{Prabhakaran}},
  \bibinfo{author}{\bibfnamefont{S.-K.} \bibnamefont{Mo}},
  \bibinfo{author}{\bibfnamefont{Z.~X.} \bibnamefont{Shen}},
  \bibinfo{author}{\bibfnamefont{Z.}~\bibnamefont{Fang}},
  \bibinfo{author}{\bibfnamefont{X.}~\bibnamefont{Dai}}, \bibnamefont{et~al.},
  \bibinfo{journal}{Science}  (\bibinfo{year}{2014}{\natexlab{b}}), ISSN
  \bibinfo{issn}{0036-8075},
  \urlprefix\url{http://science.sciencemag.org/content/early/2014/01/15/science.1245085}.

\bibitem[{\citenamefont{Borisenko et~al.}(2014)\citenamefont{Borisenko, Gibson,
  Evtushinsky, Zabolotnyy, B\"uchner, and Cava}}]{Dirac1}
\bibinfo{author}{\bibfnamefont{S.}~\bibnamefont{Borisenko}},
  \bibinfo{author}{\bibfnamefont{Q.}~\bibnamefont{Gibson}},
  \bibinfo{author}{\bibfnamefont{D.}~\bibnamefont{Evtushinsky}},
  \bibinfo{author}{\bibfnamefont{V.}~\bibnamefont{Zabolotnyy}},
  \bibinfo{author}{\bibfnamefont{B.}~\bibnamefont{B\"uchner}},
  \bibnamefont{and} \bibinfo{author}{\bibfnamefont{R.~J.} \bibnamefont{Cava}},
  \bibinfo{journal}{Phys. Rev. Lett.} \textbf{\bibinfo{volume}{113}},
  \bibinfo{pages}{027603} (\bibinfo{year}{2014}),
  \urlprefix\url{http://link.aps.org/doi/10.1103/PhysRevLett.113.027603}.

\bibitem[{\citenamefont{Weinberg et~al.}(1995)\citenamefont{Weinberg, S, and
  de~campos}}]{Weinberg}
\bibinfo{author}{\bibfnamefont{S.}~\bibnamefont{Weinberg}},
  \bibinfo{author}{\bibfnamefont{W.}~\bibnamefont{S}}, \bibnamefont{and}
  \bibinfo{author}{\bibfnamefont{T.}~\bibnamefont{de~campos}},
  \emph{\bibinfo{title}{The Quantum Theory of Fields}}, no. \bibinfo{number}{v.
  1} in \bibinfo{series}{Quantum Theory of Fields, Vol. 2: Modern Applications}
  (\bibinfo{publisher}{Cambridge University Press}, \bibinfo{year}{1995}), ISBN
  \bibinfo{isbn}{9780521550017},
  \urlprefix\url{https://books.google.be/books?id=doeDB3\_WLvwC}.

\bibitem[{\citenamefont{Alvarez-Gaum\'{e} and Witten}(1984)}]{AG1984}
\bibinfo{author}{\bibfnamefont{L.}~\bibnamefont{Alvarez-Gaum\'{e}}}
  \bibnamefont{and} \bibinfo{author}{\bibfnamefont{E.}~\bibnamefont{Witten}},
  \bibinfo{journal}{Nucl. Phys. B} \textbf{\bibinfo{volume}{234}},
  \bibinfo{pages}{269} (\bibinfo{year}{1984}).

\bibitem[{\citenamefont{Redlich}(1984)}]{Redlich1984}
\bibinfo{author}{\bibfnamefont{A.~N.} \bibnamefont{Redlich}},
  \bibinfo{journal}{Phys. Rev. Lett.} \textbf{\bibinfo{volume}{52}},
  \bibinfo{pages}{18} (\bibinfo{year}{1984}).

\bibitem[{\citenamefont{Jackiw}(1984)}]{Jackiw1984}
\bibinfo{author}{\bibfnamefont{R.}~\bibnamefont{Jackiw}},
  \bibinfo{journal}{Phys. Rev. D} \textbf{\bibinfo{volume}{29}},
  \bibinfo{pages}{2375} (\bibinfo{year}{1984}).

\bibitem[{\citenamefont{Qi et~al.}(2008)\citenamefont{Qi, Hughes, and
  Zhang}}]{Qi2008}
\bibinfo{author}{\bibfnamefont{X.~L.} \bibnamefont{Qi}},
  \bibinfo{author}{\bibfnamefont{T.~L.} \bibnamefont{Hughes}},
  \bibnamefont{and} \bibinfo{author}{\bibfnamefont{S.-C.} \bibnamefont{Zhang}},
  \bibinfo{journal}{Phys. Rev. B} \textbf{\bibinfo{volume}{78}},
  \bibinfo{pages}{195424} (\bibinfo{year}{2008}).

\bibitem[{\citenamefont{Mulligan and Burnell}(2013)}]{Mulligan2012}
\bibinfo{author}{\bibfnamefont{M.}~\bibnamefont{Mulligan}} \bibnamefont{and}
  \bibinfo{author}{\bibfnamefont{F.~J.} \bibnamefont{Burnell}},
  \bibinfo{journal}{Phys. Rev. B} \textbf{\bibinfo{volume}{88}},
  \bibinfo{pages}{085104} (\bibinfo{year}{2013}).

\bibitem[{\citenamefont{Alvarez-Gaum\'{e}
  et~al.}(1985)\citenamefont{Alvarez-Gaum\'{e}, Della~Pietra, and
  Moore}}]{AG1985}
\bibinfo{author}{\bibfnamefont{L.}~\bibnamefont{Alvarez-Gaum\'{e}}},
  \bibinfo{author}{\bibfnamefont{S.}~\bibnamefont{Della~Pietra}},
  \bibnamefont{and} \bibinfo{author}{\bibfnamefont{G.}~\bibnamefont{Moore}},
  \bibinfo{journal}{Ann. Phys.} \textbf{\bibinfo{volume}{163}},
  \bibinfo{pages}{288} (\bibinfo{year}{1985}).

\bibitem[{\citenamefont{Nielsen and Ninomiya}(1981)}]{NielsenNinomiya1}
\bibinfo{author}{\bibfnamefont{H.~B.} \bibnamefont{Nielsen}} \bibnamefont{and}
  \bibinfo{author}{\bibfnamefont{M.}~\bibnamefont{Ninomiya}},
  \bibinfo{journal}{Nucl. Phys. B} \textbf{\bibinfo{volume}{185}},
  \bibinfo{pages}{20} (\bibinfo{year}{1981}).

\bibitem[{\citenamefont{Fu et~al.}(2007)\citenamefont{Fu, Kane, and
  Mele}}]{Kane3DTI}
\bibinfo{author}{\bibfnamefont{L.}~\bibnamefont{Fu}},
  \bibinfo{author}{\bibfnamefont{C.~L.} \bibnamefont{Kane}}, \bibnamefont{and}
  \bibinfo{author}{\bibfnamefont{E.~J.} \bibnamefont{Mele}},
  \bibinfo{journal}{Phys. Rev. Lett.} \textbf{\bibinfo{volume}{98}},
  \bibinfo{pages}{106803} (\bibinfo{year}{2007}),
  \urlprefix\url{http://link.aps.org/doi/10.1103/PhysRevLett.98.106803}.

\bibitem[{\citenamefont{Yang and Nagaosa}(2014)}]{Yang2014}
\bibinfo{author}{\bibfnamefont{B.-J.} \bibnamefont{Yang}} \bibnamefont{and}
  \bibinfo{author}{\bibfnamefont{N.}~\bibnamefont{Nagaosa}},
  \bibinfo{journal}{Nat. Comm.} \textbf{\bibinfo{volume}{5}},
  \bibinfo{pages}{4898} (\bibinfo{year}{2014}).

\bibitem[{\citenamefont{Bradley and Cracknell}(1972)}]{BigBook}
\bibinfo{author}{\bibfnamefont{C.~J.} \bibnamefont{Bradley}} \bibnamefont{and}
  \bibinfo{author}{\bibfnamefont{A.~P.} \bibnamefont{Cracknell}},
  \emph{\bibinfo{title}{The Mathematical Theory of Symmetry in Solids}}
  (\bibinfo{publisher}{Clarendon Press Oxford}, \bibinfo{address}{Oxford,
  United Kingdom}, \bibinfo{year}{1972}), ISBN \bibinfo{isbn}{0199582580}.

\bibitem[{\citenamefont{Bernevig and Hughes}(2013)}]{bernevigbook}
\bibinfo{author}{\bibfnamefont{B.~A.} \bibnamefont{Bernevig}} \bibnamefont{and}
  \bibinfo{author}{\bibfnamefont{T.~L.} \bibnamefont{Hughes}},
  \emph{\bibinfo{title}{Topological Insulators and Topological
  Superconductors}} (\bibinfo{publisher}{Princeton University Press},
  \bibinfo{address}{Princeton, NJ}, \bibinfo{year}{2013}).

\bibitem[{\citenamefont{Kane and Mele}(2005)}]{Kane05}
\bibinfo{author}{\bibfnamefont{C.~L.} \bibnamefont{Kane}} \bibnamefont{and}
  \bibinfo{author}{\bibfnamefont{E.~J.} \bibnamefont{Mele}},
  \bibinfo{journal}{Phys. Rev. Lett.} \textbf{\bibinfo{volume}{95}},
  \bibinfo{pages}{146802} (\bibinfo{year}{2005}).

\bibitem[{\citenamefont{Ma\~nes}(2012)}]{manes}
\bibinfo{author}{\bibfnamefont{J.~L.} \bibnamefont{Ma\~nes}},
  \bibinfo{journal}{Phys. Rev. B} \textbf{\bibinfo{volume}{85}},
  \bibinfo{pages}{155118} (\bibinfo{year}{2012}),
  \urlprefix\url{https://link.aps.org/doi/10.1103/PhysRevB.85.155118}.

\bibitem[{\citenamefont{Chang et~al.}(2017{\natexlab{b}})\citenamefont{Chang,
  Xu, Wieder, Sanchez, Huang, Belopolski, Chang, Zhang, Bansil, Lin
  et~al.}}]{RhSi}
\bibinfo{author}{\bibfnamefont{G.}~\bibnamefont{Chang}},
  \bibinfo{author}{\bibfnamefont{S.-Y.} \bibnamefont{Xu}},
  \bibinfo{author}{\bibfnamefont{B.~J.} \bibnamefont{Wieder}},
  \bibinfo{author}{\bibfnamefont{D.~S.} \bibnamefont{Sanchez}},
  \bibinfo{author}{\bibfnamefont{S.-M.} \bibnamefont{Huang}},
  \bibinfo{author}{\bibfnamefont{I.}~\bibnamefont{Belopolski}},
  \bibinfo{author}{\bibfnamefont{T.-R.} \bibnamefont{Chang}},
  \bibinfo{author}{\bibfnamefont{S.}~\bibnamefont{Zhang}},
  \bibinfo{author}{\bibfnamefont{A.}~\bibnamefont{Bansil}},
  \bibinfo{author}{\bibfnamefont{H.}~\bibnamefont{Lin}}, \bibnamefont{et~al.},
  \bibinfo{journal}{Phys. Rev. Lett.} \textbf{\bibinfo{volume}{119}},
  \bibinfo{pages}{206401} (\bibinfo{year}{2017}{\natexlab{b}}),
  \urlprefix\url{https://link.aps.org/doi/10.1103/PhysRevLett.119.206401}.

\bibitem[{\citenamefont{Fulga and Stern}(2017)}]{AdyTriple}
\bibinfo{author}{\bibfnamefont{I.~C.} \bibnamefont{Fulga}} \bibnamefont{and}
  \bibinfo{author}{\bibfnamefont{A.}~\bibnamefont{Stern}},
  \bibinfo{journal}{Phys. Rev. B} \textbf{\bibinfo{volume}{95}},
  \bibinfo{pages}{241116} (\bibinfo{year}{2017}),
  \urlprefix\url{https://link.aps.org/doi/10.1103/PhysRevB.95.241116}.

\bibitem[{\citenamefont{Wang et~al.}(2012)\citenamefont{Wang, Sun, Chen,
  Franchini, Xu, Weng, Dai, and Fang}}]{Wang12}
\bibinfo{author}{\bibfnamefont{Z.}~\bibnamefont{Wang}},
  \bibinfo{author}{\bibfnamefont{Y.}~\bibnamefont{Sun}},
  \bibinfo{author}{\bibfnamefont{X.-Q.} \bibnamefont{Chen}},
  \bibinfo{author}{\bibfnamefont{C.}~\bibnamefont{Franchini}},
  \bibinfo{author}{\bibfnamefont{G.}~\bibnamefont{Xu}},
  \bibinfo{author}{\bibfnamefont{H.}~\bibnamefont{Weng}},
  \bibinfo{author}{\bibfnamefont{X.}~\bibnamefont{Dai}}, \bibnamefont{and}
  \bibinfo{author}{\bibfnamefont{Z.}~\bibnamefont{Fang}},
  \bibinfo{journal}{Phys. Rev. B} \textbf{\bibinfo{volume}{85}},
  \bibinfo{pages}{195320} (\bibinfo{year}{2012}),
  \urlprefix\url{http://link.aps.org/doi/10.1103/PhysRevB.85.195320}.

\bibitem[{\citenamefont{Young and Wieder}(2017)}]{SteveMagnet}
\bibinfo{author}{\bibfnamefont{S.~M.} \bibnamefont{Young}} \bibnamefont{and}
  \bibinfo{author}{\bibfnamefont{B.~J.} \bibnamefont{Wieder}},
  \bibinfo{journal}{Phys. Rev. Lett.} \textbf{\bibinfo{volume}{118}},
  \bibinfo{pages}{186401} (\bibinfo{year}{2017}),
  \urlprefix\url{https://link.aps.org/doi/10.1103/PhysRevLett.118.186401}.

\bibitem[{\citenamefont{Teo and Kane}(2009)}]{TeoQPC}
\bibinfo{author}{\bibfnamefont{J.~C.~Y.} \bibnamefont{Teo}} \bibnamefont{and}
  \bibinfo{author}{\bibfnamefont{C.~L.} \bibnamefont{Kane}},
  \bibinfo{journal}{Phys. Rev. B} \textbf{\bibinfo{volume}{79}},
  \bibinfo{pages}{235321} (\bibinfo{year}{2009}),
  \urlprefix\url{http://link.aps.org/doi/10.1103/PhysRevB.79.235321}.

\bibitem[{\citenamefont{Castro~Neto et~al.}(2009)\citenamefont{Castro~Neto,
  Guinea, Peres, Novoselov, and Geim}}]{BLG1}
\bibinfo{author}{\bibfnamefont{A.~H.} \bibnamefont{Castro~Neto}},
  \bibinfo{author}{\bibfnamefont{F.}~\bibnamefont{Guinea}},
  \bibinfo{author}{\bibfnamefont{N.~M.~R.} \bibnamefont{Peres}},
  \bibinfo{author}{\bibfnamefont{K.~S.} \bibnamefont{Novoselov}},
  \bibnamefont{and} \bibinfo{author}{\bibfnamefont{A.~K.} \bibnamefont{Geim}},
  \bibinfo{journal}{Rev. Mod. Phys.} \textbf{\bibinfo{volume}{81}},
  \bibinfo{pages}{109} (\bibinfo{year}{2009}),
  \urlprefix\url{http://link.aps.org/doi/10.1103/RevModPhys.81.109}.

\bibitem[{\citenamefont{McCann and Fal'ko}(2006)}]{BLG2}
\bibinfo{author}{\bibfnamefont{E.}~\bibnamefont{McCann}} \bibnamefont{and}
  \bibinfo{author}{\bibfnamefont{V.~I.} \bibnamefont{Fal'ko}},
  \bibinfo{journal}{Phys. Rev. Lett.} \textbf{\bibinfo{volume}{96}},
  \bibinfo{pages}{086805} (\bibinfo{year}{2006}),
  \urlprefix\url{http://link.aps.org/doi/10.1103/PhysRevLett.96.086805}.

\bibitem[{\citenamefont{Killi et~al.}(2010)\citenamefont{Killi, Wei, Affleck,
  and Paramekanti}}]{GrapheneLuttinger}
\bibinfo{author}{\bibfnamefont{M.}~\bibnamefont{Killi}},
  \bibinfo{author}{\bibfnamefont{T.-C.} \bibnamefont{Wei}},
  \bibinfo{author}{\bibfnamefont{I.}~\bibnamefont{Affleck}}, \bibnamefont{and}
  \bibinfo{author}{\bibfnamefont{A.}~\bibnamefont{Paramekanti}},
  \bibinfo{journal}{Phys. Rev. Lett.} \textbf{\bibinfo{volume}{104}},
  \bibinfo{pages}{216406} (\bibinfo{year}{2010}),
  \urlprefix\url{http://link.aps.org/doi/10.1103/PhysRevLett.104.216406}.

\bibitem[{\citenamefont{Wieder et~al.}(2015)\citenamefont{Wieder, Zhang, and
  Kane}}]{BLGWieder}
\bibinfo{author}{\bibfnamefont{B.~J.} \bibnamefont{Wieder}},
  \bibinfo{author}{\bibfnamefont{F.}~\bibnamefont{Zhang}}, \bibnamefont{and}
  \bibinfo{author}{\bibfnamefont{C.~L.} \bibnamefont{Kane}},
  \bibinfo{journal}{Phys. Rev. B} \textbf{\bibinfo{volume}{92}},
  \bibinfo{pages}{085425} (\bibinfo{year}{2015}),
  \urlprefix\url{http://link.aps.org/doi/10.1103/PhysRevB.92.085425}.

\bibitem[{\citenamefont{Ju et~al.}(2015)\citenamefont{Ju, Shi, Nair, Lv, Jin,
  Velasco~Jr, Ojeda-Aristizabal, Bechtel, Martin, Zettl et~al.}}]{BLGNature}
\bibinfo{author}{\bibfnamefont{L.}~\bibnamefont{Ju}},
  \bibinfo{author}{\bibfnamefont{Z.}~\bibnamefont{Shi}},
  \bibinfo{author}{\bibfnamefont{N.}~\bibnamefont{Nair}},
  \bibinfo{author}{\bibfnamefont{Y.}~\bibnamefont{Lv}},
  \bibinfo{author}{\bibfnamefont{C.}~\bibnamefont{Jin}},
  \bibinfo{author}{\bibfnamefont{J.}~\bibnamefont{Velasco~Jr}},
  \bibinfo{author}{\bibfnamefont{C.}~\bibnamefont{Ojeda-Aristizabal}},
  \bibinfo{author}{\bibfnamefont{H.~A.} \bibnamefont{Bechtel}},
  \bibinfo{author}{\bibfnamefont{M.~C.} \bibnamefont{Martin}},
  \bibinfo{author}{\bibfnamefont{A.}~\bibnamefont{Zettl}},
  \bibnamefont{et~al.}, \bibinfo{journal}{Nature}
  \textbf{\bibinfo{volume}{520}}, \bibinfo{pages}{650} (\bibinfo{year}{2015}),
  ISSN \bibinfo{issn}{0028-0836},
  \urlprefix\url{http://dx.doi.org/10.1038/nature14364}.

\bibitem[{\citenamefont{Li et~al.}(2016)\citenamefont{Li, Wang, McFaul, Zern,
  Ren, Watanabe, Taniguchi, Qiao, and Zhu}}]{BLGNano}
\bibinfo{author}{\bibfnamefont{J.}~\bibnamefont{Li}},
  \bibinfo{author}{\bibfnamefont{K.}~\bibnamefont{Wang}},
  \bibinfo{author}{\bibfnamefont{K.~J.} \bibnamefont{McFaul}},
  \bibinfo{author}{\bibfnamefont{Z.}~\bibnamefont{Zern}},
  \bibinfo{author}{\bibfnamefont{Y.}~\bibnamefont{Ren}},
  \bibinfo{author}{\bibfnamefont{K.}~\bibnamefont{Watanabe}},
  \bibinfo{author}{\bibfnamefont{T.}~\bibnamefont{Taniguchi}},
  \bibinfo{author}{\bibfnamefont{Z.}~\bibnamefont{Qiao}}, \bibnamefont{and}
  \bibinfo{author}{\bibfnamefont{J.}~\bibnamefont{Zhu}}, \bibinfo{journal}{Nat
  Nano} \textbf{\bibinfo{volume}{advance online publication}}
  (\bibinfo{year}{2016}), ISSN \bibinfo{issn}{1748-3395},
  \bibinfo{note}{letter},
  \urlprefix\url{http://dx.doi.org/10.1038/nnano.2016.158}.

\bibitem[{\citenamefont{Zhang and Kane}(2014)}]{Fan14}
\bibinfo{author}{\bibfnamefont{F.}~\bibnamefont{Zhang}} \bibnamefont{and}
  \bibinfo{author}{\bibfnamefont{C.~L.} \bibnamefont{Kane}},
  \bibinfo{journal}{Phys. Rev. B} \textbf{\bibinfo{volume}{90}},
  \bibinfo{pages}{020501} (\bibinfo{year}{2014}),
  \urlprefix\url{http://link.aps.org/doi/10.1103/PhysRevB.90.020501}.

\bibitem[{\citenamefont{Alexandradinata
  et~al.}(2014{\natexlab{b}})\citenamefont{Alexandradinata, Fang, Gilbert, and
  Bernevig}}]{Alexandradinata14a}
\bibinfo{author}{\bibfnamefont{A.}~\bibnamefont{Alexandradinata}},
  \bibinfo{author}{\bibfnamefont{C.}~\bibnamefont{Fang}},
  \bibinfo{author}{\bibfnamefont{M.~J.} \bibnamefont{Gilbert}},
  \bibnamefont{and} \bibinfo{author}{\bibfnamefont{B.~A.}
  \bibnamefont{Bernevig}}, \bibinfo{journal}{Phys. Rev. Lett.}
  \textbf{\bibinfo{volume}{113}}, \bibinfo{pages}{116403}
  (\bibinfo{year}{2014}{\natexlab{b}}),
  \urlprefix\url{http://link.aps.org/doi/10.1103/PhysRevLett.113.116403}.

\bibitem[{\citenamefont{Jain et~al.}(2013)\citenamefont{Jain, Ong, Hautier,
  Chen, Richards, Dacek, Cholia, Gunter, Skinner, Ceder et~al.}}]{MatProject}
\bibinfo{author}{\bibfnamefont{A.}~\bibnamefont{Jain}},
  \bibinfo{author}{\bibfnamefont{S.~P.} \bibnamefont{Ong}},
  \bibinfo{author}{\bibfnamefont{G.}~\bibnamefont{Hautier}},
  \bibinfo{author}{\bibfnamefont{W.}~\bibnamefont{Chen}},
  \bibinfo{author}{\bibfnamefont{W.~D.} \bibnamefont{Richards}},
  \bibinfo{author}{\bibfnamefont{S.}~\bibnamefont{Dacek}},
  \bibinfo{author}{\bibfnamefont{S.}~\bibnamefont{Cholia}},
  \bibinfo{author}{\bibfnamefont{D.}~\bibnamefont{Gunter}},
  \bibinfo{author}{\bibfnamefont{D.}~\bibnamefont{Skinner}},
  \bibinfo{author}{\bibfnamefont{G.}~\bibnamefont{Ceder}},
  \bibnamefont{et~al.}, \bibinfo{journal}{APL Materials}
  \textbf{\bibinfo{volume}{1}}, \bibinfo{pages}{011002} (\bibinfo{year}{2013}),
  \eprint{https://doi.org/10.1063/1.4812323},
  \urlprefix\url{https://doi.org/10.1063/1.4812323}.

\bibitem[{\citenamefont{{Inorganic Crystal Structure Database
  (ICSD)}}(Fachinformationszentrum Karlsruhe, Karlsruhe, Germany, 2015)}]{ICSD}
\bibinfo{author}{\bibnamefont{{Inorganic Crystal Structure Database (ICSD)}}}
  (\bibinfo{year}{Fachinformationszentrum Karlsruhe, Karlsruhe, Germany,
  2015}).

\bibitem[{\citenamefont{Bl\"ochl}(1994)}]{PAW}
\bibinfo{author}{\bibfnamefont{P.~E.} \bibnamefont{Bl\"ochl}},
  \bibinfo{journal}{Phys. Rev. B} \textbf{\bibinfo{volume}{50}},
  \bibinfo{pages}{17953} (\bibinfo{year}{1994}),
  \urlprefix\url{http://link.aps.org/doi/10.1103/PhysRevB.50.17953}.

\bibitem[{\citenamefont{Kresse and Furthm\"uller}(1996)}]{VASP}
\bibinfo{author}{\bibfnamefont{G.}~\bibnamefont{Kresse}} \bibnamefont{and}
  \bibinfo{author}{\bibfnamefont{J.}~\bibnamefont{Furthm\"uller}},
  \bibinfo{journal}{Phys. Rev. B} \textbf{\bibinfo{volume}{54}},
  \bibinfo{pages}{11169} (\bibinfo{year}{1996}),
  \urlprefix\url{http://link.aps.org/doi/10.1103/PhysRevB.54.11169}.

\bibitem[{\citenamefont{Perdew et~al.}(1996)\citenamefont{Perdew, Burke, and
  Ernzerhof}}]{Perdew96p3865}
\bibinfo{author}{\bibfnamefont{J.~P.} \bibnamefont{Perdew}},
  \bibinfo{author}{\bibfnamefont{K.}~\bibnamefont{Burke}}, \bibnamefont{and}
  \bibinfo{author}{\bibfnamefont{M.}~\bibnamefont{Ernzerhof}},
  \bibinfo{journal}{Phys. Rev. Lett.} \textbf{\bibinfo{volume}{77}},
  \bibinfo{pages}{3865} (\bibinfo{year}{1996}),
  \urlprefix\url{http://link.aps.org/doi/10.1103/PhysRevLett.77.3865}.

\bibitem[{\citenamefont{Marzari et~al.}(2012)\citenamefont{Marzari, Mostofi,
  Yates, Souza, and Vanderbilt}}]{ZJWannier}
\bibinfo{author}{\bibfnamefont{N.}~\bibnamefont{Marzari}},
  \bibinfo{author}{\bibfnamefont{A.~A.} \bibnamefont{Mostofi}},
  \bibinfo{author}{\bibfnamefont{J.~R.} \bibnamefont{Yates}},
  \bibinfo{author}{\bibfnamefont{I.}~\bibnamefont{Souza}}, \bibnamefont{and}
  \bibinfo{author}{\bibfnamefont{D.}~\bibnamefont{Vanderbilt}},
  \bibinfo{journal}{Rev. Mod. Phys.} \textbf{\bibinfo{volume}{84}},
  \bibinfo{pages}{1419} (\bibinfo{year}{2012}),
  \urlprefix\url{http://link.aps.org/doi/10.1103/RevModPhys.84.1419}.

\bibitem[{\citenamefont{Marzari and Vanderbilt}(1997)}]{Marzari97p12847}
\bibinfo{author}{\bibfnamefont{N.}~\bibnamefont{Marzari}} \bibnamefont{and}
  \bibinfo{author}{\bibfnamefont{D.}~\bibnamefont{Vanderbilt}},
  \bibinfo{journal}{Phys. Rev. B} \textbf{\bibinfo{volume}{56}},
  \bibinfo{pages}{12847} (\bibinfo{year}{1997}),
  \urlprefix\url{http://link.aps.org/doi/10.1103/PhysRevB.56.12847}.

\bibitem[{\citenamefont{Mostofi et~al.}(2008)\citenamefont{Mostofi, Yates, Lee,
  Souza, Vanderbilt, and Marzari}}]{Mostofi08p685}
\bibinfo{author}{\bibfnamefont{A.~A.} \bibnamefont{Mostofi}},
  \bibinfo{author}{\bibfnamefont{J.~R.} \bibnamefont{Yates}},
  \bibinfo{author}{\bibfnamefont{Y.-S.} \bibnamefont{Lee}},
  \bibinfo{author}{\bibfnamefont{I.}~\bibnamefont{Souza}},
  \bibinfo{author}{\bibfnamefont{D.}~\bibnamefont{Vanderbilt}},
  \bibnamefont{and} \bibinfo{author}{\bibfnamefont{N.}~\bibnamefont{Marzari}},
  \bibinfo{journal}{Computer physics communications}
  \textbf{\bibinfo{volume}{178}}, \bibinfo{pages}{685} (\bibinfo{year}{2008}).

\bibitem[{\citenamefont{Souza et~al.}(2001)\citenamefont{Souza, Marzari, and
  Vanderbilt}}]{Souza01p035109}
\bibinfo{author}{\bibfnamefont{I.}~\bibnamefont{Souza}},
  \bibinfo{author}{\bibfnamefont{N.}~\bibnamefont{Marzari}}, \bibnamefont{and}
  \bibinfo{author}{\bibfnamefont{D.}~\bibnamefont{Vanderbilt}},
  \bibinfo{journal}{Phys. Rev. B} \textbf{\bibinfo{volume}{65}},
  \bibinfo{pages}{035109} (\bibinfo{year}{2001}),
  \urlprefix\url{http://link.aps.org/doi/10.1103/PhysRevB.65.035109}.

\bibitem[{\citenamefont{Giannozzi et~al.}(2009)}]{Giannozzi09p395502}
\bibinfo{author}{\bibfnamefont{P.}~\bibnamefont{Giannozzi}}
  \bibnamefont{et~al.}, \bibinfo{journal}{J. Phys. Condens. Mat.}
  \textbf{\bibinfo{volume}{21}}, \bibinfo{pages}{395502}
  (\bibinfo{year}{2009}).

\bibitem[{\citenamefont{Rappe et~al.}(1990)\citenamefont{Rappe, Rabe, Kaxiras,
  and Joannopoulos}}]{Rappe90p1227}
\bibinfo{author}{\bibfnamefont{A.~M.} \bibnamefont{Rappe}},
  \bibinfo{author}{\bibfnamefont{K.~M.} \bibnamefont{Rabe}},
  \bibinfo{author}{\bibfnamefont{E.}~\bibnamefont{Kaxiras}}, \bibnamefont{and}
  \bibinfo{author}{\bibfnamefont{J.~D.} \bibnamefont{Joannopoulos}},
  \bibinfo{journal}{Phys. Rev. B} \textbf{\bibinfo{volume}{41}},
  \bibinfo{pages}{1227} (\bibinfo{year}{1990}).

\bibitem[{\citenamefont{Ramer and Rappe}(1999)}]{Ramer99p12471}
\bibinfo{author}{\bibfnamefont{N.~J.} \bibnamefont{Ramer}} \bibnamefont{and}
  \bibinfo{author}{\bibfnamefont{A.~M.} \bibnamefont{Rappe}},
  \bibinfo{journal}{Phys. Rev. B} \textbf{\bibinfo{volume}{59}},
  \bibinfo{pages}{12471} (\bibinfo{year}{1999}).

\bibitem[{\citenamefont{Sancho et~al.}(1984)\citenamefont{Sancho, Sancho, and
  Rubio}}]{Sancho84p1205}
\bibinfo{author}{\bibfnamefont{M.~L.} \bibnamefont{Sancho}},
  \bibinfo{author}{\bibfnamefont{J.~L.} \bibnamefont{Sancho}},
  \bibnamefont{and} \bibinfo{author}{\bibfnamefont{J.}~\bibnamefont{Rubio}},
  \bibinfo{journal}{Journal of Physics F: Metal Physics}
  \textbf{\bibinfo{volume}{14}}, \bibinfo{pages}{1205} (\bibinfo{year}{1984}).

\bibitem[{\citenamefont{Sancho et~al.}(1985)\citenamefont{Sancho, Sancho,
  Sancho, and Rubio}}]{Sancho85p851}
\bibinfo{author}{\bibfnamefont{M.~L.} \bibnamefont{Sancho}},
  \bibinfo{author}{\bibfnamefont{J.~L.} \bibnamefont{Sancho}},
  \bibinfo{author}{\bibfnamefont{J.~L.} \bibnamefont{Sancho}},
  \bibnamefont{and} \bibinfo{author}{\bibfnamefont{J.}~\bibnamefont{Rubio}},
  \bibinfo{journal}{Journal of Physics F: Metal Physics}
  \textbf{\bibinfo{volume}{15}}, \bibinfo{pages}{851} (\bibinfo{year}{1985}).

\bibitem[{\citenamefont{Vafek and Vishwanath}(2014)}]{Vafek14}
\bibinfo{author}{\bibfnamefont{O.}~\bibnamefont{Vafek}} \bibnamefont{and}
  \bibinfo{author}{\bibfnamefont{A.}~\bibnamefont{Vishwanath}},
  \bibinfo{journal}{Ann. Rev. Condensed Matter Phys.}
  \textbf{\bibinfo{volume}{5}}, \bibinfo{pages}{83} (\bibinfo{year}{2014}).

\bibitem[{\citenamefont{Stokes and Hatch}(2005)}]{findsym}
\bibinfo{author}{\bibfnamefont{H.}~\bibnamefont{Stokes}} \bibnamefont{and}
  \bibinfo{author}{\bibfnamefont{D.~M.} \bibnamefont{Hatch}},
  \bibinfo{journal}{J. Appl. Cryst.} \textbf{\bibinfo{volume}{38}},
  \bibinfo{pages}{237} (\bibinfo{year}{2005}), ISSN \bibinfo{issn}{1600-5767}.

\end{thebibliography}
\end{document}